\documentclass[11pt]{article}
\RequirePackage[numbers]{natbib}
\usepackage[english]{babel}
\usepackage{amsthm,amsmath,amsfonts,amssymb,mathbbol,float,verbatim,latexsym,xcolor,comment,lmodern,graphicx,xr,hyperref,algpseudocode,algorithm,mathtools}

\graphicspath{{./fig/}}

\theoremstyle{plain}
\newtheorem{theorem}{Theorem}
\newtheorem{proposition}[theorem]{Proposition}%
\newtheorem{remark}{Remark}%
\newtheorem*{definition*}{Definition}

\newtheorem{lemma}{Lemma}
\newtheorem{Hassumption}{Assumption}
\newtheorem{Passumption}{Assumption}

\newcommand{\cdummy}{\cdot}
\newcommand{\nin}{\not\in}
\newcommand{\nosymbol}{}
\newcommand{\tmmathbf}[1]{{\boldsymbol{#1}}}
\newcommand{\tmop}[1]{{\operatorname{#1}}}
\newcommand{\tmtextbf}[1]{\text{{\bfseries{#1}}}}

\newcommand{\infixand}{\text{ and }}

\makeatletter
\newcommand*{\addFileDependency}[1]{
  \typeout{(#1)}
  \@addtofilelist{#1}
  \IfFileExists{#1}{}{\typeout{No file #1.}}
}
\makeatother

\usepackage{etoolbox}
\usepackage{ragged2e}
\usepackage[affil-it]{authblk}

\usepackage[paper=a4paper,dvips,top=2cm,left=2cm,right=2cm,
foot=1cm,bottom=2cm]{geometry}
\allowdisplaybreaks

\title{A Consistent Bayesian Adaptive Quotient Graph Framework for Graph-Informed Random Partition Models}

\author[1]{Kun Huang}
\author[2]{Xiyu Peng}
\author[3]{Huiyan Sang}

\affil[1]{Department of Statistics, Texas A\&M University, College Station, US. Email: k-huang@tamu.edu}
\affil[2]{Department of Statistics, Texas A\&M University, College Station, US. Email: pengx@stat.tamu.edu}
\affil[3]{Department of Statistics, Texas A\&M University, College Station, US. Email: huiyan@stat.tamu.edu}

\date{}

\begin{document}

\maketitle

\begin{abstract}
   Bayesian graph-informed random partition models are widely used to model heterogeneity in structured data. A prominent approach assigns a prior directly on the space of contiguous partitions of a pre-determined graph $\mathcal{G}$; however, such models are not data-adaptive and can be computationally infeasible for large graphs. 
To address these limitations, we propose a {Bayesian adaptive quotient graph framework} that induces partitions through a two-level hierarchical prior: a prior is first imposed on quotient graphs of $\mathcal{G}$, followed by a conditional prior on a reduced contiguous partition space induced by each quotient graph.  We develop high-level assumptions that guide the construction of the two-level prior such that the reduced space retains desirable partitions. We further establish high-level assumptions for posterior concentration on desirable quotient graphs and partitions. 
The framework can be readily embedded into many existing Bayesian hierarchical models with latent cluster structures as a contiguous random partition prior. As a specific example, we apply the proposed framework to the spatial random partition model proposed in \cite{huang2025consistent}. The resulting model adapts to the unknown number of clusters, achieves a faster clustering-consistency contraction rate, and yields substantially lower partition error and computation time in simulations. Finally, we apply the resulting model to a large spatial transcriptomics dataset to demonstrate its practical utility.
\end{abstract}








%

\newpage

\section{Introduction}

\subsection{Motivation}

Many real-world datasets exhibit intrinsic structure arising from relationships among observational units. Such relationships may reflect spatial proximity, temporal dependence, social interactions, or biological connections. Graphs provide a flexible representation of these relationships, giving rise to graph-structured data in a broad range of applications; see \cite{Battaglia2018,wu2020comprehensive}. An important task in analyzing such data is to partition the observational units into clusters whose members share similar latent characteristics or model parameters while respecting the underlying graph structure. This has motivated a broad class of graph-informed partitioning approaches, which use the graph information to constrain or guide the induced partitions; e.g., \cite{shen2012simultaneous,zhang2024learning}.

In particular, Bayesian graph-informed random partition models incorporate graph information into prior distributions on partitions, enabling posterior inference and uncertainty quantification \citep{yan2024bayesian,chakrabarti2024graphical}. Given a graph, a natural and widely used way to incorporate graph information is through a contiguity constraint. Let $\mathcal{G}=(\mathcal{V},\mathcal{E})$ denote a graph, where $\mathcal{V}$ and $\mathcal{E}$ denote the vertex set and edge set, respectively. A partition of $\mathcal{V}$ is said to be contiguous with respect to (w.r.t.) $\mathcal{G}$ if every cluster is contiguous, i.e., it induces a connected subgraph of $\mathcal{G}$. Numerous Bayesian approaches impose priors on the space of contiguous partitions for a pre-determined graph $\mathcal{G}$ and perform posterior inference over this space; see \cite{teixeira2019bayesian,come2024bayesian} for clustering models accommodating a range of response distributions, \cite{luo2023nonstationary} for nonstationary Gaussian process models, \cite{tonellato2026bayesian} for generalized linear dynamic models, and \cite{luo2021bayesian} for varying coefficient models, among others.




Such inference becomes increasingly challenging as modern data-collection technologies produce datasets with ever larger numbers of observational units. For example, spatial transcriptomic studies, where identifying spatially homogeneous clusters is an important goal, may involve approximately $10$ million spatially resolved cells \citep{zhang2023molecularly}. When these cells are represented as vertices of a graph, the corresponding space of contiguous partitions can be extremely large, with its size typically growing exponentially in vertex count \citep{vince2017counting}. This poses a fundamental challenge for Bayesian modeling and inference, as posterior exploration over the full partition space may become computationally prohibitive or even impractical. Motivated by this challenge, we propose a Bayesian framework that performs inference over a smaller, data-adaptive partition space while retaining the desirable partition. The idea is illustrated below.

\begin{figure}[htbp]
		\centering
		\begin{tabular}{cc}
			{\includegraphics[width=0.35\linewidth,height=0.18\textheight]{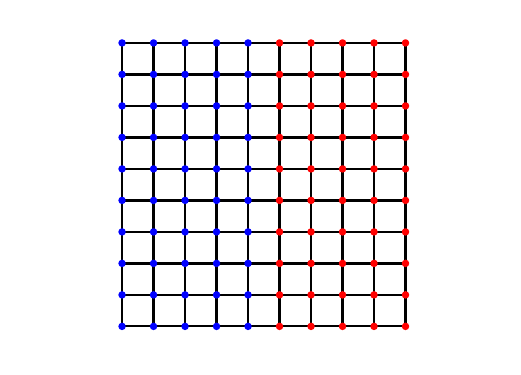}}&
			{\includegraphics[width=0.35\linewidth,height=0.18\textheight]{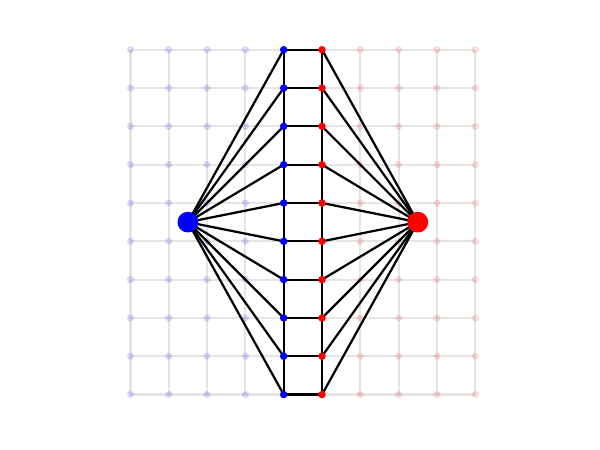}}
            \\
            {\small (a)} & {\small (b)}\\
		\end{tabular}
		\caption{{A motivating example.} (a) A mesh grid graph $\mathcal{G}$, in which the two true clusters are shown in blue and red, respectively. (b) A simple quotient graph constructed from $\mathcal{G}$. The vertices in the first four and last four columns are contracted into two quotient vertices, shown as the large blue and red vertices, respectively. The vertices in the fifth and sixth columns remain unchanged. The connectivity of the quotient graph is inherited from the original graph $\mathcal{G}$.}
		\label{FIG:motivation}
        \vspace{-10pt}
	\end{figure}

For simplicity, suppose that $\mathcal{G}$ is the mesh grid graph shown in Figure \ref{FIG:motivation}(a), and the model's goal is to partition its vertices into two contiguous clusters. The true clusters are represented by the two colors in Figure \ref{FIG:motivation}(a). Imposing a prior over the entire contiguous partition space may be computationally inefficient because most vertices lie within cluster interiors, whereas only vertices at the cluster boundaries determine how the graph is partitioned. Once these boundaries are identified, the cluster memberships of all graph vertices are determined. This motivates the idea of focusing the clustering analysis on vertices that are likely at the boundary. Figure \ref{FIG:motivation}(b) illustrates this idea using a quotient graph of $\mathcal{G}$, where vertices away from the boundary are contracted into two vertices, while vertices at the boundary remain unchanged. Partitioning vertices of this quotient graph can recover the same boundary as partitioning the original vertices of graph $\mathcal{G}$, while searching over a much smaller partition space.

Since the true boundary is unknown, the quotient graph must be learned from the data. This leads to the idea of using a two-level hierarchical prior for inducing partitions from $\mathcal{G}$. Specifically, at the first level, we impose a prior on quotient graphs of $\mathcal{G}$, based on which we   define a graph partition prior at the second level. The quotient graph at the first level contract original vertices, thereby reducing the partition space at the second level and allowing the inference to focus on identifying cluster boundaries. Since the posterior distribution of the quotient graph is informed from the data, it is data-adaptive, and, in turn, enables adaptive inference for partitions.


The adaptivity of the two-level partition construction also facilitates desirable theoretical guarantees. In some settings, the hyperparameter choices required to establish such guarantees depend on unknown quantities of the data-generating process \citep{zheng2024consistency,huang2025consistent}. Inference based on a fixed, pre-determined graph $\mathcal{G}$ cannot adapt to these unknown quantities and may therefore fail to satisfy the required conditions. By contrast, learning a graph from the data provides a way to achieve this adaptivity. We demonstrate this theoretical benefit for a concrete Bayesian spatial domain partition model in Section~\ref{SEC:AOS theory}.


\subsection{Related work}

The problem of learning graph structure from observed data has been extensively studied; see, e.g., \cite{voorman2014graph,peterson2015bayesian}. For clustering purposes, several methods have incorporated graph learning into the clustering procedure. Specifically, \cite{nie2014clustering} and \cite{jia2020clustering} jointly learn a similarity graph and the corresponding cluster structure, whereas \cite{banerjee2019spectral} adopts a Bayesian framework to first estimate the edge structure of an unknown graph, and then perform clustering based on its Laplacian embedding. These methods retain the original vertex set and primarily learn relationships among the original vertices. In contrast, our two-level approach learns which original vertices should be contracted to form a quotient graph, thereby reducing the partition space and computational burden for the subsequent clustering.

Although relatively underexplored in statistical modeling, coarsening a graph by contracting vertices has long been used in computer science, particularly in multilevel graph-partitioning algorithms, to improve computational efficiency. These methods contract vertices to construct a smaller graph, partition the smaller graph, and then map and refine the resulting partition on the original graph \citep{karypis1998fast}; see \cite{catalyurek2023more} for a recent review. Their goal is to solve a non-probabilistic graph-partitioning problem, with the coarsened graphs constructed as intermediate objects in an algorithmic procedure. This differs fundamentally from our Bayesian model-based clustering framework, in which the coarsened, or quotient, graph is treated as a random object and is itself an inferential target.

\subsection{Contribution}

We propose a {Bayesian adaptive quotient graph framework for contiguous random partition models}, built on the two-level hierarchical prior for inducing partitions. The framework is general in that this two-level prior can be readily incorporated into existing random partition models that induce contiguous partitions from a pre-determined graph $\mathcal{G}$. The framework enables data-adaptive inference while reducing the partition space and the associated computational burden. To the best of our knowledge, data-adaptive quotient graphs have not been previously studied for Bayesian partition inference.


Developing an effective framework presents two nontrivial challenges. First, one must specify a prior over quotient graphs such that partitions subsequently induced on these graphs remain contiguous w.r.t. the original graph $\mathcal{G}$, while still permitting efficient posterior sampling. To address this challenge, we construct quotient graphs from intermediate contiguous partitions obtained by cutting edges of a pre-determined spanning tree. This construction ensures that any contiguous partition w.r.t. the quotient graph corresponds to a contiguous partition w.r.t. the original graph $\mathcal{G}$. It also leads to a simple algorithm for drawing posterior samples of the quotient graph.

The second challenge is that, although a quotient graph reduces the partition space, the reduction must not be so aggressive that it excludes the desirable partition. It is therefore important to characterize how aggressively vertices can be contracted while retaining this partition. We formally study this issue and establish high-level assumptions under which the desirable partition is guaranteed to remain in the reduced space. These assumptions provide practical guidance for selecting hyperparameters when applying the proposed framework to a specific model.

Furthermore, we study the theoretical properties of the proposed framework, focusing on posterior concentration for both the quotient graph and resulting partitions. We formally define desirable sets for these two components and establish high-level assumptions under which their posterior concentrates on the corresponding desirable sets. These high-level assumptions provide a convenient route for establishing theoretical guarantees when applying the proposed framework to a specific model.

As a specific example to illustrate the proposed framework, we consider the spatial domain partition model (Spat-RPM)  \citep{huang2025consistent}, a prominent model-based spatial clustering method. Spatial clustering methods have been developed for broad applications such as economics and environmental science \citep{shenweining}. We develop an adaptive quotient graph version, referred to as adaptive Spat-RPM (A-Spat-RPM), and demonstrate substantial improvements over Spat-RPM. Using the established high-level assumptions, we show that A-Spat-RPM achieves partition consistency at a faster posterior contraction rate and adapts to the unknown number of clusters, a property not shared by Spat-RPM. Computationally, simulation studies show that A-Spat-RPM is approximately $20$ times faster than Spat-RPM while achieving significantly better clustering performance. 

The rest of the paper is organized as follows. Section~\ref{SEC:general graph} introduces some graph basics and formally introduces the proposed framework. Section~\ref{SEC:AOScaes} introduces Spat-RPM and demonstrate how the proposed framework can be incorporated into it to obtain A-Spat-RPM. Section~\ref{SEC:AOS theory} develops theoretical properties of A-Spat-RPM using high-level assumptions in Section \ref{SEC:general graph}. Section~\ref{SEC:simu} presents simulation studies that examine the performance of A-Spat-RPM and compare it with Spat-RPM. Section~\ref{SEC:realdata} applies A-Spat-RPM to a large spatial transcriptomic dataset to demonstrate its practical utility. Supplementary Material contains all technical proofs and additional model details.

\section{An adaptive graph construction framework and theory}
\label{SEC:general graph}

\subsection{Graph basics and notation for partitions}
\label{SEC:graphbasic}
Let $\mathcal{G} = (\mathcal{V}, \mathcal{E})$ be an undirected graph. A path from $u\in\mathcal{V}$ to $v\in\mathcal{V}$ is a sequence of vertices $v_0,v_1,\ldots,v_r$ in $\mathcal{V}$, such that $v_0=u$, $v_r=v$, and $\{v_{j-1},v_j\}\in\mathcal{E}$ for every $1\leq j\leq r$. A path is called a cycle if
${v}_{0} ={v}_{r}$ and all other vertices
are distinct. A subgraph $(\mathcal{V}_0, \mathcal{E}_0)$, where $\mathcal{V}_{0}
\subseteq \mathcal{V}$ and $\mathcal{E}_0 \subseteq \mathcal{E}$, is called a
connected component of $\mathcal{G}$ if there is a path between any two vertices and there is no path between any vertex in $\mathcal{V}_0$ and any vertex in
$\mathcal{V} \setminus \mathcal{V}_0$, the difference between $\mathcal{V}$ and $\mathcal{V}_0$. 

  

Throughout this paper, we assume that $\mathcal{G}=(\mathcal{V},\mathcal{E})$ is connected and simple; that is, it contains neither self-loops nor parallel edges. A spanning tree of $\mathcal{G}$ is a subgraph $\mathcal{T}=(\mathcal{V},\mathcal{E}_{\mathcal{T}})$, where 
$\mathcal{E}_{\mathcal{T}} \subseteq \mathcal{E}$ connects all vertices without any cycles. See Figure \ref{FIG:pidaggerillustration}(b) for an example of spanning trees of the graph in Figure \ref{FIG:pidaggerillustration}(a). Every spanning tree contains $|\mathcal{V}|-1$ edges, where $|\cdot|$ denotes cardinality.  One nice property of spanning trees is that, after cutting $k-1$ edges, it induces $k$ connected components, each of which can be treated as a contiguous cluster w.r.t. $\mathcal{G}$. Correspondingly, we say that a spanning tree $\mathcal{T}$ is compatible with a partition if it can be induced from $\mathcal{T}$ by cutting some of its edges.

The nice property of spanning trees has motivated the construction of random partition priors in many studies \citep{luo2021bayesian,teixeira2019bayesian,teixeira2015generative}. We next formally define the quotient graph, an important component in our proposed framework.
\begin{definition*}
  (Quotient Graph) Given $\mathcal{G} = (\mathcal{V}, \mathcal{E})$,
  let $\pi^{\dagger} = \{ C_1, \ldots, C_{k^{\dagger}} \}$ be a 
  partition of $\mathcal{V}$. The quotient multigraph of $\mathcal{G}$
  w.r.t. $\pi^{\dagger}$, denoted by $\mathcal{G}_{\tmop{multi}}/\pi^{\dagger}$,
  is the graph obtained by contracting each cluster $C_j$ into a vertex and retaining all edges between distinct clusters, including parallel edges. Edges whose endpoints lie in the same cluster are removed in $\mathcal{G}_{\tmop{multi}}/\pi^{\dagger}$. The simple quotient graph of $\mathcal{G}$
  w.r.t. $\pi^{\dagger}$, hereafter referred to as the quotient graph and denoted by $\mathcal{G} / \pi^{\dagger}$,
  is the graph obtained by replacing all parallel edges in $\mathcal{G}_{\tmop{multi}} / \pi^{\dagger}$ with a single edge. The vertices of $\mathcal{G}_{\tmop{multi}} / \pi^{\dagger}$ and $\mathcal{G} / \pi^{\dagger}$ are referred to as quotient vertices.
\end{definition*}

We conclude this section by introducing some notation regarding partitions. We use $\preceq$ to denote the nesting relation between partitions: for two partitions $\pi_1$ and $\pi_2$, $\pi_1 \preceq \pi_2$ means that $\pi_1$ is nested in $\pi_2$; that is, every cluster of $\pi_1$ is contained in some cluster of $\pi_2$. We use $\succeq$ to denote the reverse relation. For $\pi_1 \succeq \pi_2$, we use the notation $\pi_1 / \pi_2$ to denote the partition of the clusters of $\pi_2$ such that two clusters of $\pi_2 $ belong to the same cluster of $\pi_1 / \pi_2$ if and only if they are contained in the same cluster of $\pi_1$.


\subsection{A two-level hierarchical prior}
\label{SEC:twostep approach}
Given a graph $\mathcal{G}=(\mathcal{V},\mathcal{E})$, let $\pi^\ast$ denote a contiguous partition of $\mathcal{V}$ w.r.t. $\mathcal{G}$, which is the primary object of interest in Bayesian random partition models. Suppose an existing random partition model imposes a prior $\mathbb{P}_{\tmop{direct}}(\pi^\ast \mid  \mathcal{G})$ to induce partitions directly from $\mathcal{G}$. We next introduce the proposed two-level hierarchical prior for $\pi^\ast$ and demonstrate how it can be incorporated into this existing model.

\begin{figure}[htbp]
		\centering
		\begin{tabular}{cccc}
			\hspace{-8mm}{\includegraphics[width=0.31\linewidth,height=0.15\textheight]{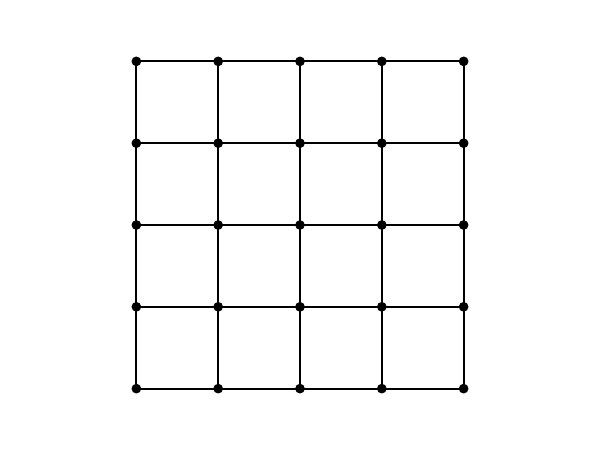}}&
			\hspace{-10mm}{\includegraphics[width=0.31\linewidth,height=0.15\textheight]{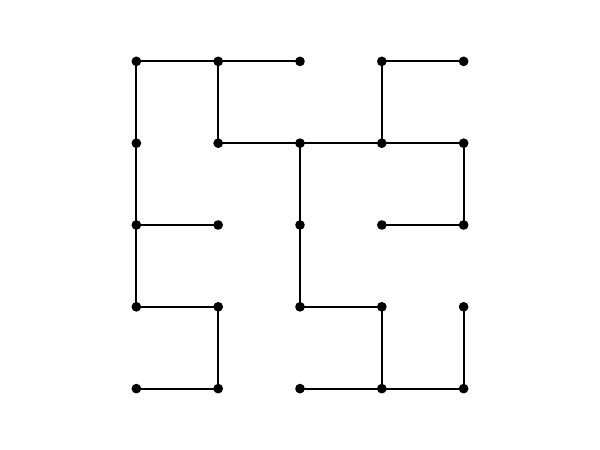}}&
            \hspace{-12mm}{
            {\includegraphics[width=0.31\linewidth,height=0.15\textheight]{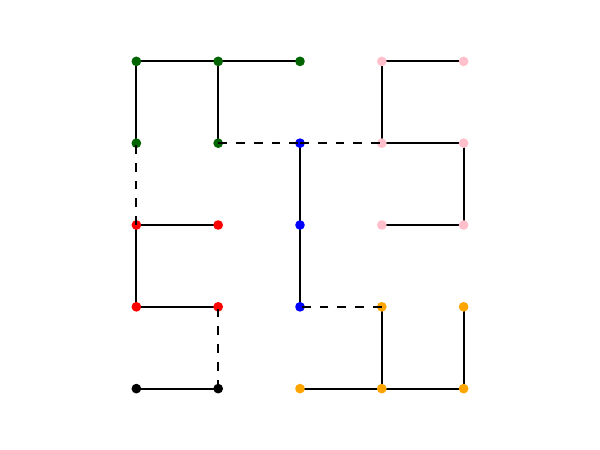}}}&
            \hspace{-14mm}{\includegraphics[width=0.31\linewidth,height=0.15\textheight]{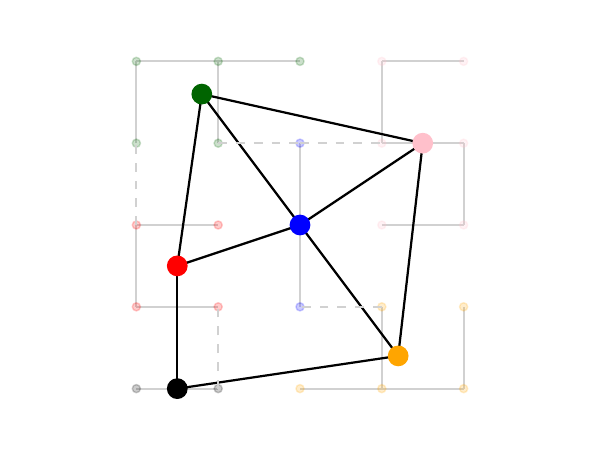}}
            \\
            \hspace{-8mm}{\small (a)} & \hspace{-10mm}{\small (b)} & \hspace{-12mm}{\small (c)} & \hspace{-14mm}{\small (d)}\\
		\end{tabular}
		\caption{An illustration example of inducing $\pi^\dagger$ and the corresponding quotient graph $\mathcal{G}/\pi^\dagger$. (a) An example of $\mathcal{G}=(\mathcal{V},\mathcal{E})$, where the vertices in $\mathcal{V}$ are represented by black dots and the edges in $\mathcal{E}$ by solid lines connecting them. (b) An example of a spanning tree $\mathcal{T}^\dagger$ of $\mathcal{G}$. (c) An example of obtaining $\pi^\dagger$ by cutting edges (dashed lines) of $\mathcal{T}^\dagger$. The cuts produce $6$ connected components (denoted by different colors), each of which is treated as a cluster in $\pi^\dagger$. (d) The quotient graph $\mathcal{G}/\pi^\dagger$. Each cluster in $\pi^\dagger$ is contracted into a quotient vertex. Two quotient vertices are connected if an edge in $\mathcal{E}$ connects their corresponding clusters.}
		\label{FIG:pidaggerillustration}
        \vspace{-10pt}
	\end{figure}

At the first level, we impose a prior, say $\mathbb{P}(\pi^\dagger \mid \mathcal{G})$, to induce an intermediate contiguous partition $\pi^\dagger$ and the corresponding quotient graph $\mathcal{G}/\pi^\dagger$. Specifically, we pre-determine a spanning tree of $\mathcal{G}$, denoted by $\mathcal{T}^\dagger$, and an integer $k^\dagger$, and then randomly cut $k^{\dagger} - 1$ edges from $\mathcal{T}^{\dagger}$. Following the property of spanning trees, this cuts result in a contiguous partition (i.e., $\pi^{\dagger}$) with $k^\dagger$ clusters. The corresponding prior $\mathbb{P} (\pi^{\dagger} \mid \mathcal{G}) \triangleq \mathbb{P}(\pi^{\dagger} \mid \mathcal{G},\mathcal{T}^\dagger,k^\dagger)$ is written as
\begin{equation}
    \mathbb{P} (\pi^{\dagger} \mid \mathcal{G}) \propto \mathbb{I}(\pi^\dagger \text{ can be induced by cutting $k^\dagger-1$ edges from $\mathcal{T}^\dagger$}),
    \label{EQ:PpidaggermidG}
\end{equation}
where $\mathbb{I}(\cdot)$ is the indicator function. As defined in Section \ref{SEC:graphbasic}, $\pi^\dagger$ naturally induces a quotient graph $\mathcal{G}/\pi^\dagger$. See Figure \ref{FIG:pidaggerillustration} for an illustration example of inducing $\pi^\dagger$ and the corresponding $\mathcal{G}/\pi^\dagger$ at the first level.

Note that we pre-determine and fix $(\mathcal{T}^\dagger,k^\dagger)$ at the first level, so the support of $\mathbb{P}(\pi^{\dagger}\mid\mathcal{G})$ is restricted to a subset of all contiguous partitions w.r.t. $\mathcal{G}$. This reduces the space of possible $\pi^\dagger$, thereby alleviating the computational burden. Importantly, this space reduction need not exclude the desirable partition: a properly chosen $(\mathcal{T}^\dagger,k^\dagger)$ can ensure that it remains in the reduced space, as established in Lemma \ref{PP:goodcondition}. We provide further details on the choice of $(\mathcal{T}^\dagger,k^\dagger)$ in Section \ref{SEC:choiceofTk}.

Conditional on $\pi^\dagger$ (hence $\mathcal{G}/\pi^\dagger$), we then impose a prior, say $\mathbb{P}(\pi^\ast \mid \mathcal{G},\pi^\dagger)=\mathbb{P}(\pi^\ast \mid \mathcal{G}/\pi^\dagger)$, to induce $\pi^\ast$ through a random contiguous partition model for the quotient vertices at the second level. Note that each quotient vertex of $\mathcal{G}/\pi^\dagger$ corresponds to a collection of vertices in the original vertex set $\mathcal{V}$ that are contracted into a single quotient vertex. Therefore, any partition of the quotient vertices of $\mathcal{G}/\pi^\dagger$ naturally induces a partition of $\mathcal{V}$ (i.e., $\pi^\ast$): all original vertices associated with quotient vertices in the same cluster are assigned the same cluster membership.

Recall the notation $\cdot/\cdot$ and $\preceq$ introduced in Section \ref{SEC:graphbasic}. By construction, every $\pi^\ast$ in the support of $\mathbb{P}(\pi^\ast \mid \mathcal{G},\pi^\dagger)$ satisfies $\pi^{\dagger}
\preceq \pi^{\ast}$, and $\pi^{\ast} / \pi^{\dagger}$ is a contiguous partition of the quotient vertices w.r.t. $\mathcal{G} / \pi^{\dagger}$. Since $\pi^\dagger$ is contiguous and the quotient graph $\mathcal{G}/\pi^\dagger$ preserves connectivity among the clusters of $\pi^\dagger$, the resulting $\pi^\ast$ is contiguous w.r.t. $\mathcal{G}$.

The choice of $\mathbb{P}(\pi^\ast \mid \mathcal{G},\pi^\dagger)$ is context-specific, and we do not specify its exact form in this section to preserve generality. One way to incorporate the existing prior $\mathbb{P}_{\tmop{direct}}(\cdot \mid  \cdot)$ is to set  {$\mathbb{P}(\pi^\ast \mid \mathcal{G},\pi^\dagger)=\mathbb{P}_{\tmop{direct}}(\pi^\ast/\pi^\dagger \mid \mathcal{G}/\pi^\dagger)$ if $\pi^{\ast}
\succeq \pi^{\dagger}$, and $0$ otherwise.} That is, use the prior $\mathbb{P}_{\tmop{direct}}(\cdot \mid \mathcal{G}/\pi^\dagger)$ to induce partitions of quotient vertices of $\mathcal{G}/\pi^\dagger$. In Section \ref{SEC:AOScaes}, we provide a specific choice of $\mathbb{P}(\pi^\ast \mid \mathcal{G},\pi^\dagger)$ to demonstrate how the proposed two-level prior can be incorporated into Spat-RPM.

Conditional on $\pi^\ast$, the likelihood is specified as in the existing model. Combining this likelihood with the proposed prior yields the posterior distribution of $\pi^\ast$.

\begin{remark}
\label{RM:spacereduction}
We provide further details on how the proposed two-level prior reduces the partition space in this remark. Since $\pi^\dagger$ is obtained by cutting $k^\dagger-1$ edges from $\mathcal{T}^\dagger$, the number of possible $\pi^\dagger$'s is upper bounded by
$\binom{|\mathcal{V}|-1}{k^\dagger-1}
\leq \exp\{k^\dagger \log(|\mathcal{V}|)\}$. Conditional on $\pi^\dagger$, the number of possible partitions of the quotient vertices is upper bounded by $
|\pi^\dagger|^{|\pi^\dagger|}=\exp\{k^\dagger \log(k^\dagger)\}$. Therefore, since $k^\dagger\leq |\mathcal{V}|$, the joint space of $(\pi^\dagger,\pi^\ast)$ is upper bounded by $\exp\{2k^\dagger\log(|\mathcal{V}|)\}$. In contrast, inducing $\pi^\ast$ from $\mathbb{P}_{\tmop{direct}}(\pi^\ast\mid \mathcal{G})$ generally leads to a much larger partition space, whose size may grow exponentially, or even super-exponentially, with $|\mathcal{V}|$; see, e.g., \cite{huang2025consistent,luo2021bayesian}. Thus, when $k^\dagger$ is chosen to be appreciably smaller than $|\mathcal{V}|$, the proposed two-level approach can substantially reduce the partition space, thereby reducing both model uncertainty and computational burden. 
\end{remark}

\subsection[]{The choice of $(\mathcal{T}^\dagger,k^\dagger)$ and the prior model $\mathbb{P}(\pi^\ast \mid \mathcal{G},\pi^\dagger)$}
\label{SEC:choiceofTk}

As discussed in Remark \ref{RM:spacereduction}, the proposed two-level prior can reduce the partition space by choosing a small $k^\dagger$. However, this reduction entails an important caveat: the space should not be reduced excessively that it excludes all desirable partitions. In this section, we formally study this issue, which is closely related to the choice of $(\mathcal{T}^\dagger,k^\dagger)$ and the prior model $\mathbb{P}(\pi^\ast \mid \mathcal{G},\pi^\dagger)$.

Let $\Xi^{\ast}_{\tmop{direct}}(\mathcal{G})$ denote the support of $\mathbb{P}_{\mathrm{direct}}(\pi^{\ast} \mid \mathcal{G})$, and suppose there exists a desirable partition $\pi_0^{\ast} \in \Xi_{\tmop{direct}}^{\ast}(\mathcal{G})$ that the model aims to recover. The guiding principle to choose $(\mathcal{T}^\dagger,k^\dagger)$ and $\mathbb{P}(\pi^\ast \mid \mathcal{G},\pi^\dagger)$ is to ensure that the resulting partition space is sufficiently reduced, while still retaining $\pi_0^{\ast}$. Let $\Xi^{\ast}(\pi^{\dagger})$ denote the support of $\mathbb{P}(\pi^\ast \mid \mathcal{G},\pi^\dagger)$. We next introduce high-level assumptions required to ensure $\pi_0^\ast$ remains in our reduced partition space.

{\Hassumption{\label{AS:pidaggerspace}We assume that $\Xi^{\ast}
(\pi^{\dagger}) \supseteq \{ \pi^{\ast} \in \Xi_{\tmop{direct}}^{\ast} (\mathcal{G}):\pi^{\ast}
\succeq \pi^{\dagger} \}$.}}

Assumption \ref{AS:pidaggerspace} concerns the prior model $\mathbb{P}(\pi^\ast \mid \mathcal{G},\pi^\dagger)$, and requires that $\Xi^{\ast}(\pi^{\dagger})$ is not overly restrictive: it contains every partition $\pi^{\ast} \in \Xi_{\tmop{direct}}^{\ast}(\mathcal{G})$ satisfying $\pi^{\ast} \succeq \pi^{\dagger}$. Recall that $\mathbb{P}(\pi^\ast \mid \mathcal{G},\pi^\dagger)$ is induced by imposing a random contiguous partition model on the quotient vertices of $\mathcal{G}/\pi^\dagger$. The next lemma provides an easily verifiable sufficient condition on the imposed model on the quotient vertices, under which Assumption \ref{AS:pidaggerspace} is satisfied. An example of such a model is given later in Section \ref{SEC:Ppiastconditional}.


\begin{lemma}
\label{LM:verifyAS1_H}
If, for every $\pi^\dagger \in \Xi^\dagger$, the support of the imposed random contiguous partition model on the quotient vertices of $\mathcal{G}/\pi^\dagger$ contains all contiguous partitions w.r.t. $\mathcal{G}/\pi^\dagger$, then the induced $\mathbb{P}(\pi^\ast \mid \mathcal{G},\pi^\dagger)$ satisfies Assumption \ref{AS:pidaggerspace}.

\end{lemma}

To include $\pi_0^\ast$ in our reduced partition space, we adopt a multi-resolution idea: we construct a series of $q \geqslant 2$ nested contiguous partitions w.r.t. $\mathcal{G}$, say $\{\pi^{\dagger\dagger}_{p}\}_{p=1}^{q}$, such that $ \pi_1^{\dagger \dagger} \succeq \pi_2^{\dagger \dagger} \succeq \cdots \succeq \pi_q^{\dagger \dagger}$, where 
   $\pi_1^{\dagger\dagger}=\{\mathcal{V}\}$ is the trivial partition with $\mathcal{V}$ as its only cluster, and $\pi_q^{\dagger\dagger}=\{\{v\}:v\in\mathcal{V}\}$ has each vertex in $\mathcal{V}$ as a cluster. The construction is intended to include $\pi_0^\ast$ in the partition space induced by $\mathcal{G}/\pi^{\dagger \dagger}_{p^\dagger}$ for some resolution level $1\leq p^{\dagger}\leq q$. Given such a $p^{\dagger}$, additional conditions on $(\mathcal{T}^\dagger,k^\dagger)$ are assumed to ensure that $\pi_0^\ast$ is further included in the reduced partition space of our two-level prior. The specific conditions are stated in the following two assumptions.


{\Hassumption{\label{AS:pi0existence}We assume that $\pi^{\ast}_0 \succeq \pi^{\dagger \dagger}_{p^{\dagger}}$ for some $1 \leqslant p^{\dagger} \leqslant q$.}}

{\Hassumption{\label{AS:Tdaggercompatible} For the same $p^\dagger$ in Assumption~\ref{AS:pi0existence}, we assume that $\mathcal{T}^{\dagger}$ is compatible with $\mathcal{}
\pi_{p^\dagger}^{\dagger \dagger}$, and $k^{\dagger} - 1 \geqslant |
\Omega_{\pi_0^{\ast}} (\pi^{\dagger \dagger}_{p^\dagger}) |$,
where $\Omega_{\pi_0^{\ast}}(\pi^{\dagger \dagger}_{p^\dagger})$ is the set of edges in $\mathcal{G}/\pi^{\dagger \dagger}_{p^\dagger}$ that connect distinct clusters of the partition $\pi_0^{\ast}/\pi^{\dagger \dagger}_{p^\dagger}$.
}}

Under Assumption \ref{AS:pi0existence}, $\pi^{\ast}_0/ \pi^{\dagger \dagger}_{p^{\dagger}}$, which is a partition of quotient vertices w.r.t. $\mathcal{G}/\pi^{\dagger \dagger}_{p^{\dagger}}$, is well defined. Hence, Assumption \ref{AS:pi0existence} enables $\pi_0^\ast$ to be included in the partition space induced by $\mathcal{G}/\pi^{\dagger \dagger}_{p^{\dagger}}$. On the other hand, by definition, $\pi_q^{\dagger\dagger}$ separates all vertices in $\mathcal{V}$.
Therefore, Assumption~\ref{AS:pi0existence} is immediately satisfied with $p^\dagger=q$.



Assumption~\ref{AS:Tdaggercompatible} concerns the choice of $(\mathcal{T}^\dagger,k^\dagger)$ and provides guidance for selecting these components. 
Assumption~\ref{AS:Tdaggercompatible} indicates that some prior information on $|\Omega_{\pi_0^{\ast}}(\pi^{\dagger \dagger}_{p^\dagger})|$ is needed to make the condition $k^{\dagger}-1\geq |\Omega_{\pi_0^{\ast}}(\pi^{\dagger \dagger}_{p^\dagger})|$ operational. By definition, $|\Omega_{\pi_0^{\ast}}(\pi^{\dagger \dagger}_{p^\dagger})|$ is the number of edges crossing the boundaries of $\pi_0^\ast/\pi^{\dagger \dagger}_{p^\dagger}$ in the quotient graph $\mathcal{G}/\pi^{\dagger \dagger}_{p^\dagger}$. Regularity conditions on these boundaries are often imposed, which can help determine the order of $|\Omega_{\pi_0^{\ast}}(\pi^{\dagger \dagger}_{p^\dagger})|$. For example, in the spatial clustering literature, the boundary of the true partition is typically assumed to have finite length \citep{huang2025consistent}, from which the order of $|\Omega_{\pi_0^{\ast}}(\pi^{\dagger \dagger}_{p^\dagger})|$ can be derived. We provide further details in deciding this order in the case study in Section~\ref{SEC:AOScaes}.


\begin{remark}
    The construction of nested partitions $\{\pi_p^{\dagger\dagger}\}_{p=1}^q$ yields more flexibility in our model, because Assumptions \ref{AS:pi0existence} - \ref{AS:Tdaggercompatible} need to be satisfied at only one resolution level $p^\dagger$. In general, the resolution level $p^\dagger$ reflects a trade-off. A larger $p^\dagger$ corresponds to a finer partition, which is more likely to nest in $\pi_0^\ast$. However, the finer resolution may increase the number of edges crossing the boundaries of $\pi_0^\ast/\pi^{\dagger \dagger}_{p^\dagger}$, making the condition $k^{\dagger} - 1 \geqslant |
\Omega_{\pi_0^{\ast}} (\pi^{\dagger \dagger}_{p^\dagger}) |$ in Assumption \ref{AS:Tdaggercompatible} more difficult to satisfy. Since $\pi_0^\ast$ is unknown and depends on the data-generating process, incorporating multiple resolutions allows the assumptions to hold at the appropriate level without requiring $p^\dagger$ to be specified in advance, thereby achieving adaptivity. In Section \ref{SEC:AOS pdagger}, we provide a concrete example of how to construct $\{\pi_p^{\dagger\dagger}\}_{p=1}^q$ for A-Spat-RPM.

\end{remark}

Let $\Xi^{\dagger}$ be the support of $\mathbb{P}(\pi^\dagger\mid \mathcal{G})$. We clarify that the nested partitions $\{\pi_p^{\dagger \dagger}\}_{p=1}^{q}$ are not required to be included in $\Xi^{\dagger}$. The purpose of Assumptions \ref{AS:pi0existence} - \ref{AS:Tdaggercompatible} is to provide a benchmark for assessing the conditions on $(\mathcal{T}^\dagger,k^\dagger)$ required for our model to retain $\pi_0^\ast$, as stated in the following lemma.





\begin{lemma}
  \label{PP:goodcondition}Under Assumptions \ref{AS:pidaggerspace} - \ref{AS:Tdaggercompatible}, there exists a
  $\pi^{\dagger}_0\in \Xi^\dagger$, such that $\pi_0^{\ast} \in \Xi^{\ast}
  (\pi^{\dagger}_0)$. 
\end{lemma}

Lemma~\ref{PP:goodcondition} guarantees that $\pi_0^\ast$ is contained in our model space. Recall that Assumption~\ref{AS:Tdaggercompatible} requires $\mathcal{T}^\dagger$ to be compatible with $\pi_{p^\dagger}^{\dagger\dagger}$. To ensure this compatibility, we propose Algorithm \ref{AL:spanningtree} for constructing a spanning tree $\mathcal{T}^{\dagger}$ that is compatible with each partition in the sequence $\{ \pi^{\dagger \dagger}_p \}_{p = 1}^q$. The details of the algorithm are provided in Section \ref{SEC:Tdagger construction} of Supplementary Material. The following two lemmas justify the validity of the constructed $\mathcal{T}^\dagger$ by Algorithm \ref{AL:spanningtree}.

\begin{lemma}
  \label{PP:Tistree}The constructed $\mathcal{T}^{\dagger}$ by Algorithm
  \ref{AL:spanningtree} is a spanning tree of $\mathcal{G}$. 
\end{lemma}

\begin{lemma}
\label{LM:Tdaggercompatible}
  The constructed $\mathcal{T}^{\dagger}$ by Algorithm
  \ref{AL:spanningtree} is compatible with each partition in the sequence $\{ \pi^{\dagger \dagger}_p \}_{p = 1}^q$.
\end{lemma}

Summarizing the results in this section, we provide a specific choice of the model components as follows. We choose $\mathbb{P}(\pi^\ast\mid \mathcal{G},\pi^\dagger)$ to ensure that Assumption \ref{AS:pidaggerspace} is satisfied. To construct the spanning tree $\mathcal{T}^\dagger$, we first pre-determine a sequence of nested partitions $\{\pi^{\dagger\dagger}_{p}\}_{p=1}^{q}$ and then apply Algorithm \ref{AL:spanningtree}. We choose $k^\dagger$ according to the conditions in Assumption \ref{AS:Tdaggercompatible}, assuming that some prior information about $|\Omega_{\pi_0^{\ast}}(\pi^{\dagger \dagger}_{p^\dagger})|$ is available. Under Assumptions \ref{AS:pidaggerspace} - \ref{AS:Tdaggercompatible}, the resulting model space is then guaranteed to contain the desirable partition $\pi^\ast_0$.

\subsection{Consistency results}
\label{SEC:highlevelAS}
In this section, we study theoretical properties concerning the consistency of the proposed two-level framework. We introduce high-level assumptions under which consistency can be established, and in later sections, we demonstrate how these results apply to a specific model.

A consistent random partition model typically concentrates on a neighborhood of the true partition, denoted by $\Pi^\ast_0$. Correspondingly, we define a desirable set for $\pi^\dagger$, denoted by $\Pi^\dagger_0$, as the set of $\pi^\dagger$'s that can induce at least one partition in $\Pi^\ast_0$; that is,
\begin{equation}
    \Pi^{\dagger}_0 = \{ \pi^{\dagger} : \text{there exists } \pi^{\ast} \in \Pi_0^{\ast} \text{ such that } \pi^{\ast} \in
\Xi^{\ast} (\pi^{\dagger}) \} .
\label{EQ:Pi0dagger}
\end{equation}
Since $\pi^\dagger$ determines the quotient graph $\mathcal{G}/ \pi^\dagger$, $\Pi^{\dagger}_0$ also identifies a collection of desirable quotient graphs for inducing partitions in $\Pi_0^\ast$. Our goal is to establish consistency for both $\Pi^\dagger_0$ and $\Pi_0^\ast$.


Recall that $\pi_0^\ast$ is the desirable partition under the prior $\mathbb{P}_{\tmop{direct}}(\pi^\ast\mid \mathcal{G})$. Typically, $\Pi^\ast_0$ is expected to contain $\pi_0^\ast$.
 Recall that, under Assumptions~\ref{AS:pidaggerspace} - \ref{AS:Tdaggercompatible}, Lemma~\ref{PP:goodcondition} implies that there exists at least one $\pi_0^\dagger \in \Xi^{\dagger}$ such that $\pi_0^\ast \in \Xi^{\ast}(\pi_0^\dagger)$. Consequently, the proposed two-level prior assigns positive support to at least one pair $(\pi_0^\dagger,\pi_0^\ast)$ in the desirable sets $\Pi^{\dagger}_0$ and $\Pi^{\ast}_0$, respectively. 
 
 We next introduce a high-level assumption. Let $\mathfrak{D}$ denote the observed data and $\mathbb{P} (\mathfrak{D} \mid \pi^{\ast})=\mathbb{P} (\mathfrak{D} \mid \mathcal{G},\pi^\dagger, \pi^{\ast})$ denote the data likelihood conditional on the partition $\pi^{\ast}$.

{\Hassumption{\label{AS:ansum}With probability tending to $1$,
\[ \sum_{\pi^{\dagger} \in \Xi^{\dagger}} \sum_{\pi^{\ast} \in  \Xi^{\ast}
   (\pi^{\dagger}) \cap (\Pi_0^{\ast})^c } \frac{\mathbb{P} (\pi^{\ast} \mid \mathcal{G},
   \pi^{\dagger}) \mathbb{P} (\mathfrak{D} \mid  \pi^{\ast})}{\mathbb{P}
   (\pi^{\ast}_0 \mid \mathcal{G}, \pi^{\dagger}_0) \mathbb{P} (\mathfrak{D} \mid
   \pi^{\ast}_0)} \rightarrow 0. \]}}

Assumption~\ref{AS:ansum} implies that the sum of $\mathbb{P}
(\pi^{\ast} \mid \mathcal{G}, \pi^{\dagger}) \mathbb{P} (\mathfrak{D} \mid
\pi^{\ast})$ over all $(\pi^\dagger,\pi^\ast)$ with $\pi^\ast \in (\Pi_0^{\ast})^c$ is negligible relative to $\mathbb{P}
(\pi^{\ast}_0 \mid \mathcal{G}, \pi^{\dagger}_0) \mathbb{P} (\mathfrak{D} \mid
\pi^{\ast}_0)$. We have the following result.

\begin{lemma}
  \label{LM:pidaggerconsistency_general}Under Assumptions
  \ref{AS:pidaggerspace} - \ref{AS:ansum}, with probability tending to 1, we
  have $\mathbb{P} (\pi^{\dagger} \in \Pi^{\dagger}_0 \mid \mathcal{G}, \mathfrak{D})
  \rightarrow 1$, and $\mathbb{P}
  (\pi^{\ast} \in \Pi^{\ast}_0 \mid \mathcal{G}, \mathfrak{D}) \rightarrow 1$.
\end{lemma}

Lemma~\ref{LM:pidaggerconsistency_general} establishes consistency for both $\Pi^\dagger_0$ and $\Pi^\ast_0$. For a given $\pi^\dagger \in \Xi^\dagger \cap \Pi_0^\dagger$, we next study the conditional consistency of $\pi^\ast \mid \mathcal{G}, \mathfrak{D},\pi^\dagger$. This perspective highlights the role of such a $\pi^\dagger$: once $\pi^\dagger$ is fixed in $\Xi^\dagger \cap \Pi_0^\dagger$, inducing partitions from the corresponding quotient graph $\mathcal{G}/\pi^\dagger$ is sufficient to achieve partition consistency. Moreover, by Lemma~\ref{LM:pidaggerconsistency_general}, a posterior draw of $\pi^\dagger$ belongs to $\Xi^\dagger\cap\Pi_0^\dagger$ with probability tending to 1. This suggests a potential computational strategy for sampling $\pi^\ast$ while retaining partition consistency: a posterior draw of $\pi^\dagger$ is first fixed and only $\pi^\ast\mid\mathcal G,\mathfrak D,\pi^\dagger$ is subsequently sampled. The resulting computational savings from fixing $\pi^\dagger$ can be particularly substantial when sampling $\pi^\dagger$ is expensive.

For a given $\pi^{\dagger} \in\Xi^\dagger \cap \Pi_0^\dagger$, let $\pi^{\ast}_0
(\pi^{\dagger})$ be the induced partition such that $\pi^{\ast}_0 (\pi^{\dagger})
\in \Pi^{\ast}_0$. We make the following assumption.

{\Hassumption{\label{AS:anfurther}With probability tending to 1, we have
\begin{align*}
\sup_{\pi^{\dagger} \in \Xi^\dagger \cap \Pi^{\dagger}_0} \left\{ \frac{\mathbb{P}
   (\pi^{\ast}_0 \mid \mathcal{G}, \pi^{\dagger}_0)}{\mathbb{P} (\pi^{\ast}_0
   (\pi^{\dagger}) \mid \mathcal{G}, \pi^{\dagger})} \frac{\mathbb{P} (\mathfrak{D} \mid
   \pi^{\ast}_0)}{\mathbb{P} (\mathfrak{D} \mid \pi^{\ast}_0 (\pi^{\dagger}))}
   \times \sum_{\pi^{\ast} } \frac{\mathbb{P} (\pi^{\ast} \mid \mathcal{G},\pi^{\dagger})
   \mathbb{P} (\mathfrak{D} \mid \pi^{\ast})}{\mathbb{P} (\pi^{\ast}_0 \mid \mathcal{G},
   \pi^{\dagger}_0) \mathbb{P} (\mathfrak{D} \mid \pi^{\ast}_0)} \right\}
   \rightarrow 0,
\end{align*} 
where the summation $\sum_{\pi^\ast}$ is taken over $\pi^\ast \in \Xi^\ast(\pi^\dagger)\cap(\Pi_0^\ast)^c$.

}}
With Assumption~\ref{AS:anfurther}, the following lemma establishes the conditional consistency of $\pi^\ast \mid \mathcal{G},\mathfrak{D},\pi^\dagger$. Specifically, it shows that, for any $\pi^\dagger \in \Xi^\dagger \cap \Pi_0^\dagger$, the conditional posterior of $\pi^\ast$ concentrates on $\Pi^\ast_0$.

\begin{lemma}
  \label{LM:piastconditionalpidagger}Under Assumptions \ref{AS:pidaggerspace}
  - \ref{AS:Tdaggercompatible}, and \ref{AS:anfurther}, with probability
  tending to 1, we have ${\sup_{\pi^{\dagger} \in \Xi^\dagger \cap \Pi^{\dagger}_0}} 
  \mathbb{P} (\pi^{\ast} \nin \Pi^{\ast}_0 \mid \mathcal{G}, \mathfrak{D}, \pi^{\dagger})
  \rightarrow 0$.

\end{lemma}

\subsection{Bayesian computation}
\label{SEC:Bayesian computation}
We use a Markov chain Monte Carlo (MCMC) algorithm to draw samples from the posterior distribution $\mathbb{P}(\pi^\dagger,\pi^\ast \mid \mathcal{G}, \mathfrak{D})$. At each iteration, we randomly choose to update either $\pi^{\ast}\mid \mathcal{G}, \mathfrak{D}, \pi^{\dagger}$ or $\pi^{\dagger}\mid \mathcal{G}, \mathfrak{D},\pi^{\ast}$, with probabilities $r^\ast$ and $r^\dagger$, respectively. Updating $\pi^{\ast}\mid \mathcal{G}, \mathfrak{D}, \pi^{\dagger}$ involves proposing a new partition $\pi^\ast_{\tmop{new}}/ \pi^\dagger$ on the current quotient graph $\mathcal{G}/\pi^\dagger$. This can typically be accomplished by applying the same graph-based proposal mechanism used to propose a new partition $\pi^\ast_{\tmop{new}}$ of vertices of $\mathcal{G}$ under the prior model $\mathbb{P}_{\tmop{direct}}(\pi^\ast \mid \mathcal{G})$, while replacing $(\mathcal{G},\pi^\ast_{\tmop{new}})$ with $(\mathcal{G}/\pi^\dagger,\pi^\ast_{\tmop{new}}/\pi^\dagger)$, respectively.


Updating $\pi^{\dagger}\mid \mathcal{G}, \mathfrak{D},\pi^{\ast}$ requires proposing a new intermediate partition $\pi^\dagger_{\tmop{new}}$ such that $\pi^\ast \in \Xi^\ast(\pi^\dagger_{\tmop{new}})$, which requires $\pi^\ast \succeq \pi^\dagger_{\tmop{new}}$. Recall that under our model, we fix $(\mathcal{T}^\dagger,k^\dagger)$, and $\pi^\dagger_{\tmop{new}}$ is obtained by cutting $k^\dagger-1$ edges of $\mathcal{T}^\dagger$. Fixing $(\mathcal{T}^\dagger,k^\dagger)$ restricts $\pi^\dagger_{\tmop{new}}$ to a small space, which simplifies MCMC sampling and reduces the computational burden. Let $\mathcal{E} (\mathcal{T}^{\dagger},
\pi^\ast)$ denote the set of edges in $\mathcal{T}^\dagger$ whose endpoints belong to distinct clusters in $\pi^\ast$. To ensure $\pi^\ast \succeq \pi^\dagger_{\tmop{new}}$, all edges in $\mathcal{E} (\mathcal{T}^{\dagger},
\pi^\ast)$ must be cut. To obtain exactly $k^\dagger-1$ cuts, we further cut $k^\dagger - 1 - |\mathcal{E} (\mathcal{T}^{\dagger},
\pi^\ast)|$ edges uniformly at random from the remaining edges of $\mathcal{T}^\dagger$. This produces a partition $\pi^\dagger_{\tmop{new}}$ which has $k^\dagger$ clusters and satisfies $\pi^\ast \succeq \pi^\dagger_{\tmop{new}}$. We compute the Metropolis-Hastings (M-H) acceptance ratio as $\min \left\{ 1, {\mathbb{P}(\pi^\ast\mid \mathcal{G},\pi^\dagger_{\tmop{new}})}/{\mathbb{P}(\pi^\ast\mid \mathcal{G},\pi^\dagger)} \right\}$.

\section{A case study of Spat-RPM}
\label{SEC:AOScaes}
In this section, we demonstrate how the proposed two-level prior can be incorporated into the Spat-RPM of \cite{huang2025consistent} to obtain A-Spat-RPM. We first provide a brief recap of Spat-RPM in Section~\ref{SEC:AOS recap}. We then specify the priors $\mathbb{P} (\pi^{\dagger} \mid
\mathcal{G})$ and $\mathbb{P} (\pi^{\ast} \mid \mathcal{G},
\pi^{\dagger})$, followed by a specific computational strategy, in Sections~\ref{SEC:AOS pdagger} -  \ref{SEC:AOS computation}.

\subsection{A recap of Spat-RPM}
\label{SEC:AOS recap}

Let $\mathfrak{D}=\{\tmmathbf{s}_i,
\tmmathbf{x}(\tmmathbf{s}_i), y (\tmmathbf{s}_i)\}_{i = 1}^n$ be the spatial
data observed at $n$ locations $\tmmathbf{s}_1, \ldots, \tmmathbf{s}_n \in
\mathcal{D} \subset \mathbb{R}^2$, where $\tmmathbf{x}(\tmmathbf{s}_i)$ is $d$-dimensional covariate and $y(\tmmathbf{s}_i)$ is the response variable at $\tmmathbf{s}_i$. Conditional on $\{
\tmmathbf{x}(\tmmathbf{s}_i)\}_{i = 1}^n$, the likelihood function of $\{{y}(\tmmathbf{s}_i)\}_{i=1}^{n}$ is written as $\prod_{i =
1}^n \mathbb{P}_{\tmmathbf{\theta} (\tmmathbf{s}_i)} \{y (\tmmathbf{s}_i) \mid
\tmmathbf{x}(\tmmathbf{s}_i)\}$, where $\mathbb{P}_{\tmmathbf{\theta}
(\tmmathbf{s}_i)} \{\cdummy \mid \tmmathbf{x}(\tmmathbf{s}_i)\}$ is the conditional probability density function of $y (\tmmathbf{s}_i)$ with unknown
parameter $\tmmathbf{\theta} (\tmmathbf{s}_i)$. To account for spatial heterogeneity, Spat-RPM assumes $\tmmathbf{\theta} (\cdot)$ to be a piecewise constant function, i.e., the domain $\mathcal{D}$ can be partitioned into several clusters, such that $\tmmathbf{\theta} (\tmmathbf{s}) =\tmmathbf{\theta} (\tmmathbf{s}')$ if $\tmmathbf{s}$ and $\tmmathbf{s}'$ belong to the same cluster, and
$\tmmathbf{\theta} (\tmmathbf{s}) \neq \tmmathbf{\theta}
(\tmmathbf{s}')$ otherwise.

\begin{figure}[htbp]
		\centering
		\begin{tabular}{ccc}
			{\includegraphics[width=0.28\linewidth,height=0.15\textheight]{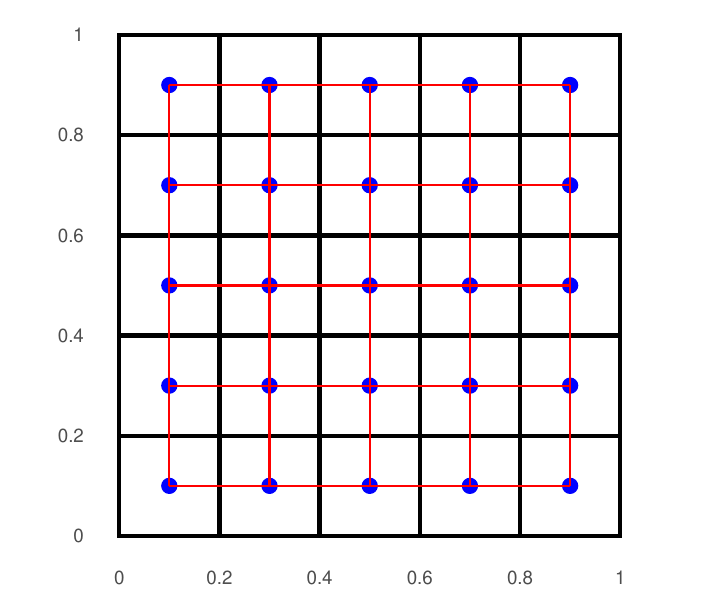}}&
			{\includegraphics[width=0.28\linewidth,height=0.15\textheight]{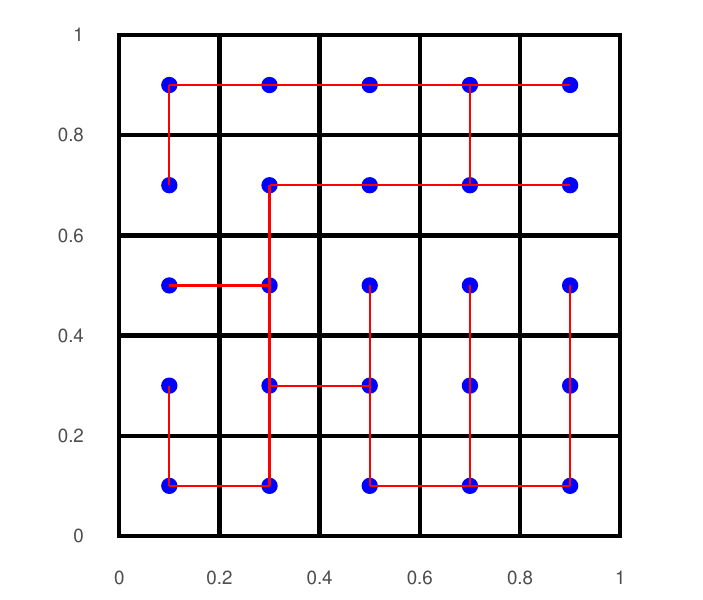}}&
			{\includegraphics[width=0.28\linewidth,height=0.154\textheight]{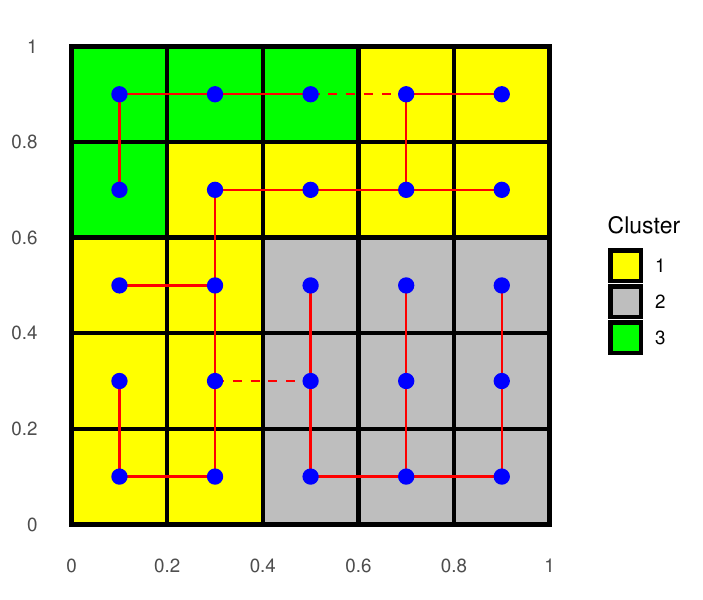}}
			\\
			{\small (a)} & {\small (b)} &{\small (c) } 
		\end{tabular}
		\caption{Illustration of inducing domain partitions in Spat-RPM. (a) The graph $\mathcal{G} = (\mathcal{V}, \mathcal{E})$, where $\mathcal{V}=\{B_m\}_{m=1}^{K^2}$ and $\mathcal{E}$ contains edges connecting adjacent blocks, shown as red lines. (b) A spanning tree $\mathcal{T}$ induced from $\mathcal{G}$ in (a). (c) The domain partition $\pi^\ast$ induced by cutting two edges (dashed lines) of the spanning tree shown in (b). This edge cutting produces three connected components, each of which is treated as a cluster and denoted by a different color. All locations within the same block share the same cluster membership.}
		\label{FIG:illustrationaos}
        \vspace{-10pt}
	\end{figure}

 Without loss of generality, Spat-RPM assumes that $\mathcal{D}=[0,1]^2$, while the method can be readily extended to more general domains that are homeomorphic to $[0,1]^2$ under the Euclidean metric via a bi-Lipschitz homeomorphism. The goal of Spat-RPM is to recover the latent partition of domain $\mathcal{D}$. To this end, Spat-RPM proposes a prior for domain partitions. 
Spat-RPM first segments $\mathcal{D}$ into $K^2$ disjoint blocks, denoted by $\{B_m\}_{m=1}^{K^2}$, where $K$ is a pre-specified integer. Based on these blocks, an undirected spatial mesh grid graph $\mathcal{G}=(\mathcal{V},\mathcal{E})$ is constructed, where $\mathcal{V}=\{B_m\}_{m=1}^{K^2}$ is the vertex set and $\mathcal{E}$ contains edges connecting adjacent blocks. Figure \ref{FIG:illustrationaos}(a) shows an example of $\mathcal{G}$ with $K=5$. Spat-RPM then partitions the vertices of $\mathcal{V}$, which naturally induces a partition of $\mathcal{D}$.

Specifically, given the graph $\mathcal{G}$, Spat-RPM induces contiguous partitions of $\mathcal{V}$ by using spanning trees. Spat-RPM first assumes a uniform spanning tree (UST, \citealp{aldous1990random}) prior on $\mathcal{T}\in \Delta(\mathcal{G})$, i.e., 
\begin{equation}
    \mathbb{P}(\mathcal{T}\mid \mathcal{G})\propto \mathbb{I}\{\mathcal{T}\in \Delta(\mathcal{G})\},
    \label{EQ:spanningtreeprior}
\end{equation}
where $\Delta(\mathcal{G})$ denotes the space of all spanning trees of $\mathcal{G}$. Spat-RPM then assumes a Poisson distribution on the number of clusters $k$ with mean parameter $\lambda$:
\begin{equation}
 k\sim \tmop{Poisson}(\lambda) \cdot \mathbb{I}(1\leq k\leq K^2).
    \label{EQ:kprior}
\end{equation}

Conditional on $\mathcal{T}$ and $k$, a contiguous partition of $\mathcal{V}$ w.r.t. $\mathcal{G}$, say $\pi^\ast$, is induced by randomly cutting $k-1$ edges from $\mathcal{T}$. $\pi^\ast$ naturally induces a domain partition by assigning all locations within the same block to the same cluster. Thus, in what follows, we view $\pi^\ast$ as playing a dual role: depending on the context, it may represent either a partition of the vertices or the induced domain partition. See Figures \ref{FIG:illustrationaos}(b - c) for an example of a spanning tree $\mathcal{T}$ induced from $\mathcal{G}$ and a resulting $\pi^\ast$, respectively. 


Conditional on the domain partition $\pi^\ast=\{\mathcal{D}_{1}^{\ast}, \ldots,
\mathcal{D}_k^{\ast} \}$, where each $\mathcal{D}_j^\ast \subseteq \mathcal{D}$ is a sub-domain representing a cluster, $\tmmathbf{\theta}(\cdot)$ is assumed to be a constant wihtin each $\mathcal{D}_j^\ast$. Let $\tmmathbf{\theta}_j$ be the constant regression coefficient within cluster $\mathcal{D}_{j}^{\ast}$, and write $\tmmathbf{\theta}= \{\tmmathbf{\theta}_j\}_{j =
1}^k$. Spat-RPM assumes independent priors on $\mathbb{P}(\tmmathbf{\theta})=\prod_{j=1}^{k}\mathbb{P}(\tmmathbf{\theta}_j) $, and derive the likelihood afterwards. In this paper, for demonstration, we focus on a linear regression model, which is also the model considered in the theoretical analysis in \cite{huang2025consistent}. Specifically, we consider the model 
\begin{equation}
    y(\tmmathbf{s}_i)=\mu(\tmmathbf{s}_i)+\epsilon(\tmmathbf{s}_i),\text{ } 1\leq i \leq n,
    \label{EQ:model}
\end{equation}
where $\mu(\tmmathbf{s}_i)=\tmmathbf{x}^T(\tmmathbf{s}_i)\tmmathbf{\theta}(\tmmathbf{s}_i)$ is the regression mean of $y(\tmmathbf{s}_i)$,  and $\{\epsilon
(\tmmathbf{s}_i)\}_{i = 1}^n$ are identically and independently distributed ($i.i.d.$) mean zero Gaussian noises. For Model (\ref{EQ:model}), Spat-RPM assumes an independent conjugate Zellner’s g-prior on the regression coefficient at each cluster, and writes the hierarchical model as 
\begin{gather}
  \tmmathbf{\theta} \mid \pi^\ast , k, \mathcal{T},
  \mathcal{S}, \mathcal{X} \sim \prod_{j = 1}^k \mathbb{P}_{\tmop{Gaussian}}
  \bigg[ \tmmathbf{\theta}_j ; \tmmathbf{0}, \gamma n \sigma^2 \bigg\{
  \sum_{\tmmathbf{s}_i \in \mathcal{D}_j^\ast}
  \tmmathbf{x}(\tmmathbf{s}_i)\tmmathbf{x}^T (\tmmathbf{s}_i)
  \bigg\}^{\dagger} \bigg],\text{ and} \label{EQ:priortheta}\\
  \tmmathbf{y} \mid \pi^\ast , k, \mathcal{T}, \mathcal{S},
  \mathcal{X}, \tmmathbf{\theta} \sim \prod_{j = 1}^k \prod_{\tmmathbf{s}_i
  \in \mathcal{D}_j^\ast} \mathbb{P}_{\tmop{Gaussian}} \{y (\tmmathbf{s}_i) ;
  \tmmathbf{x}^T (\tmmathbf{s}_i)\tmmathbf{\theta}_j, \sigma^2 \},
  \label{EQ:priory}
\end{gather}
where $\mathcal{S}= \{\tmmathbf{s}_i\}_{i = 1}^n$, $\mathcal{X}= \{\tmmathbf{x} (\tmmathbf{s}_i)\}_{i = 1}^n$, $\tmmathbf{y}= \{y (\tmmathbf{s}_i)\}_{i = 1}^n$, $\mathbb{P}_{\tmop{Gaussian}} (\cdummy ; \tmmathbf{a},
\tmmathbf{\Sigma})$ denotes the probability density function of the Gaussian distribution with mean $\tmmathbf{a}$ and covariance $\tmmathbf{\Sigma}$, $\gamma>0$ is a hyperparameter controlling the prior variance, $\cdot^{\dagger}$ denotes the pseudoinverse, and $\sigma^2$ denotes the unknown variance of $\epsilon (\tmmathbf{s}_i)$. Spat-RPM fixes $\sigma^2$ at a pre-determined value, and the consistency result remains valid even if this value is misspecified.

The posterior distribution of $\pi^\ast$ is then derived and the consistency theory for the domain partition is established in \cite{huang2025consistent}. Instead of inducing $\pi^\ast$ directly from $\mathcal{G}$ as in priors of Spat-RPM, we demonstrate in the following sections how the proposed two-level hierarchical prior can be applied to induce $\pi^\ast$.

\subsection[]{Prior specification for $\mathbb{P} (\pi^{\dagger} \mid
\mathcal{G})$}
\label{SEC:AOS pdagger}

By Equation~(\ref{EQ:PpidaggermidG}), specifying $\mathbb{P}(\pi^\dagger \mid \mathcal{G})$ reduces to specifying the spanning tree $\mathcal{T}^\dagger$ and the number of clusters $k^\dagger$. Following Section~\ref{SEC:choiceofTk}, we construct $\mathcal{T}^\dagger$ by first constructing a sequence of nested partitions $\{\pi^{\dagger \dagger}_p\}_{p=1}^{q}$, and then applying Algorithm~\ref{AL:spanningtree}. Specifically, we construct $\{\pi^{\dagger \dagger}_p\}_{p=1}^{q}$ following a recursive refinement approach. We start with a trivial partition $\pi_1^{\dagger\dagger}$ with $\mathcal{V}$ as its only cluster, i.e., $\pi_1^{\dagger\dagger}=\{\mathcal{V}\}$. At each resolution level $p\geq 1$, every cluster in $\pi_p^{\dagger\dagger}$ that can still be split is divided into approximately equal-sized sub-clusters, while clusters that cannot be further split are kept unchanged. These sub-clusters, together with the unchanged clusters, form the next-level partition $\pi_{p+1}^{\dagger\dagger}$.
This procedure is repeated until no cluster can be further divided. The detailed construction is given in Algorithm~\ref{AL:nestedpartition} in Section \ref{SEC:ALnestedpartitions} of Supplementary Material. Note that the mesh grid graph $\mathcal{G}$ makes the construction easy to demonstrate, because each cluster $C\in \pi_p^{\dagger\dagger}$ in Algorithm~\ref{AL:nestedpartition} is a rectangle. For a more general graph, the same recursive refinement idea can still be applied to construct $\{\pi^{\dagger \dagger}_p\}_{p=1}^{q}$.

For A-Spat-RPM, each cluster of $\pi^{\dagger \dagger}_p$ represents a subset of vertices in $\mathcal{V}$, or equivalently, a union of blocks in $\{B_m\}_{m=1}^{K^2}$. As $p$ increases, each cluster of $\pi^{\dagger \dagger}_p$ contains fewer vertices of $\mathcal{V}$. Consequently, the corresponding union of blocks contains fewer blocks from $\{B_m\}_{m=1}^{K^2}$ and therefore has a smaller size. Thus, $\{\Xi^\ast(\pi^{\dagger \dagger}_p)\}_{p=1}^{q}$ at different resolution levels $p$ corresponds to partition spaces induced using unions of blocks of different sizes. In Spat-RPM, the hyperparameter $K$ controls the block size, with each block having side length $K^{-1}$, and its choice depends on the unknown true number of clusters. Hence, $\{\pi^{\dagger \dagger}_p\}_{p=1}^{q}$ enables A-Spat-RPM to adapt to the unknown true number of clusters through different sizes of unions of blocks. We demonstrate this adaptivity in greater detail through the theoretical analysis in Section \ref{SEC:AOS theory}.


According to Assumption~\ref{AS:Tdaggercompatible}, $k^\dagger$ should be selected based on the boundary information of the desirable partition $\pi_0^\ast$. We defer this choice to Section~\ref{SEC:AOS theory}, where $\pi_0^\ast$ is formally introduced.

\subsection[]{Prior specification for $\mathbb{P} (\pi^{\ast} \mid \mathcal{G}, \pi^{\dagger})$}

\label{SEC:Ppiastconditional}
Conditional on $\pi^\dagger$, we 
use the similar prior as that in Spat-RPM to induce partitions of quotient vertices of $\mathcal{G}/\pi^\dagger$.
Specifically, following the priors in (\ref{EQ:spanningtreeprior}) -  (\ref{EQ:kprior}), we assume a UST prior on the spanning trees of $\mathcal{G}/\pi^\dagger$: $    \mathbb{P}(\mathcal{T}\mid \mathcal{G/\pi^\dagger})\propto \mathbb{I}\{\mathcal{T}\in \Delta(\mathcal{G}/\pi^\dagger)\}$,
and a Poisson prior on the number of clusters $k$ with mean parameter $\lambda$: $ k\sim \tmop{Poisson}(\lambda) \cdot \mathbb{I}(1\leq k\leq k^\dagger)$.

Conditional on $\mathcal{T}$ and $k$, $\pi^\ast/\pi^\dagger$ is induced by randomly cutting $k-1$ edges from $\mathcal{T}$. It has been shown in Proposition 2 of \cite{luo2021bayesian} that such prior induces a partition space that contains all contiguous partitions w.r.t. $\mathcal{G}/\pi^\dagger$. The corresponding partition of $\mathcal{V}$, i.e., $\pi^\ast$, is then obtained by assigning the same cluster membership to all original vertices whose quotient vertices belong to the same cluster of $\pi^\ast/\pi^\dagger$.

Conditional on $\pi^\ast$, we adopt the same hierarchical model as specified in (\ref{EQ:priortheta}) - (\ref{EQ:priory}). The posterior distribution of $\pi^\ast$ can be derived afterwards.


\subsection{Detailed computational strategy specific to A-Spat-RPM}
\label{SEC:AOS computation}
Building on Section \ref{SEC:Bayesian computation}, this section presents a specific computational strategy for A-Spat-RPM. The details are deferred to Section \ref{SEC:AOS computationS} of Supplementary Material.

\section{Theoretical results of A-Spat-RPM}
\label{SEC:AOS theory}
In this section, we conduct a theoretical analysis of A-Spat-RPM introduced in Section~\ref{SEC:AOScaes}. Recall that we assume $\mathcal{D}$ can be partitioned into several heterogeneous clusters. The main theoretical objective is to establish posterior consistency for the domain partition, after which consistency for $\tmmathbf{\theta}(\cdot)$ and $\mu(\cdot)$ can be derived in a relatively standard manner.

 Let $k_0$ denote the true number of clusters. Write $\mathcal{D}_{l,0}$, $1\leq l\leq k_0$ as the $l$-th true cluster and $\tmmathbf{\theta}_{l, 0}$  the true regression coefficient within  $\mathcal{D}_{l,0}$. In the theoretical analysis, \cite{huang2025consistent} considers two different asymptotic regimes: Fixed Data-Generating (FDG) regime, where the true domain partition $\{\mathcal{D}_{l,0}\}_{l=1}^{k_0}$ and regression coefficients $\{\tmmathbf{\theta}_{l,0}\}_{l=1}^{k_0}$ are fixed as $n \rightarrow \infty$, and Varying Data-Generating (VDG) regime, where $k_0$ may grow with $n$, and both $\{\mathcal{D}_{l, 0} \}_{l = 1}^{k_0}$ and $\{\tmmathbf{\theta}_{l,0}\}_{l=1}^{k_0}$ may vary as $n \rightarrow \infty$. Since the FDG regime can be considered as a special case of the VDG regime, where the truth remains fixed as $n\rightarrow \infty$, the theoretical results are presented under the VDG regime in \cite{huang2025consistent}.

In this paper, we follow a theoretical framework similar to that in \cite{huang2025consistent} and present the results under the VDG regime. For simplicity, we assume $r_n$, the radius of the ball containing $\{\tmmathbf{\theta}_{l,0}\}_{l=1}^{k_0}$ in \cite{huang2025consistent}, to be a constant, although the results can be trivially extended to a more general $r_n$. We assume $k_0$ diverges in a way that  
\begin{equation}
    k_0 = O \left( n^{\frac{d \tau}{2 d + 2}} \right)
    \label{EQ:k0oarder}
\end{equation}
for some $\tau<1/2$, where we recall that $d$ is the dimension of $\tmmathbf{x}(\cdot)$.

One drawback of Spat-RPM is that, under the VDG regime, the hyperparameter condition on $(K,\lambda)$ required for consistency (see Assumption 5 of \cite{huang2025consistent}), involves the order of the unknown quantity $k_0$. Simply assuming $k_0$ to be a constant and applying the hyperparameter condition derived under the FDG regime typically does not yield domain partition consistency, when the data-generating process follows the VDG regime. To address this issue, \cite{huang2025consistent} proposes to first estimate $k_0$ and then plug this estimator into Assumption 5 for hyperparameter selection. However, this strategy uses the data twice, which is undesirable from a Bayesian perspective, and also requires additional computation.

By contrast, A-Spat-RPM provides a more elegant approach to address this issue. Specifically, we show that, for A-Spat-RPM, the hyperparameter condition selected under the FDG regime (see Assumption \ref{AS:hyperpara} later) is sufficient to achieve domain partition consistency even when the data-generating process follows the VDG regime. In other words, A-Spat-RPM automatically adapts to the data-generating process, including the unknown quantity $k_0$. Furthermore, the resulting rate under the VDG regime improves upon that in \cite{huang2025consistent}, attributable to the reduced model space induced by A-Spat-RPM.

To establish the consistency theory, we verify the high-level assumptions in Section~\ref{SEC:general graph} for A-Spat-RPM. This section is organized as follows.  Section~\ref{SEC:SpecificAS} presents primitive assumptions specific to A-Spat-RPM. Under these assumptions, Section~\ref{SEC:existenceofgoodpartition} verifies Assumptions~\ref{AS:pidaggerspace} - \ref{AS:Tdaggercompatible} and applies Lemma~\ref{PP:goodcondition} to establish the existence of the desirable partition. Section \ref{SEC:pidaggerpiastconcertration} verifies Assumptions \ref{AS:ansum} - \ref{AS:anfurther}, and establishes the corresponding consistency results. We refer to Section \ref{SEC:notation2} in Supplementary Material for notation used throughout the paper.

\subsection{Primitive assumptions}
\label{SEC:SpecificAS}

We adopt the same Assumptions 1 - 4 from \cite{huang2025consistent}, with $r_n$ treated as a constant as discussed above. These assumptions concern the spatial distribution of the locations $\{\tmmathbf{s}_i\}_{i=1}^{n}$, regularity conditions on the true domain partition, separability among $\{\tmmathbf{\theta}_{l,0}\}_{l=1}^{k_0}$, and non-collinearity of the covariates $\{\tmmathbf{x}(\tmmathbf{s}_i)\}_{i = 1}^n$. These assumptions have been justified in \cite{huang2025consistent}. For reference, we restate them as Assumptions \ref{AS:distributionloc} - \ref{AS:covariates} in Section \ref{SEC:AS1TO4} of Supplementary Material. 

The next assumption is on hyperparameters of A-Spat-RPM.

\setcounter{Passumption}{4}
{\Passumption{\label{AS:hyperpara}We assume that hyperparameters $(K,k^\dagger,\lambda)$ satisfy
\begin{align}
  K \sim  & n^{1 / 2} \log^{- (1 + \alpha_b) / 2} (n), \label{EQ:K}\\
  k^{\dagger} \sim  & K \log^{\alpha_{\dagger}} (n) \label{EQ:kdaggerorder}\\
  \shortintertext{for some $0<\alpha_{\dagger}\leqslant\alpha_b$, and}
  \lambda = o(1) & \text{ with }  \log (\lambda^{- 1}) \sim n^{\frac{3 d + 2}{4 d + 4}} \label{EQ:lambdarate}
\end{align}
}
}

As discussed above, we choose hyperparameters according to the FDG regime. The order of $K$ specified in Equation~(\ref{EQ:K}) coincides with its optimal order in Spat-RPM, so that the optimal contraction rate can be achieved under the FDG regime; see Corollary 2 of \cite{huang2025consistent}. Following Proposition 1 of \cite{huang2025consistent}, the desirable partition $\pi_0^\ast$, which will be formally introduced in Section~\ref{SEC:existenceofgoodpartition}, satisfies $| {\Omega }_{\pi_0^{\ast}}(\pi^{\dagger \dagger}_{q})| \leqslant c K$ under the FDG regime. Hence, Equation~(\ref{EQ:kdaggerorder}) ensures that $k^{\dagger}$ is larger than $| {\Omega }_{\pi_0^{\ast}}
(\pi^{\dagger \dagger}_{q})|$ in an asymptotic sense, so that Assumption~\ref{AS:Tdaggercompatible} can be satisfied with $p^\dagger=q$. The Poisson hyperparameter $\lambda$ penalizes the number of clusters. Since the likelihood tends to favor larger numbers of clusters to better fit the data, we take $\lambda=o(1)$ to strengthen the penalty and counterbalance this effect. However, $\lambda$ should not converge to zero too rapidly; otherwise, the posterior may underestimate the number of clusters. Equation (\ref{EQ:lambdarate}) specifies the order condition on $\log(\lambda^{-1})$ needed to balance this trade-off.




For simplicity, we assume that $K$ grows at the order specified in (\ref{EQ:K}) and that $\log_2 (K)$ is an integer. This constraint is only to simplify the theoretical analysis; the unconstrained case can be handled by a trivial extension. Under this constraint, $q$, the number of constructed nested partitions $\lbrace \pi^{\dagger\dagger}_p\rbrace_{p=1}^q$ by Algorithm \ref{AL:nested_grid_partition}, satisfies $q = 1 +
\log_2 (K)$. Furthermore, for each $1 \leqslant p \leqslant q$, the quotient graph $\mathcal{G} / \pi^{\dagger \dagger}_p$ is a $2^{p-1}$ by $2^{p-1}$ mesh grid, where each quotient vertex represents a union of $2^{q-p}\times 2^{q-p}$ blocks in $\{B_m\}_{m=1}^{K^2}$.

\subsection[]{The desirable partition $\pi_0^{\ast}$}
\label{SEC:existenceofgoodpartition}

Recall that Spat-RPM uses $K^2$ blocks to cover the domain $\mathcal{D}$, and induces the domain partition through partitions of these blocks. However, some blocks may intersect the boundary of the true domain partition $\{\mathcal{D}_{l,0}\}_{l=1}^{k_0}$. Such boundary-crossing blocks contain locations from different true clusters, while locations within the same block are assigned to the same cluster membership by the model. Consequently, there is a discrepancy between the block-induced partition space and the true domain partition. The desirable partition $\pi_0^\ast$ is therefore not the true domain partition itself, but rather the partition in the block-induced partition space that closely approximates $\{\mathcal{D}_{l,0}\}_{l=1}^{k_0}$. The next lemma formally states the desirable partition $\pi_0^\ast\in \Xi_{\tmop{direct}}^\ast(\mathcal{G})$ under Spat-RPM, and verifies Assumption \ref{AS:pi0existence}.

\begin{lemma}
\label{LM:goodpartition}
 Let $p^{\dagger} = 1+ \lfloor \log_2 (k_0^{- 1 / 2} K) \rfloor $. Under Assumption \ref{AS:lengthofboundary}, there exists a contiguous partition $\pi_0^{\ast} \succeq \pi_{p^\dagger}^{\dagger\dagger}$, say $\pi^{\ast}_0 =
  \{\mathcal{D}_{1, 0}^{\ast}, \ldots \mathcal{D}_{k_0, 0}^{\ast} \}$,  such that
$   | \mathcal{W}_{\pi_0^{\ast}} | = \sum_{j = 1}^{k_0} | \{B_m : B_m
     \subseteq \mathcal{D}_{j, 0}^{\ast}, B_m \subsetneq \mathcal{D}_{j, 0} \}
     | \leqslant c k_0 K$ 
  for some constant $c$, where $\mathcal{W}_{\pi_0^{\ast}} = \cup_{j =
  1}^{k_0} \{B_m : B_m \subseteq \mathcal{D}_{j, 0}^{\ast}, B_m \subsetneq
  \mathcal{D}_{j, 0} \}$ is the set of blocks inside of $\mathcal{D}_{j,0}^\ast$, but not fully included in $\mathcal{D}_{j,0}$.
\end{lemma}


In the construction of $\pi_0^\ast$, each cluster $\mathcal{D}_{j,0}^{\ast}\in \pi^\ast_0$ is designed to approximate $\mathcal{D}_{j,0}$. Thus, $\mathcal{W}_{\pi_0^{\ast}}$ represents the set of blocks containing misclustered locations. The area of $\mathcal{W}_{\pi_0^{\ast}}$, upper bounded by $c k_0 K \times K^{-2}=O(k_0 K^{-1})$, represents the ``approximation error" between the block-induced partition space and the true domain partition. According to Equations (\ref{EQ:k0oarder}) and (\ref{EQ:kdaggerorder}), this ``approximation error" converges to $0$ as $n \rightarrow \infty$.

Note that the ``approximation error" increases with $k_0$. This is because, under Assumption~\ref{AS:lengthofboundary}, the total boundary length of the true partition increases with $k_0$, which in turn increases the number of boundary-crossing blocks, and hence the number of blocks containing misclustered locations. Besides, since the appropriate resolution level $p^\dagger$ depends on the unknown true number of clusters $k_0$, introducing the nested partitions $\{\pi_p^{\dagger \dagger}\}_{p=1}^q$ ensures that $p^\dagger$ is represented in the model space, allowing A-Spat-RPM to adapt to $k_0$ without explicitly estimating it.






Following Lemma \ref{PP:goodcondition}, it remains to verify Assumptions \ref{AS:pidaggerspace} and \ref{AS:Tdaggercompatible} to ensure that $\pi_0^\ast$ is contained in the partition space induced by A-Spat-RPM. Since the support of prior model specified in Section \ref{SEC:Ppiastconditional} contains all contiguous partitions of quotient vertices in $\mathcal{G}/\pi^\dagger$, Assumption \ref{AS:pidaggerspace} is satisfied following Lemma \ref{LM:verifyAS1_H}. The following lemma verifies Assumption \ref{AS:Tdaggercompatible}.


  

\begin{lemma}
  \label{LM:goodpidagger} Let $p^{\dagger}$ and $\pi^\ast_0$ be as introduced in Lemma~\ref{LM:goodpartition}. Under Assumptions \ref{AS:lengthofboundary} and \ref{AS:hyperpara}, we have $k^\dagger\gg | \Omega_{\pi_0^{\ast}}
    (\pi^{\dagger \dagger}_{p^{\dagger}}) |$.
\end{lemma}

Combining the above results and applying Lemma~\ref{PP:goodcondition}, we conclude that, for A-Spat-RPM, there exists a $\pi^{\dagger}_0\in \Xi^\dagger$ such that $\pi_0^{\ast} \in \Xi^{\ast}(\pi^{\dagger}_0)$.

\subsection[]{Consistency results}
\label{SEC:pidaggerpiastconcertration}

We first introduce a distance between two domain partitions, say $\pi_1 (\mathcal{D}) = \{\mathcal{D}_{11}, \ldots,
\mathcal{D}_{1 k_1} \}$ and $\pi_2 (\mathcal{D}) = \{\mathcal{D}_{21}, \ldots,
\mathcal{D}_{2 k_2} \}$ with $k_1$ and $k_2$ denoting the number of clusters.
We write their distance metric as
\[ \epsilon \{\pi_1 (\mathcal{D}), \pi_2 (\mathcal{D})\} = 2 -
   |\mathcal{D}|^{- 1} \left[ \sum_{j = 1}^{k_1} \max_{l \in \{1, \ldots, k_2
   \}} |\mathcal{D}_{1 j} \cap \mathcal{D}_{2 l} | + \sum_{l = 1}^{k_2}
   \max_{j \in \{1, \ldots, k_1 \}} |\mathcal{D}_{1 j} \cap \mathcal{D}_{2 l}
   | \right] . \]
   
It has been shown in \cite{huang2025consistent} (Proposition 2) that $\epsilon(\cdot,\cdot)$ is a valid distance, and we refer to the content after Equation (15) of \cite{huang2025consistent} for a detailed understanding of $\epsilon(\cdot,\cdot)$. Based on $\epsilon(\cdot,\cdot)$, we define a neighborhood of the true partition $\{\mathcal{D}_{l,0}\}_{l=1}^{k_0}$ as
\[ \Pi_0^{\ast} = \left\{ \pi^{\ast} \in \Xi^{\ast} : | \pi^{\ast} | = k_0,
   \text{ and } \epsilon (\pi^{\ast}, \{ \mathcal{D}_{l, 0} \}_{l = 1}^{k_0})
   \leqslant 2 k_0^{1 + 2 / d} n^{- 1 / 2} \log^{\alpha_0 + (1 + \alpha_b) /
   2} (n) \right\} \]
for some $\alpha_0 > 0$, where $\Xi^{\ast}=\cup_{\pi^\dagger\in \Xi^\dagger}\Xi^\ast(\pi^\dagger)$ is the space of $\pi^\ast$ under A-Spat-RPM.

Under Equation (\ref{EQ:k0oarder}), we have $k_0^{1 + 2 / d} n^{- 1 / 2} \log^{\alpha_0 + (1 + \alpha_b) /
   2} (n)\rightarrow 0 $. Thus, $\Pi_0^\ast$ is considered as the desirable domain partition set: each partition in this set has the correct number of clusters, and its distance from the true domain partition $\{ \mathcal{D}_{l, 0} \}_{l = 1}^{k_0}$ converges to $0$. Furthermore, for $\pi_0^\ast$ introduced in Lemma \ref{LM:goodpartition}, it can be shown that $\pi_0^\ast \in \Pi_0^\ast$ when $n \rightarrow \infty$ following Assumption \ref{AS:hyperpara}.  Recall the definition of the desirable set $\Pi_0^\dagger$ in (\ref{EQ:Pi0dagger}). We have the following results.

\begin{theorem}
  \label{TH:piast}Under Assumptions \ref{AS:distributionloc} - \ref{AS:hyperpara}, with probability tending to 1, we have $     \mathbb{P} (\pi^{\dagger} \in \Pi_0^{\dagger} \mid \mathcal{G}, \mathfrak{D})
     \rightarrow 1, \text{ and }\mathbb{P} (\pi^{\ast} \in \Pi_0^{\ast} \mid \mathcal{G}, \mathfrak{D}) \rightarrow
     1.$
  \begin{proof}
      The proof is conducted by verifying Assumption \ref{AS:ansum} and applying Lemma \ref{LM:pidaggerconsistency_general}, with details deferred to Section \ref{SEC:proofsforaosconsistency} in Supplementary Material.
  \end{proof}
\end{theorem}

Theorem~\ref{TH:piast} establishes the consistency of A-Spat-RPM: with probability tending to $1$, the posterior distributions of $\pi^\dagger$ and $\pi^\ast$ concentrate on their respective desirable sets. By the definition of $\Pi_0^\ast$, the posterior contraction rate for the domain partition is $k_0^{1 + 2 / d} n^{- 1 / 2} \log^{\alpha_0 + (1 + \alpha_b) /2} (n)$. This rate is slightly faster than the optimal contraction rate established in \cite{huang2025consistent} under the VDG regime, which, according to Corollary 1 therein, is $k_0^{1 + 3 / d} n^{- 1 / 2} \log^{\alpha_0 + (1 + \alpha_b) /2} (n)$. This improvement arises from the reduced partition space of A-Spat-RPM, which still retains the desirable partition $\pi_0^\ast$. Furthermore, although the contraction rate depends on $k_0$, which may diverge, the hyperparameter choices in Assumption~\ref{AS:hyperpara} are based on the FDG regime and independent of $k_0$. Therefore, the consistency result establishes that A-Spat-RPM automatically adapts to different values of $k_0$, a property not directly available in Spat-RPM.

Given the consistency result for domain partition, the consistency results for estimating $\{\tmmathbf{\theta}_{l,0}\}_{l=1}^{k_0}$ and predicting $\mu(\cdot)$ can be established using arguments similar to those in \cite{huang2025consistent}. For brevity, we do not present these results in the paper.

For $\pi^\dagger \in \Xi^\dagger \cap\Pi_0^\dagger$, Equation~(\ref{EQ:kdaggerorder}) shows that the number of quotient vertices in $\mathcal{G}/\pi^\dagger$ is of order $K \log^{\alpha_\dagger}(n)$, which is substantially smaller than the $K^2$ vertices in $\mathcal{G}$ when $K$ is large. This reduction lowers both the computational burden and model uncertainty when inducing partitions conditional on $\pi^\dagger$. Accordingly, one may consider fixing a $\pi^\dagger \in \Xi^\dagger \cap\Pi_0^\dagger$ and inducing partitions conditional on it. The next theorem provides theoretical justification for this conditional strategy by showing that conditioning on any $\pi^\dagger \in \Xi^\dagger \cap \Pi_0^\dagger$ suffices to achieve posterior consistency for the domain partition.

\begin{theorem}
  \label{TH:piastpidagger}Under Assumptions \ref{AS:distributionloc} - \ref{AS:hyperpara}, with
  probability tending to 1, we have $\sup_{\pi^\dagger \in \Xi^\dagger \cap \Pi_0^\dagger}\mathbb{P} (\pi^{\ast} \nin \Pi_0^{\ast} \mid \mathcal{G}, \mathfrak{D},
     \pi^{\dagger}) \rightarrow 0.$
    \begin{proof}
      The proof is conducted by verifying Assumption \ref{AS:anfurther} and applying Lemma \ref{LM:piastconditionalpidagger}, with details deferred to Section \ref{SEC:proofsforaosconsistency} in Supplementary Material.
  \end{proof}
\end{theorem}

\section{Simulation}
\label{SEC:simu}
The simulation study consists of three parts. In Section \ref{SEC:asymptotic}, we investigate the asymptotic properties of A-Spat-RPM. In Section \ref{SEC:n10000case}, we demonstrate the effectiveness of A-Spat-RPM by presenting a fitting result under $n=50000$. In Section \ref{SEC:comparison}, we compare the performance between A-Spat-RPM and Spat-RPM.

\subsection{Asymptotic analysis}
\label{SEC:asymptotic}

\begin{figure}[htbp]
		\centering
		\begin{tabular}{ccc}
        			\raisebox{0mm}{\hspace{-2mm}{\includegraphics[width=0.38\linewidth,height=0.2\textheight]{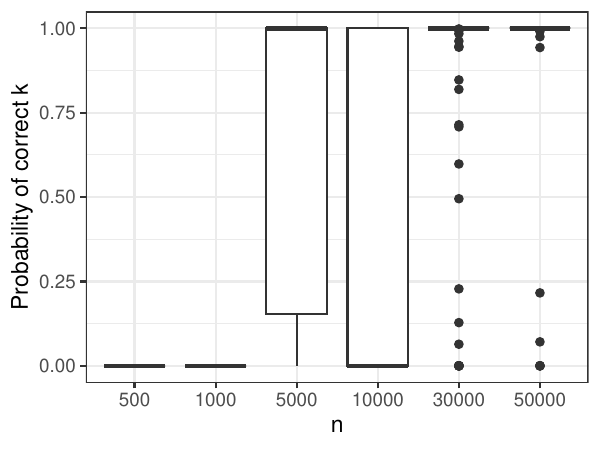}}}

            &\raisebox{2mm}{\hspace{-8mm}{\includegraphics[width=0.39\linewidth,height=0.2\textheight]{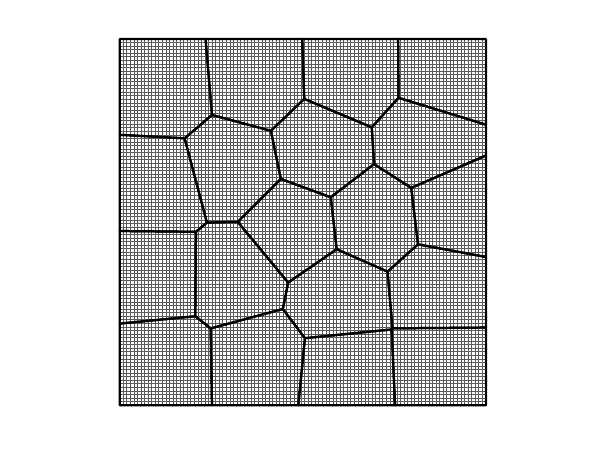}}}
            &\raisebox{-1.3mm}{\hspace{-20mm}{\includegraphics[width=0.38\linewidth,height=0.215\textheight]{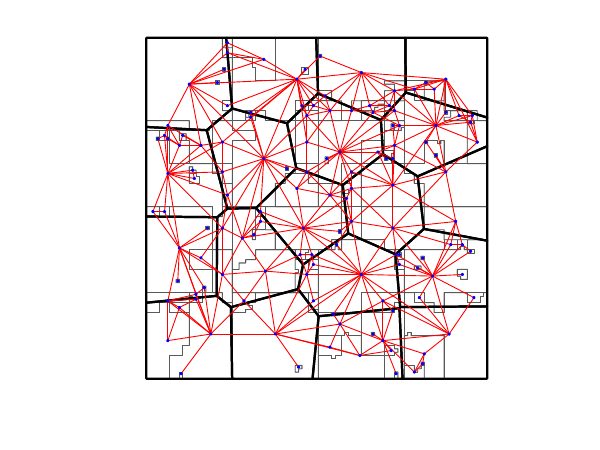}}}
            \\
            \hspace{-2mm}{\small (a)} & \hspace{-8mm}{\small (c)} & \hspace{-18mm} {\small (e)}
			\\
            
            \raisebox{-2mm}{\hspace{-2mm}{\includegraphics[width=0.38\linewidth,height=0.2\textheight]{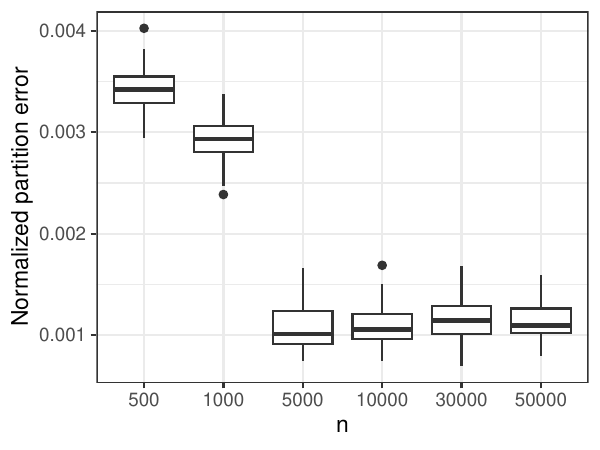}}}

            &\raisebox{2mm}{\hspace{-8mm}{\includegraphics[width=0.38\linewidth,height=0.19\textheight]{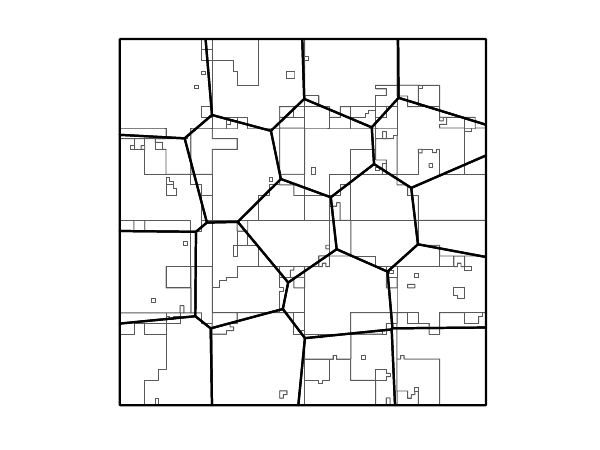}}}
            &\raisebox{2mm}{\hspace{-17mm}{\includegraphics[width=0.36\linewidth,height=0.19\textheight]{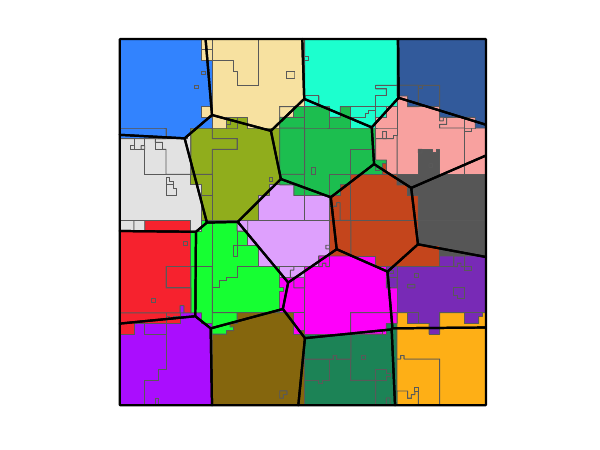}}}
            \\
            \hspace{-2mm}{\small (b)} & \hspace{-8mm}{\small (d)} & \hspace{-18mm} {\small (f)}\\
		\end{tabular}
		\caption{ (a) Boxplots of the posterior probability of the correct number of clusters. (b) Boxplots of the posterior mean of normalized $\epsilon(\pi^\ast,\{\mathcal{D}_{l,0}\}_{l=1}^{k_0})$. (c) The true domain partition $\{\mathcal{D}_{l,0}\}_{l=1}^{k_0}$, where the boundary is denoted by thick black lines, and the constructed blocks $\{B_m\}_{m=1}^{K^2}$ under $n=50000$. Owing to the large number of blocks, $\{B_m\}_{m=1}^{K^2}$ are clearly visible only at higher magnification. (d) A posterior sample of the intermediate partition $\pi^\dagger$, whose clusters are unions of the blocks shown in (c). The cluster boundaries in $\pi^\dagger$ are represented by thin lines. 
      (e) Quotient graph $\mathcal{G}/\pi^\dagger$ constructed from (d). Each quotient vertex corresponds to a cluster of $\pi^\dagger$ and is represented by blue dots. Edges are represented by red lines. (f) The posterior domain partition sample induced by the graph $\mathcal{G}/\pi^\dagger$ in (e), and selected by \texttt{salso}. Different colors represent different clusters.}
		\label{FIG:piastanddagger}
        \vspace{-10pt}
	\end{figure}

Following Equation~(\ref{EQ:model}), we adopt the same data-generating process as that in the simulation study (Section 4.1) of \cite{huang2025consistent}. The hyperparameters $(K,k^\dagger,\lambda)$ are chosen according to Assumption \ref{AS:hyperpara}. Further details are provided in Section \ref{SEC:simulationdetail} of the Supplementary Material. To investigate the asymptotic properties of A-Spat-RPM, we compare posterior samples for $n = (500, 1000, 5000, 10000,30000, 50000)$, with $100$ simulation repeats for each $n$. For each repeat, we obtain posterior samples following Section \ref{SEC:AOS computation}, with $20000$ MCMC iterations, a $15000$ burn-in period, and a thinning parameter of $5$. Based on MCMC samples, we compute the posterior probability of the correct number of clusters and the posterior mean of the normalized domain partition error (defined as $\epsilon(\pi^\ast,\{\mathcal{D}_{l,0}\}_{l=1}^{k_0})$ scaled by the contraction rate established in Theorem \ref{TH:piast}). The corresponding boxplots across repeats and different values of $n$ are shown in Figures \ref{FIG:piastanddagger}(a - b), respectively. 

We can see from Figure \ref{FIG:piastanddagger}(a) that when $n$ is small, the posterior number of clusters does not concentrate on the true value, but does so once $n\geq 30000$ in most of the repeats. In addition, Figure \ref{FIG:piastanddagger}(b) shows that the normalized domain partition error first decreases rapidly with $n,$ and then remains relatively stable as expected when $n \geq 5000$. These results demonstrate the consistency of A-Spat-RPM, and align with Theorem \ref{TH:piast}.

\subsection[]{A result under $n=50000$}
\label{SEC:n10000case}

Among the fitting results obtained in Section \ref{SEC:asymptotic}, we focus on a case with $n=50000$ to demonstrate the effectiveness of A-Spat-RPM. Figure \ref{FIG:piastanddagger}(c) shows the true domain partition $\{\mathcal{D}_{l,0}\}_{l=1}^{k_0}$, whose boundaries are denoted by thick black lines, together with the constructed blocks $\{B_m\}_{m=1}^{K^2}$. The number of blocks is $K^2 = 103^2$, and the mesh grid graph $\mathcal{G}$ is constructed treating each $B_m$ as a vertex. Directly partitioning $\{B_m\}_{m=1}^{K^2}$ based on $\mathcal{G}$ results in a prohibitively large partition space. In contrast, A-Spat-RPM first induces a quotient graph $\mathcal{G}/\pi^\dagger$ with a vertex number $k^\dagger=338$, and then induces partitions based on it.

Figure \ref{FIG:piastanddagger}(f) shows the posterior domain partition sample selected by R package \texttt{salso}, where different colors represent different clusters. The posterior domain partition recovers the true partition well, except for some locations near the boundaries. Figures \ref{FIG:piastanddagger}(d - e) show the corresponding posterior $\pi^\dagger$ and $\mathcal{G}/\pi^\dagger$, respectively. Each cluster in $\pi^\dagger$, and the corresponding quotient vertex in $\mathcal{G}/\pi^\dagger$, represent a union of the blocks in $\{B_m\}_{m=1}^{K^2}$. The result illustrates the data-adaptive behavior of posterior $\pi^\dagger$ and $\mathcal{G}/\pi^\dagger$: near the true boundaries, smaller clusters and larger number of quotient vertices are retained to preserve the model’s ability to approximate the true domain partition, whereas in regions far from the boundaries, larger clusters and fewer quotient vertices are used to reduce the partition space and improve efficiency. In the next section, we further demonstrate the advantages of A-Spat-RPM by comparing it with Spat-RPM, which directly partitions blocks $\{B_m\}_{m=1}^{K^2}$ based on $\mathcal{G}$.

\subsection{Comparison with Spat-RPM}
\label{SEC:comparison}

We use the same simulation setting as in Section \ref{SEC:asymptotic} with $n=50000$. The hyperparameters of Spat-RPM (i.e., $K$ and $\lambda$), are set to the same values as those used for A-Spat-RPM. The simulation is repeated $100$ times. In each repeat, both models are fitted to the same dataset. The comparison results are as follows.

\begin{figure}[htbp]
		\centering
		\begin{tabular}{ccc}
        \hspace{-2mm}
			{\includegraphics[width=0.3\linewidth,height=0.25\textheight]{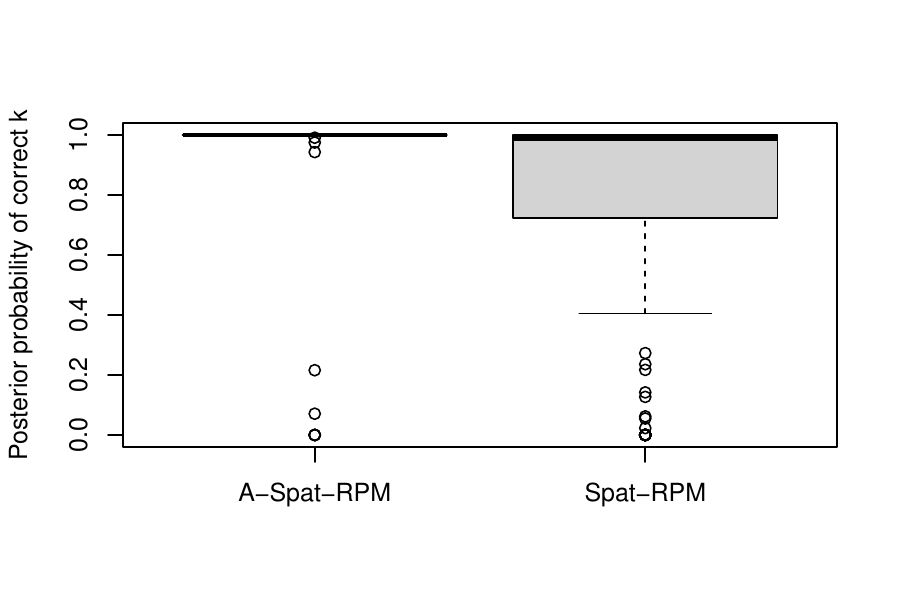}}&
            \hspace{-5mm}
			{\includegraphics[width=0.3\linewidth,height=0.25\textheight]{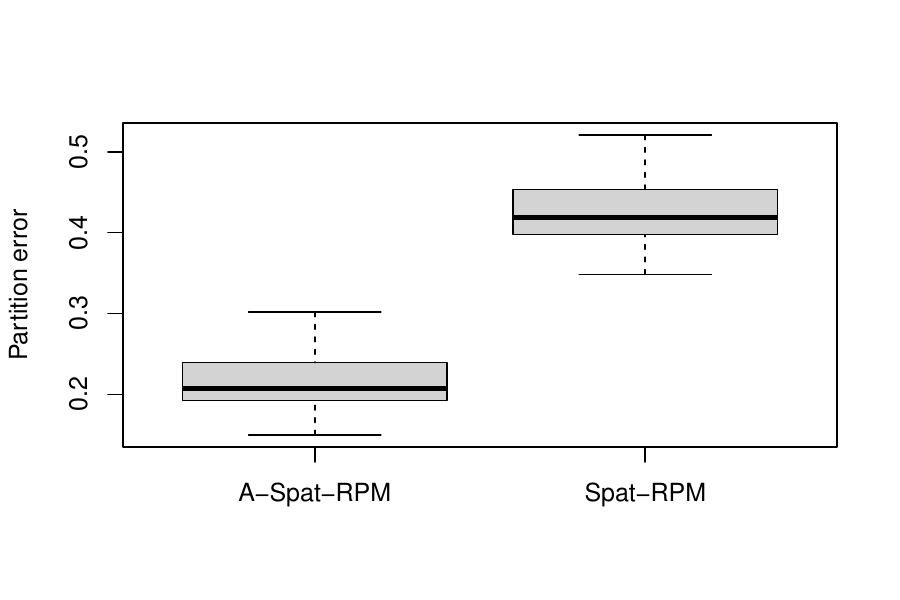}}&
            \hspace{-5mm}\raisebox{6mm}{
            {\includegraphics[width=0.4\linewidth,height=0.18\textheight]{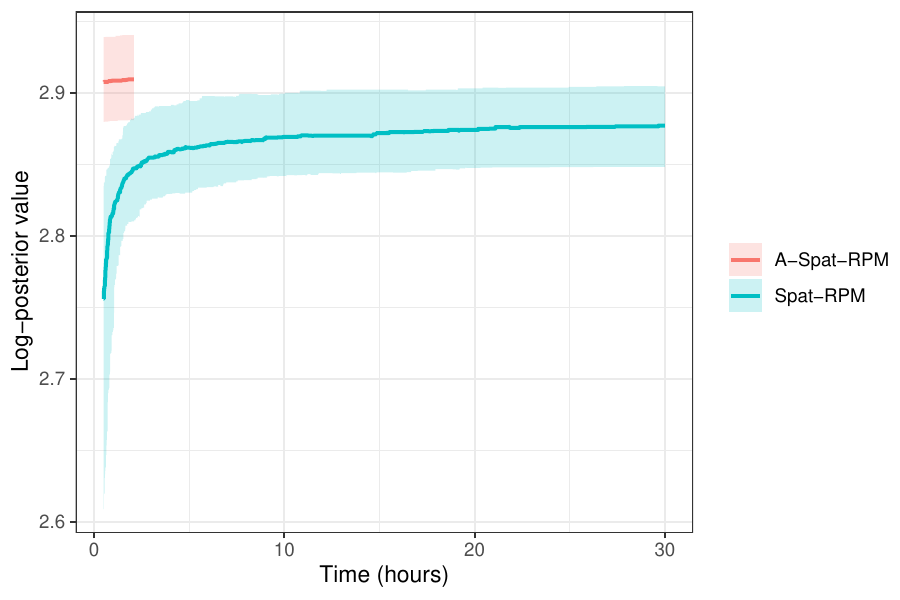}}}
            \\
            \hspace{-2mm}{\small (a)} & \hspace{-5mm}{\small (b)} & \hspace{-5mm}{\small (c)} \\
		\end{tabular}
		\caption{Comparison between A-Spat-RPM and Spat-RPM (a) Boxplots of the posterior probability of the correct number of clusters over $100$ repeats (b) Boxplots of the posterior mean of partition error over $100$ repeats. (c) Plots of $\tmop{log-post}(\pi_s^\ast)$ against $t_s$ for $t_s$ ranging from $0.5$ to $30$ hours. The middle solid line denotes the median across repeats, and the shaded region represents the area between the pointwise $2.5\%$ and $97.5\%$ quantiles. The trajectory for A-Spat-RPM is shown only over a short time interval because it completes all MCMC iterations within approximately $2$ hours.}
		\label{FIG:timecomp}
        \vspace{-10pt}
	\end{figure}

We first compare the domain partition results. Figure \ref{FIG:timecomp}(a) shows boxplots of the posterior probability of the correct number of clusters over $100$ repeats. We see that A-Spat-RPM assigns posterior probability close to one to the correct cluster number in most repeats, whereas Spat-RPM fails to do so in more than $25\%$ of the repeats. Figure \ref{FIG:timecomp}(b) shows boxplots of the posterior mean of domain partition error measured by $\epsilon(\pi^\ast,\{\mathcal{D}_{l,0}\}_{l=1}^{k_0})$, over the repeats. We see that A-Spat-RPM yields a substantially smaller partition error, reducing it by approximately $50\%$ relative to Spat-RPM. 

We next compare the computing time and sampling behavior of the two models. Over $20000$ MCMC iterations and $100$ repeats, A-Spat-RPM takes $1.80$ hours on average (SE $0.10$), compared with $39.16$ hours for Spat-RPM (SE $3.57$).
This substantial reduction in computing time arises because A-Spat-RPM updates $\pi^\ast$ (Section \ref{SEC:MCMC update piast}) conditional on the quotient graph $\mathcal{G}/\pi^\dagger$ with $k^\dagger$ vertices, whereas Spat-RPM updates $\pi^\ast$ on the mesh grid graph $\mathcal{G}$ with $K^2 \gg k^\dagger$ vertices. The larger number of vertices incurs additional computational costs when evaluating the M-H acceptance ratio and sampling spanning trees. Meanwhile, the update of $\pi^\dagger$ (Section \ref{SEC:MCMC update pidagger}) remains computationally manageable in A-Spat-RPM.

We further compare how the posterior value evolves over MCMC iterations for the two models. Note that A-Spat-RPM imposes an additional prior on $\mathcal{G}/\pi^\dagger$, so the posterior functions differ between the two models. To ensure a fair comparison, we evaluate posterior domain partition samples from both models using the same function, which is set as the posterior function under Spat-RPM, denoted by $\mathbb{P}_{\tmop{direct}}(\pi^\ast \mid \mathcal{G}, \mathfrak{D})$. $\mathbb{P}_{\tmop{direct}}(\pi^\ast \mid \mathcal{G}, \mathfrak{D})$ can be computed up to a normalizing constant. Specifically, for the $s$-th posterior domain partition sample $\pi_s^\ast$, we record the corresponding computing time $t_s$ and compute
$\tmop{log-post}(\pi_s^\ast) = n^{-1}\log\{\mathbb{P}_{\tmop{direct}}(\pi^\ast \mid \mathcal{G}, \mathfrak{D})\}$.

Figure \ref{FIG:timecomp}(c) plots $\tmop{log-post}(\pi_s^\ast)$ against $t_s$ for $t_s$ ranging from $0.5$ to $30$ hours. A-Spat-RPM reaches high log-posterior values within approximately $2$ hours, whereas Spat-RPM does not reach comparable values in most repeats even after $30$ hours. Since the space of $\pi^\ast$ under A-Spat-RPM is a subset of that under Spat-RPM, the maximum attainable value of $\tmop{log-post}(\pi_s^\ast)$ for Spat-RPM is at least as large as that for A-Spat-RPM. This suggests that Spat-RPM's failure to reach comparable posterior values is attributable to inefficient exploration of its prohibitively large partition space when $K^2$ is large. This inefficient exploration also helps explain the poorer fitting performance shown in Figures \ref{FIG:timecomp}(a - b). Overall, these results highlight the advantage of A-Spat-RPM: by restricting attention to a smaller partition space while retaining desirable partitions, it explores high-posterior regions more efficiently and achieves better performance within a practical computing time.


\section{Real data analysis}
\label{SEC:realdata}
\begin{figure}[htbp]
		\centering
		\begin{tabular}{cc}
        {\includegraphics[width=0.3\linewidth,height=0.16\textheight]{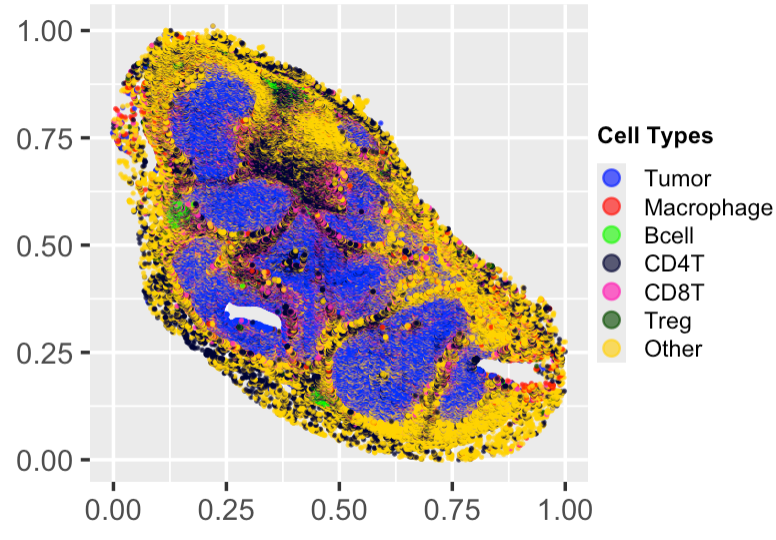}}&
			{\includegraphics[width=0.3\linewidth,height=0.16\textheight]{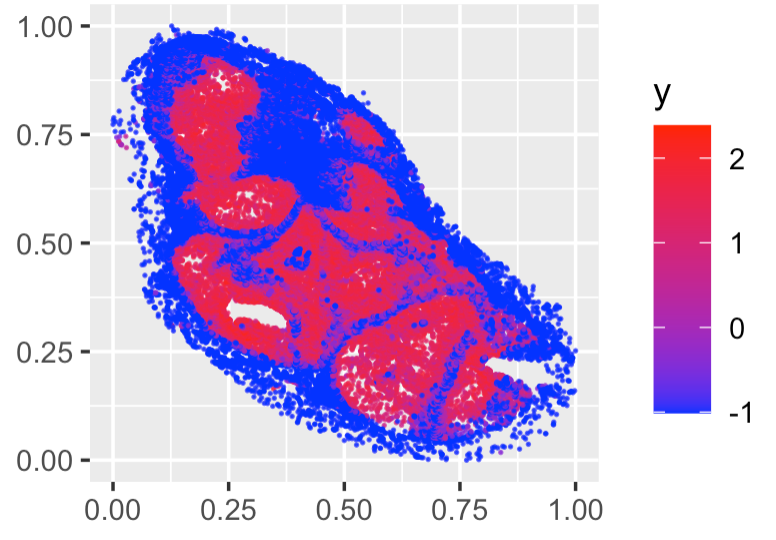}}\\
            {\small (a)} & {\small (b)}
		\end{tabular}
		\caption{(a) Spatial distribution of all 7 cell types. (b) The heatmap of the response ${y}(\tmmathbf{s}_i)$}
		\label{FIG:realdataplot}
        \vspace{-10pt}
	\end{figure}

In this section, we demonstrate the scalability of A-Spat-RPM by applying it to a large multiplexed immunofluorescence image containing millions of spatially resolved cells.
The image is a pre-treatment snapshot of a tumor tissue sample from a melanoma patient receiving anti-PD-1 immunotherapy~\citep{Realdata}.  The dataset records the spatial distribution of $7$ cell types (Figure~\ref{FIG:realdataplot}(a)), including tumor cells and several immune cell populations, such as T cells, B cells, and macrophages. For each T cell, the dataset also records the expression levels of selected functional markers, including TCF-1/7, PD-1, LAG-3, and TOX. These markers characterize distinct T-cell functional states associated with antitumor activity. The number of T cells 
is $n=123,587$. This large number of spatial locations requires computationally efficient statistical methods. 
Our scientific objective is to investigate the spatially heterogeneous interplay between the immune system and tumor~\citep{fridman_immune_2017}. At each T cell location $\tmmathbf{s}_i$, we compute the local tumor intensity as the number of tumor cells within $30\mu m$, divided by the corresponding area. We then log-transform and normalize this intensity and use the resulting value as the response ${y}(\tmmathbf{s}_i)$. The covariate $\tmmathbf{x}(\tmmathbf{s}_i)$ includes an intercept term and the scaled expression levels of T-cell functional markers. Figure \ref{FIG:realdataplot}(b) and Figures \ref{FIG:addrealdataplot}(a - d) show heatmaps of ${y}(\tmmathbf{s}_i)$ and $\tmmathbf{x}(\tmmathbf{s}_i)$, respectively, revealing substantial spatial heterogeneity. Applying A-Spat-RPM to $\{\tmmathbf{s}_i,\tmmathbf{x}(\tmmathbf{s}_i),y(\tmmathbf{s}_i)\}_{i=1}^{n}$ identifies distinct spatial domains characterized by different regression coefficients. These coefficients quantify the spatially varying relationship between marker expression and local tumor intensity, thereby providing insight into how each functional marker is associated with T-cell-mediated antitumor activity across locations.


Based on Assumption \ref{AS:hyperpara}, hyperparameters in our model are chosen as $ K = \lfloor c_b n^{{1/2}}  \log^{-(1 + \alpha_b) / 2} (n)\rfloor$, $k^{\dagger} = \lfloor c^\dagger K \log^{\alpha_{\dagger}} (n)\rfloor$,  and $\log(\lambda^{-1}) = c_p n^{(3d+2)/(4d+4)}$, where $(\alpha^\dagger,\alpha_b)=(0.1,1)$ and $d=5$. To decide values of $(c_b,c^\dagger,c_p)$, we set $(c_b,c^\dagger,c_p)\in \{1,2,3,4,5\} \times \{0.5,1,1.5,2\}\times \{0.01,0.025,0.05,0.1,0.2\}$ and select the optimal combination following the approach proposed in \cite{huang2025consistent}. We run 250,000 MCMC iterations with a burn-in period 240,000 and a thinning parameter 50. The results are shown as follows.

\begin{figure}[htbp]
		\centering
		\begin{tabular}{ccc}
		 \hspace{-2.7mm}	\raisebox{-1.4mm}{{\includegraphics[width=0.31\linewidth,height=0.157\textheight]{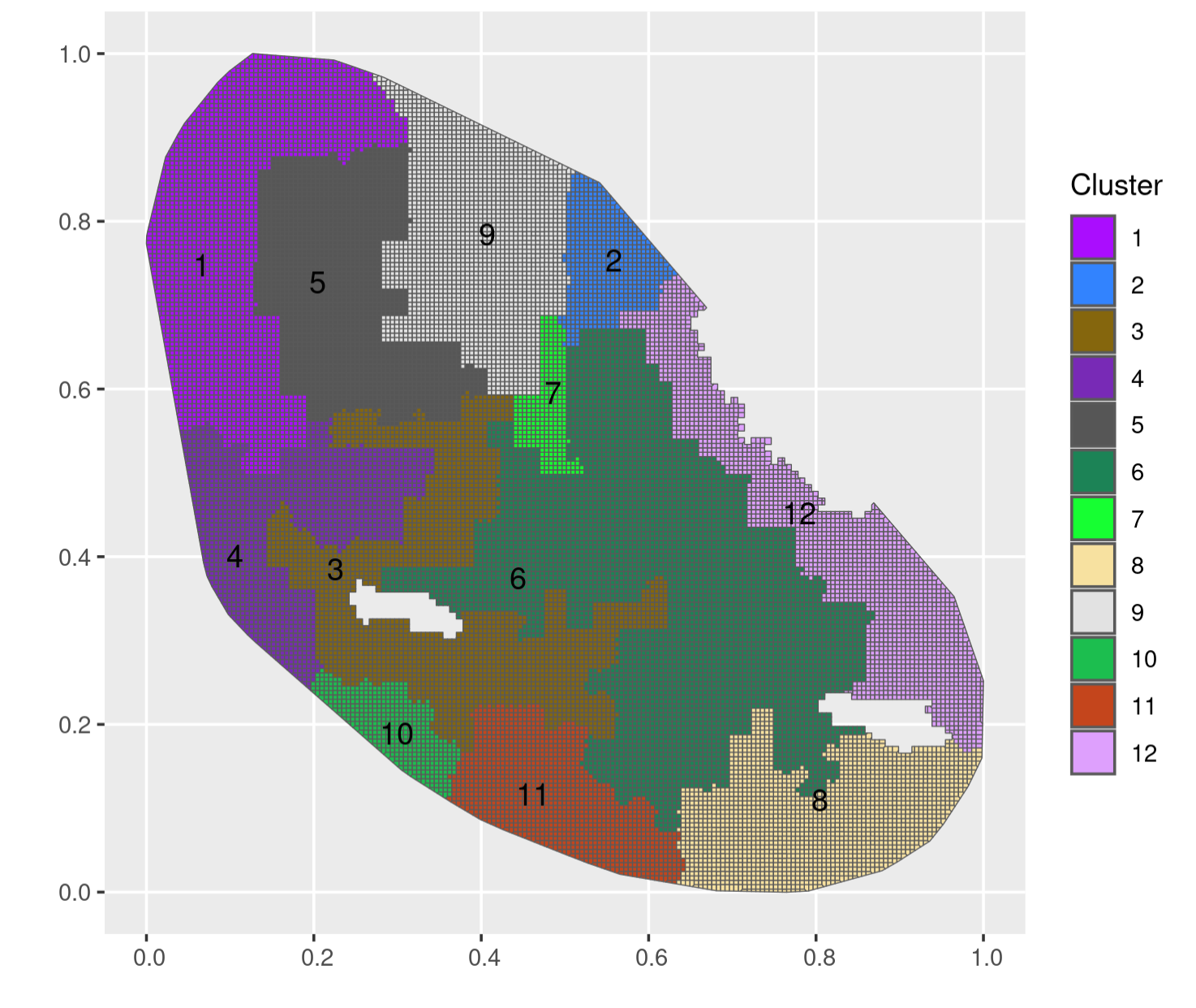}}}&
			{\includegraphics[width=0.3\linewidth,height=0.15\textheight]{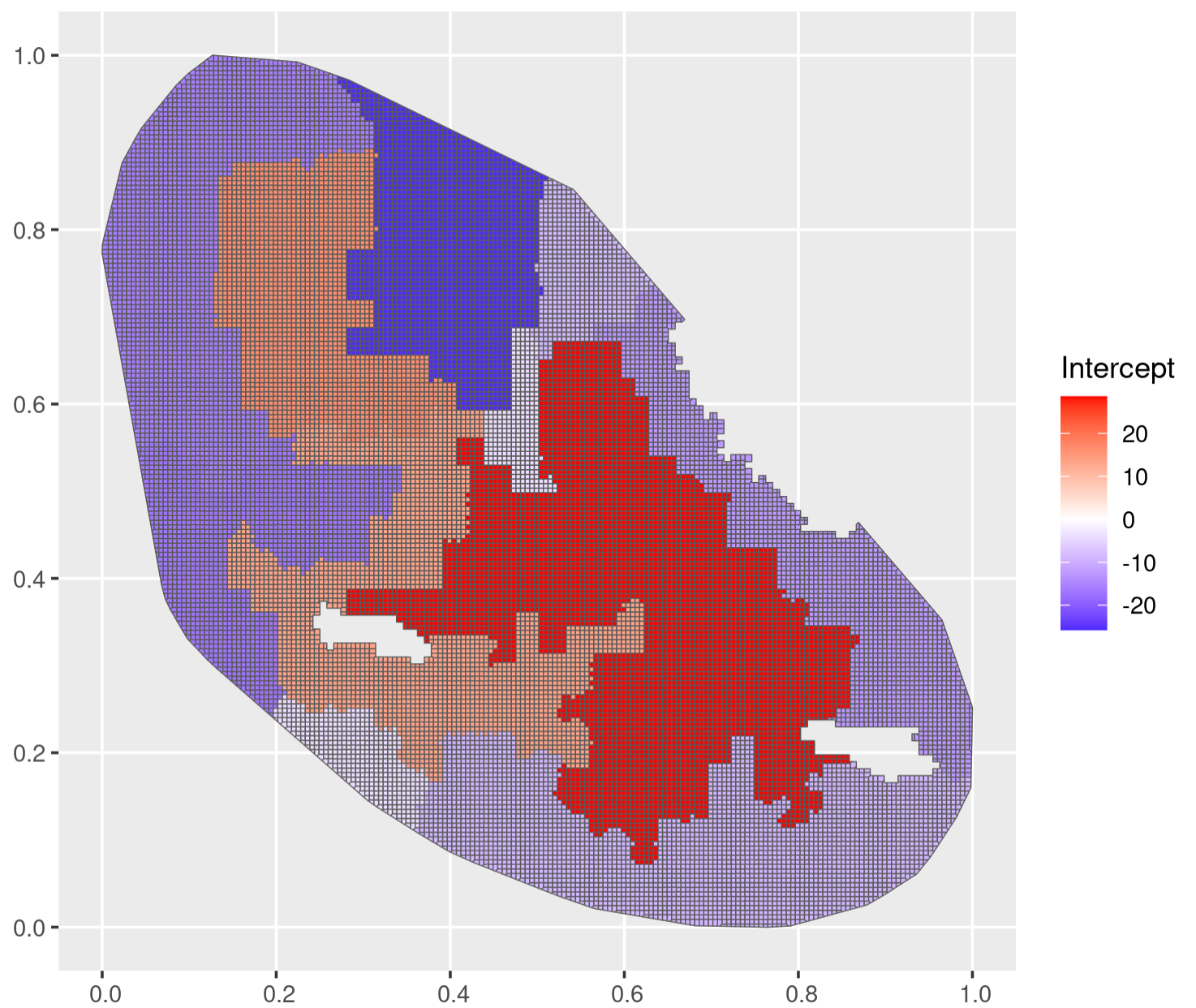}}&
			{\includegraphics[width=0.3\linewidth,height=0.15\textheight]{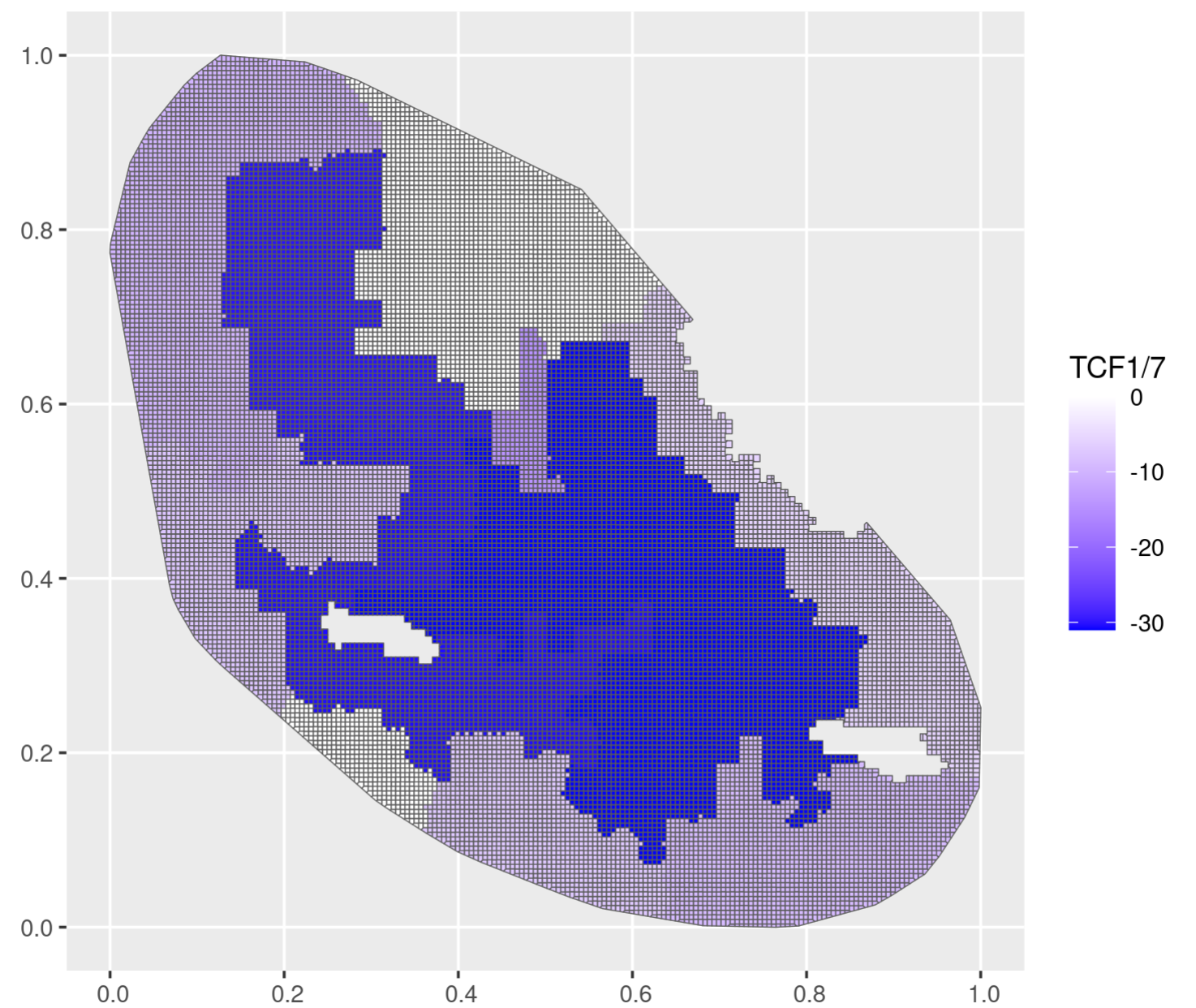}}\\
            {\small (a)} & {\small (b)} &{\small (c) }\\
            {\includegraphics[width=0.3\linewidth,height=0.15\textheight]{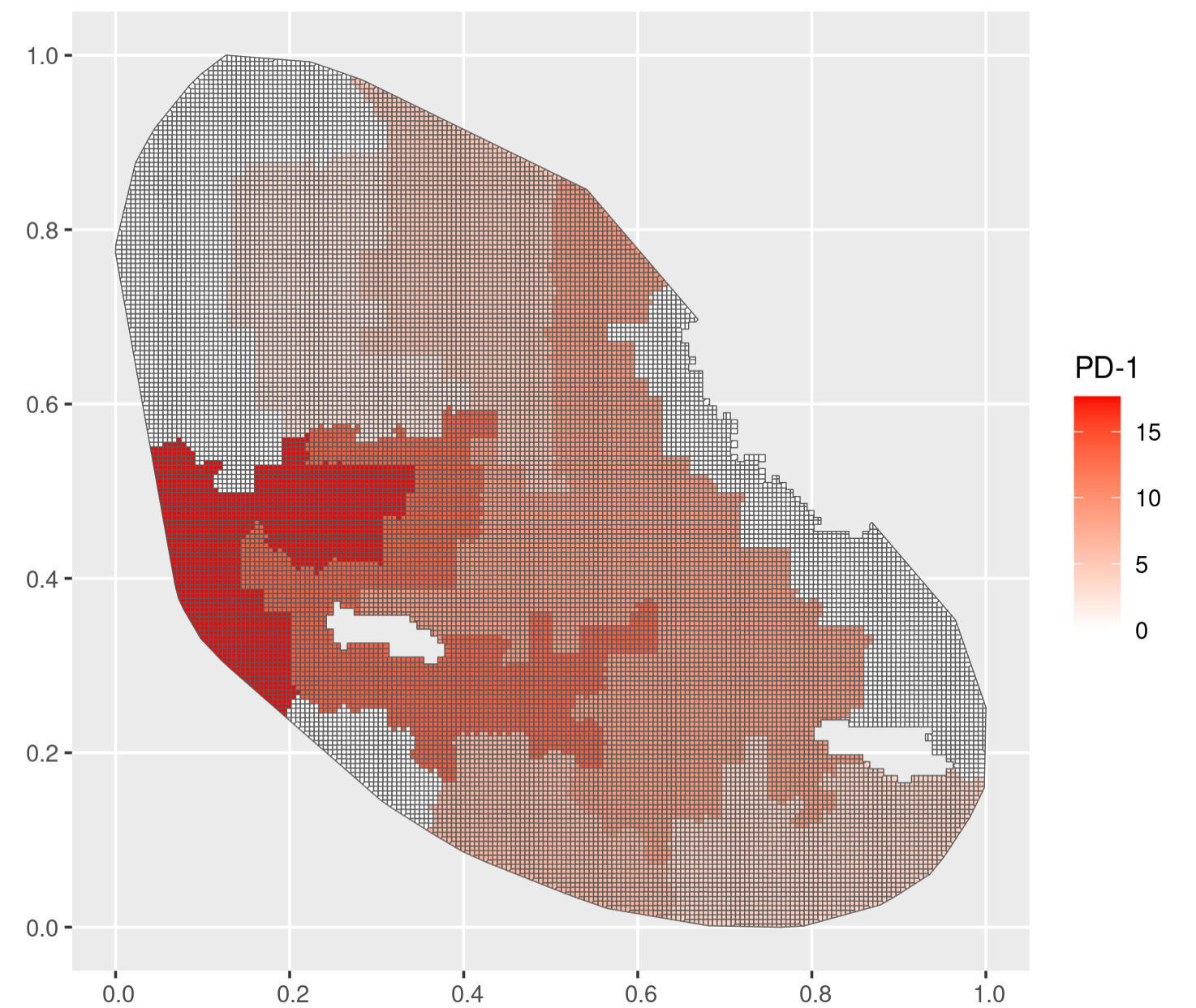}}&
            {\includegraphics[width=0.3\linewidth,height=0.15\textheight]{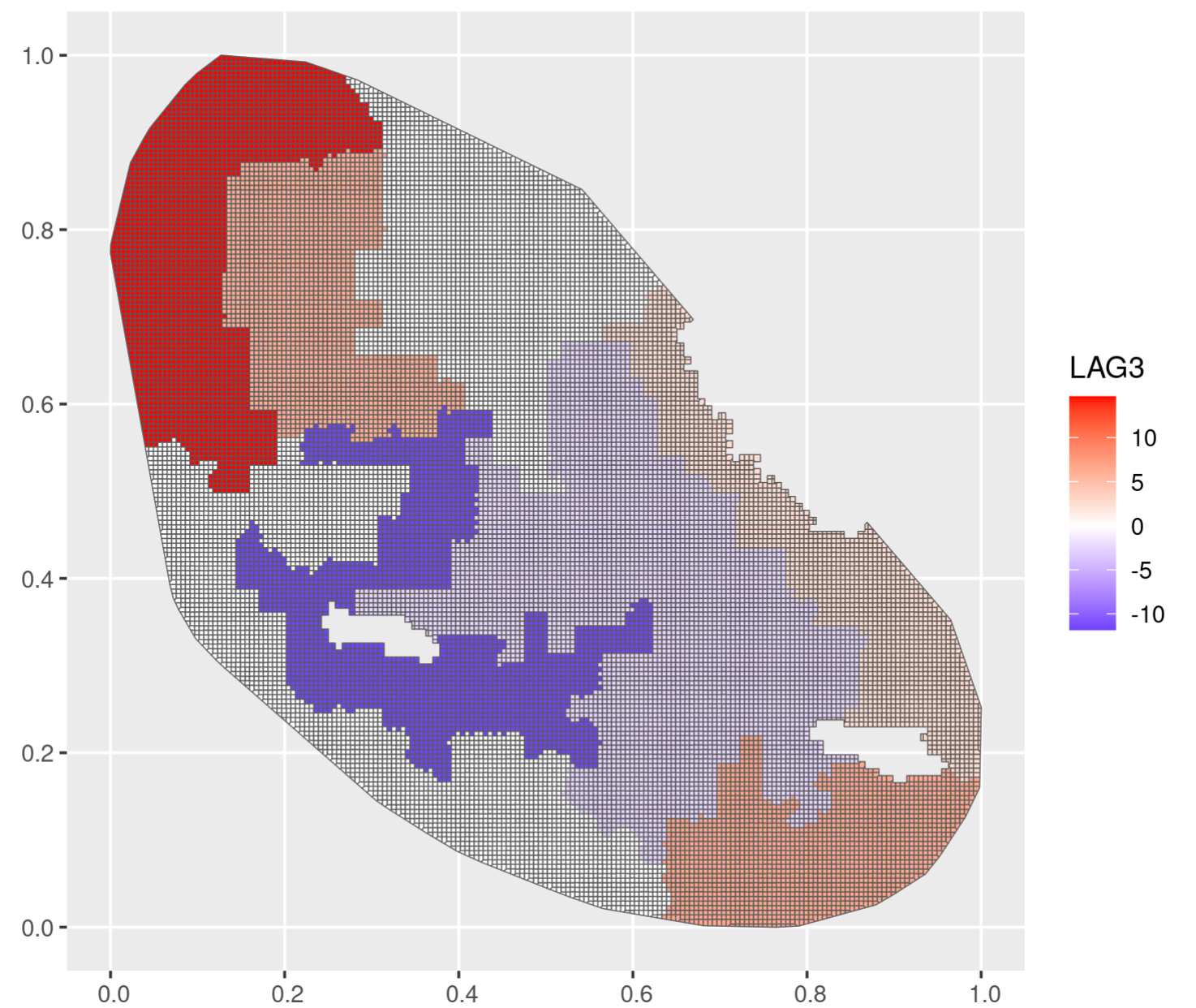}}&
            {\includegraphics[width=0.3\linewidth,height=0.15\textheight]{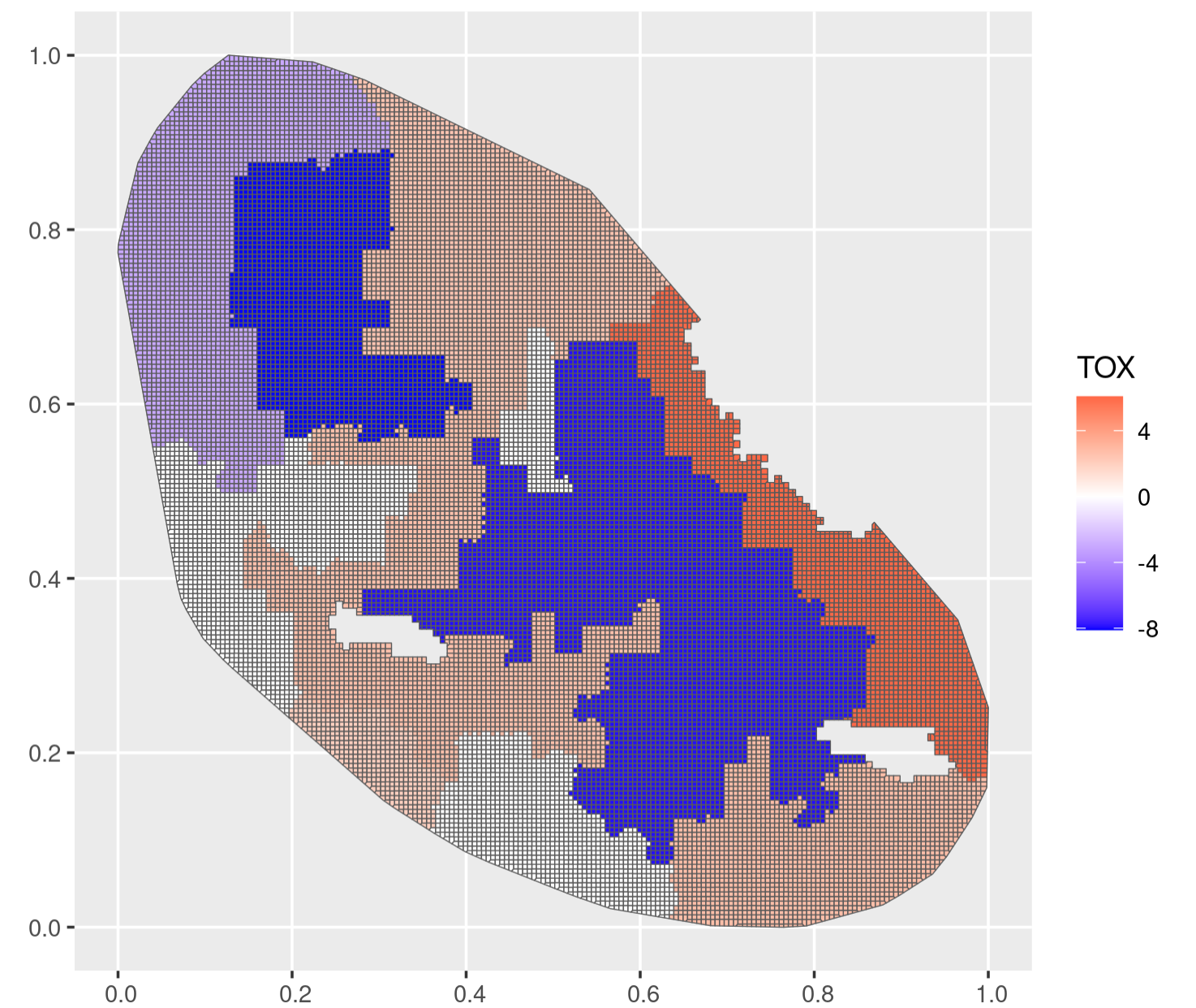}}
			\\
			{\small (d) } &{\small (e)} & {\small (f) } 
		\end{tabular}
		\caption{Model fitting result. (a) The posterior domain partition sample selected by \texttt{salso}, where different numbers indicate the corresponding cluster memberships. (b - f) Heatmaps of the posterior z-scores for the regression coefficients of each covariate (intercept, TCF-1/TCF-7, PD-1, LAG-3, TOX). 
        At a $95\%$ credibility level, a red color represents a significantly positive effect, and a blue color represents a significantly negative effect. A neutral color represents an insignificant effect.}
		\label{FIG:realdatapartition}
        \vspace{-10pt}
	\end{figure}

Figure~\ref{FIG:realdatapartition}(a) shows the posterior domain partition selected using the R package \texttt{salso}. Our model identifies 12 clusters, revealing substantial spatial heterogeneity within the tissue. To further explore the association between local tumor intensity and T-cell functional markers, we examine the posterior distributions of the regression coefficients, which indicate whether each functional marker is associated with \textit{enhanced} or \textit{reduced} T-cell-mediated antitumor activity. Specifically, conditional on the domain partition, we compute the posterior z-score for each regression coefficient within each cluster and the resulting heatmaps are shown in Figures~\ref{FIG:realdatapartition}(b - f). The posterior z-scores for the intercept in Figure~\ref{FIG:realdatapartition}(b) suggest substantial spatial heterogeneity in the baseline level of local tumor intensity.



Figures \ref{FIG:realdatapartition}(c - d) show that TCF-1/7 and PD-1 are generally negatively and positively associated with local tumor intensity, respectively.
Higher TCF-1/7 expression indicates stronger T-cell-mediated 
antitumor immune activity and, consequently, lower local tumor intensity.
This interpretation is consistent with prior studies reporting an association between higher TCF-1/7 expression and improved survival in melanoma and other cancers \citep{sade2018defining}. 
In contrast, PD-1, an immune checkpoint receptor, is commonly associated with T-cell dysfunction and exhaustion~\citep{PD1}. 
Its positive association with local tumor intensity may therefore reflect impaired or constrained T-cell-mediated tumor killing in these clusters. Notably, this patient responded well to the subsequent anti-PD-1 therapy, which acts by blocking PD-1 signaling and reactivating exhausted T cells against the tumor. This clinical response is consistent with the biological interpretation of the observed positive association, although larger cohorts are required to determine whether such associations are predictive of treatment response. 


Interestingly, both LAG-3 and TOX show associations of opposite signs with local tumor  intensity across clusters (Figures \ref{FIG:realdatapartition}(e - f)), 
suggesting that the biological interpretation of these markers depends on the local immune context.
The posterior z-scores for the LAG-3 coefficient are positive in most clusters, consistent with its role as an immune checkpoint receptor associated with dysfunctional or exhausted T cells, similar to PD-1~\citep{PD1}.
In contrast, Cluster 3 exhibits a strongly negative LAG-3 coefficient.
Although LAG-3 is commonly regarded as a marker of T-cell exhaustion, it can also be expressed on activated T cells \citep{ren2024lymphocyte}. Therefore, the negative LAG-3 association observed in Cluster 3 may reflect a distinct immune microenvironment in which LAG-3 expression marks activated T cells engaged in an ongoing antitumor response. 
Supporting this interpretation, several tertiary lymphoid structure-like structures are located adjacent to Cluster 3, as suggested by the local enrichment of B cells in Figure~\ref{FIG:realdataplot}(a), indicating that this subregion may represent an immune-active niche. 

\newpage

\renewcommand{\thealgorithm}{S.\arabic{algorithm}}
\def\theequation{S.\arabic{equation}}
\def\thefigure{S.\arabic{figure}}
\def\thesection{S.\arabic{section}}
\def\thepage{S\arabic{page}}
\def\thelemma{S.\arabic{lemma}}
\def\thecorollary{S.\arabic{corollary}}
\def\theproposition{S.\arabic{proposition}}



\begin{center}
    \large\bf Supplementary Material for ``A Consistent Bayesian Adaptive Quotient Graph Framework for Graph-Informed Random Partition Models''
\end{center}
 
\vspace{.25cm}

\vspace{.4cm}

\noindent This supplement is organized as follows. Section \ref{SEC:notation2} introduces notation used throughout the paper. Section \ref{SEC:AS1TO4} restates Assumptions 1 - 4 in \cite{huang2025consistent}. Section \ref{SEC:computation details} provides computational details. Section \ref{SEC:simulationdetail} provides detailed simulation settings. Section \ref{SEC:Addfigure_data} presents additional figures in real data analysis, and Section \ref{SEC:allproofs} presents the technical proofs of all theoretical results in the main paper.

\setcounter{section}{0} \setcounter{equation}{0} \setcounter{figure}{0} %
\setcounter{page}{1} \setcounter{assumption}{0} \setcounter{Passumption}{0} \setcounter{Hassumption}{0}

\section{Notation}
\label{SEC:notation2}
We first introduce some standard notation and conventions used throughout the paper. 
Denote $\| \cdummy \|_2$ and $\|
\cdummy \|_{\infty}$ as $L_2$ norm and $L_{\infty}$ norm, respectively.
For two locations $\tmmathbf{s}_1, \tmmathbf{s}_2$, let $d
(\tmmathbf{s}_1, \tmmathbf{s}_2) = \|\tmmathbf{s}_1 -\tmmathbf{s}_2 \|_2$ be
the Euclidean distance. For two \ spatial domains $\mathbb{D}_1$ and
$\mathbb{D}_2$, we write $d (\mathbb{D}_1, \mathbb{D}_2) =
\inf_{\tmmathbf{s}_1 \in \mathbb{D}_1, \tmmathbf{s}_2 \in \mathbb{D}_2} d
(\tmmathbf{s}_1, \tmmathbf{s}_2)$. 
The notation $|\mathbb{D}|$ is used as two ways: if $\mathbb{D}$ is a spatial
domain, $|\mathbb{D}|$ represents its area; if $\mathbb{D}$ is a set,
$|\mathbb{D}|$ denotes its cardinality. Furthermore, for a spatial domain
$\mathbb{D}$, we denote $\|\mathbb{D}\|$ as the number of observed locations within it.

We use $c$ and $C$ to denote some constants independent of $\pi^\dagger$ and $\pi^\ast$. The
values of $c$ and $C$ may change from line to line. For two positive series $a_n$ and $b_n$, we say $a_n \gg b_n$ if $a_n / b_n
\rightarrow \infty$, and $a_n \ll b_n$ if $a_n / b_n \rightarrow 0$. We write
$a_n \sim b_n$ if there exist constants $c, C > 0$, such that $c < a_n / b_n <
C, \text{ for all }n \geqslant 1$. We write $a_n = O (b_n)$, if there exits a
constant $c > 0$, such that $a_n / b_n \leqslant c$ for all $n \geqslant 1$.
We write $a_n = o (b_n)$ if $a_n \ll b_n$.

For a matrix $\tmmathbf{A}$ and a constant $c$, we say
$\tmmathbf{A}> c$, if $\lambda_{\min} (\tmmathbf{A}) > c$, and $\tmmathbf{A}<
c$ if $\lambda_{\max} (\tmmathbf{A}) < c$, where $\lambda_{\min}
(\tmmathbf{A})$ and $\lambda_{\max} (\tmmathbf{A})$ are the minimum and maximum eigenvalues of $\tmmathbf{A}$, respectively. For a positive integer $a$, we use $\tmmathbf{I}_a$ to denote an $a \times a$
identity matrix.

We next introduce notation used only in the proofs in Section \ref{SEC:proofsforaosconsistency}.
For two partitions $\pi^{\ast}_1$ and $\pi^{\ast}_2$, we write
$\pi^{\ast}_1 \cap \pi^{\ast}_2$ as the intersection partition, such that two
locations are within the same cluster if and only if they are within the same
cluster by both $\pi^{\ast}_1$ and $\pi^{\ast}_2$. Recall the definition of $\epsilon \{\pi_1 (\mathcal{D}), \pi_2 (\mathcal{D})\}$ for two domain partitions $\pi_1 (\mathcal{D}) = \{\mathcal{D}_{11}, \ldots,
\mathcal{D}_{1 k_1} \}$ and $\pi_2 (\mathcal{D}) = \{\mathcal{D}_{21}, \ldots,
\mathcal{D}_{2 k_2} \}$ in Section \ref{SEC:pidaggerpiastconcertration}. We further decompose $\epsilon \{\pi_1 (\mathcal{D}), \pi_2 (\mathcal{D})\} =
\epsilon_1 \{\pi_1 (\mathcal{D}), \pi_2 (\mathcal{D})\} + \epsilon_2 \{\pi_1
(\mathcal{D}), \pi_2 (\mathcal{D})\}$, where
\begin{equation}
  \epsilon_1 \{\pi_1 (\mathcal{D}), \pi_2 (\mathcal{D})\} = 1 -
  |\mathcal{D}|^{- 1} \left\{ \sum_{j = 1}^{k_1} \max_{l \in \{1, \ldots, k_2
  \}} |\mathcal{D}_{1 j} \cap \mathcal{D}_{2 l} | \right\}, \text{and}
  \label{EQ:epsilon1dis}
\end{equation}
\begin{equation}
  \epsilon_2 \{\pi_1 (\mathcal{D}), \pi_2 (\mathcal{D})\} = 1 -
  |\mathcal{D}|^{- 1} \left\{ \sum_{l = 1}^{k_2} \max_{j \in \{1, \ldots, k_1
  \}} |\mathcal{D}_{1 j} \cap \mathcal{D}_{2 l} | \right\} .
  \label{EQ:epsilon2dis}
\end{equation}

Recall the definition of $\pi^{\ast}_0 =
  \{\mathcal{D}_{1, 0}^{\ast}, \ldots \mathcal{D}_{k_0, 0}^{\ast} \}$ in Lemma \ref{LM:goodpartition}. For a given $\pi^{\ast} = \{\mathcal{D}_1^{\ast}, \ldots, \mathcal{D}^{\ast}_k
\}$, we write $\mathcal{M} (\mathcal{D}^{\ast}_j) = \tmop{argmax}_{l \in \{1,
\ldots, k_0 \}} |\mathcal{D}^{\ast}_j \cap \mathcal{D}^{\ast}_{l, 0} |$ as the
index of the cluster in $\{\mathcal{D}^{\ast}_{l, 0} \}_{l = 1}^{k_0}$ with
the largest intersection area with $\mathcal{D}^{\ast}_j$. Thus,
$\mathcal{D}^{\ast}_{\mathcal{M} (\mathcal{D}^{\ast}_j), 0}$ is considered as
the ``best matched" cluster in $\{\mathcal{D}^{\ast}_{l, 0} \}_{l = 1}^{k_0}$
for $\mathcal{D}^{\ast}_j$. 

Finally, Table \ref{TAB:notations} summarizes commonly used notation in the paper.

\begin{table}[h]
  \begin{center}
    \begin{tabular}{ll}
      \hline
      Notation & Meaning\\
      \hline
      $\Xi^{\dagger}$ & The space of $\pi^{\dagger}$ under our model\\
      $\Xi^{\ast} (\pi^{\dagger})$ & The support of $\mathbb{P}(\pi^\ast \mid \mathcal{G},\pi^\dagger)$, i.e., the set of $\pi^{\ast}$ induced from $\pi^\dagger$ \\
      $\Xi^{\ast}$ & $\cup_{\pi^{\dagger} \in \Xi^{\dagger}} \Xi^{\ast}
      (\pi^{\dagger})$, i.e., the space of $\pi^{\ast}$ under our model\\
            $\pi^{\ast}_0$ & The desirable partition \\
      $\pi^{\dagger}_0$ & An intermediate partition in $\Xi^\dagger$ such that $\pi_0^\ast \in \Xi^\ast(\pi^{\dagger}_0)$\\
      $\Pi_0^{\dagger}$ & The desirable set of $\pi^\dagger$\\
      $\Pi_0^{\ast}$ & The desirable set of $\pi^\ast$\\
      $\pi^{\ast}_0 (\pi^{\dagger})$ & The partition in $\Pi_0^\ast \cap \Xi^{\ast} (\pi^{\dagger})$ \\
      $\mathcal{M} (\mathcal{D}^{\ast}_j)$ & $\tmop{argmax}_{l \in \{1,
      \ldots, k_0 \}} |\mathcal{D}^{\ast}_j \cap \mathcal{D}^{\ast}_{l, 0}
      |$\\
      \hline
    \end{tabular}\label{TAB:notations}
  \end{center}
  \caption{Commonly used notation in the paper}
\end{table}

\section[]{Assumptions 1 - 4 in \cite{huang2025consistent}}
\label{SEC:AS1TO4}
We refer to Section \ref{SEC:notation2} for notation used in this paper. We restate Assumptions 1 - 4 from \cite{huang2025consistent} as Assumptions \ref{AS:distributionloc} - \ref{AS:covariates} below, treating $r_n$ as a constant as discussed in the main paper.

\begin{Passumption}
  \label{AS:distributionloc}The observed locations $\{\tmmathbf{s}_i \}_{i =
  1}^n$ are i.i.d. on spatial domain $\mathcal{D}$ with a probability density
  function $\mathbb{P}_{\mathcal{D}} (\cdummy)$ satisfying $0 <
  \inf_{\tmmathbf{s} \in \mathcal{D}} \mathbb{P}_{\mathcal{D}} (\tmmathbf{s})
  \leqslant \sup_{\tmmathbf{s} \in \mathcal{D}} \mathbb{P}_{\mathcal{D}}
  (\tmmathbf{s}) < \infty$.
\end{Passumption}

Recall that the true domain partition is denoted by $\{\mathcal{D}_{l,0}\}_{l=1}^{k_0}$. Write $\partial \mathcal{D}_{l, 0}$ as the boundary of $\mathcal{D}_{l,0}$. Let $N (\cdot, \delta, \| \cdot \|_2)$ be the $\delta$-covering number and $|
\partial \mathcal{D}_{l, 0} | = \lim_{\delta \rightarrow 0} \delta N (\partial
\mathcal{D}_{l, 0}, \delta, \| \cdot \|_2)$ be the length of $\partial
\mathcal{D}_{l, 0}$. The following assumption is on the true domain partition.

\begin{Passumption}
  \label{AS:lengthofboundary}There exist some constants $c, C > 0$, such that
  \begin{equation}
    c k_0^{- 1} \leqslant \inf_{1 \leqslant l \leqslant k_0} | \mathcal{D}_{l,
    0} | \leqslant \sup_{1 \leqslant l \leqslant k_0} | \mathcal{D}_{l, 0} |
    \leqslant C k_0^{- 1},\text{ }     c \leqslant \inf_{1 \leqslant l \leqslant k_0} \frac{| \partial
    \mathcal{D}_{l, 0} |}{| \mathcal{D}_{l, 0} |^{1 / 2}} \leqslant \sup_{1
    \leqslant l \leqslant k_0} \frac{| \partial \mathcal{D}_{l, 0} |}{|
    \mathcal{D}_{l, 0} |^{1 / 2}} \leqslant C, \label{EQ:boundary1}
  \end{equation}
  \begin{equation}
    \text{and }    N (\partial \mathcal{D}_{l, 0}, \delta,
    \| \cdot \|_2) \leqslant C |\partial \mathcal{D}_{l, 0} | \delta^{-1}, \text{ } \forall 1 \leqslant l \leqslant k_0, 0 < \delta < | \partial
    \mathcal{D}_{l, 0} |.
    \label{EQ:boundary3}
  \end{equation}
  Furthermore, for any two locations $\tmmathbf{s}, \tmmathbf{s}' \in
  \mathcal{D}_{l, 0}$, there exists a path connecting $\tmmathbf{s}$ and
  $\tmmathbf{s}'$, say $\mathcal{P} (\tmmathbf{s}, \tmmathbf{s}')$, such that
  $\mathcal{P} (\tmmathbf{s}, \tmmathbf{s}')$ is contained in $\mathcal{D}_{l,
  0}$, and
  \begin{equation}
    \label{EQ:pathdis} d \{\mathcal{P}(\tmmathbf{s}, \tmmathbf{s}'), \partial
    \mathcal{D}_{l, 0} \} = \min \{d (\tmmathbf{s}, \partial
    \mathcal{D}_{l, 0}), d (\tmmathbf{s}', \partial \mathcal{D}_{l, 0})\} \text{ or } d \{\mathcal{P}(\tmmathbf{s}, \tmmathbf{s}'), \partial
    \mathcal{D}_{l, 0} \} \geqslant C.
  \end{equation}
\end{Passumption}

The following two assumptions are on the separability among $\{\tmmathbf{\theta}_{l,0}\}_{l=1}^{k_0}$, and covariates $\{\tmmathbf{x}(\tmmathbf{s}_i)\}_{i = 1}^n$, respectively. 

\begin{Passumption}
  \label{AS:thetagap}We assume that $\{ \tmmathbf{\theta}_{l, 0} \}_{l =
  1}^{k_0}$ are well separated, i.e., $c k_0^{- 1 / d} <\min_{l \neq l'} \|\tmmathbf{\theta}_{l, 0} -\tmmathbf{\theta}_{l', 0}
  \|_2 <C $ for some constants $c,C>0$.
\end{Passumption}

\begin{Passumption}
  \label{AS:covariates}Conditional on $\{ \tmmathbf{s}_i \}_{i = 1}^n$, we
  assume $\{ \tmmathbf{x}(\tmmathbf{s}_i) \}_{i = 1}^n$ are independent
  d-dimensional bounded random variables. Furthermore, we assume
  that there exist constants $c$ and $C$, such that $0 < c
  <\mathbb{E} \{\tmmathbf{x}(\tmmathbf{s}_i)\tmmathbf{x}^T (\tmmathbf{s}_i)
  \mid \tmmathbf{s}_i \} < C$ holds for $1 \leqslant i \leqslant n$. Conditional on $\{ \tmmathbf{s}_i \}_{i = 1}^n$ and $\{ \tmmathbf{x}(\tmmathbf{s}_i) \}_{i = 1}^n$, we assume $\{ y(\tmmathbf{s}_i) \}_{i = 1}^n$ are generated from Equation (\ref{EQ:model}).
\end{Passumption}

\section{Computation details}
\label{SEC:computation details}

\subsection[]{Construction of $\mathcal{T}^\dagger$ with multiresolution compatibility}
\label{SEC:Tdagger construction}

Note that $\{ \pi^{\dagger \dagger}_p \}_{p = 1}^q$ are nested contiguous partitions. For any cluster $C \in \pi^{\dagger
\dagger}_{p - 1}$, we refer to the clusters in $\pi^{\dagger \dagger}_p$ that are contained in $C$ as the child clusters of $C$, and refer to $C$ as their parent cluster. For each resolution level $p = 2, \ldots, q$ and each parent cluster $C \in
\pi^{\dagger \dagger}_{p - 1}$, write $ \mathcal{S}_p (C) = \{C' \in \pi^{\dagger \dagger}_p : C' \subseteq C\}$ as the set of child clusters of $C$ at resolution level $p$. We then construct a quotient multigraph, denoted by $\mathcal{Q}_p (C)$, whose vertices are the child clusters in $\mathcal{S}_p (C)$. This multigraph is obtained by contracting each child cluster into a single vertex and retaining all edges of $\mathcal{G}$ whose endpoints lie in two distinct child clusters. Since $C$ is contiguous,
$\mathcal{Q}_p (C)$ is connected. We then randomly generate a spanning tree of
$\mathcal{Q}_p (C)$, denoted by $\mathcal{T}_p^{\dagger} (C)$. Each edge in
$\mathcal{T}_p^{\dagger} (C)$ corresponds to an original edge in $\mathcal{G}$. We finally obtain $\mathcal{T}^{\dagger}$ by
collecting these corresponding original edges across all resolution levels and
parent clusters. See the following Algorithm \ref{AL:spanningtree} for implementation details.


\begin{algorithm}[H]
\caption{Construction of $\mathcal{T}^\dagger$ with multiresolution compatibility}
\begin{algorithmic}[1]
\State \textbf{Input:} A simple connected graph $\mathcal{G}=(\mathcal{V},\mathcal{E})$; a sequence of nested partitions $\{ \pi^{\dagger \dagger}_p \}_{p = 1}^q$, where $\pi_1^{\dagger\dagger}=\{\mathcal{V}\}$ is the trivial partition with $\mathcal{V}$ as its only cluster, and $\pi_q^{\dagger\dagger}=\{\{v\}:v\in\mathcal{V}\}$ has each vertex in $\mathcal{V}$ as a cluster.

\State Initialize $\mathcal{T}^\dagger=\emptyset$.
\For{$p=2,\ldots,q$}
    \For{each parent cluster $C\in\pi^{\dagger \dagger}_{p-1}$}
        \State Let $\mathcal{S}_p(C)=\{C'\in\pi^{\dagger \dagger}_p:C'\subseteq C\}$.
        
        \State Construct the quotient multigraph $\mathcal{Q}_p(C)$ whose vertices are 
the child clusters in $\mathcal{S}_p(C)$, with edges induced by the edges of 
$\mathcal{G}$ whose endpoints lie in two distinct child clusters.
        \State Randomly generate a spanning tree of $\mathcal{Q}_p(C)$, say $\mathcal{T}_p^\dagger(C)$.
        \State Add to $\mathcal{T}^\dagger$ the original edges in $\mathcal{G}$ corresponding to the edges of $\mathcal{T}_p^\dagger(C)$.
    \EndFor
\EndFor
\State \textbf{Output:} $\mathcal{T}^\dagger$.
\end{algorithmic}
\label{AL:spanningtree}     
\end{algorithm}

\subsection[]{Construction of nested partitions on a $K\times K$ mesh grid graph $\mathcal{G}$}
\label{SEC:ALnestedpartitions}
\begin{algorithm}[H]
\caption{Construction of nested partitions on a $K\times K$ mesh grid graph $\mathcal{G}$}
\label{AL:nested_grid_partition}
\begin{algorithmic}[1]
\State \textbf{Input:} Mesh grid graph $\mathcal{G}=\{\mathcal{V},\mathcal{E}\}$, where $\mathcal{V}=\lbrace 1,\ldots,K\rbrace\times\lbrace 1,\ldots,K\rbrace$.
\State Set $p=1$ and $\pi^{\dagger\dagger}_1=\lbrace \mathcal{V}\rbrace$.
\While{there exists a cluster $C\in \pi_p^{\dagger\dagger}$ such that $|C|\geq 2$}
\State Initialize $\pi^{\dagger\dagger}_{p+1}=\emptyset$.
\For{each cluster $C\in \pi^{\dagger\dagger}_p$}
\State Write $C=\lbrace a,\ldots,b\rbrace\times\lbrace c,\ldots,d\rbrace$ with $1\leq a \leq b\leq K$ and $ 1 \leq c\leq d\leq K$.
\If{$b>a$ and $d>c$}
\State Let $m_1=\left\lfloor (a+b)/2 \right\rfloor$ and $m_2=\left\lfloor (c+d)/2 \right\rfloor$.
\State Add the following four sub-clusters to $\pi^{\dagger\dagger}_{p+1}$:
\[
\begin{aligned}
C_1 &= \lbrace a,\ldots,m_1\rbrace\times\lbrace c,\ldots,m_2\rbrace,
C_2 = \lbrace m_1+1,\ldots,b\rbrace\times\lbrace c,\ldots,m_2\rbrace,\\
C_3 &= \lbrace a,\ldots,m_1\rbrace\times\lbrace m_2+1,\ldots,d\rbrace,
C_4 = \lbrace m_1+1,\ldots,b\rbrace\times\lbrace m_2+1,\ldots,d\rbrace.
\end{aligned}
\]
\ElsIf{$b>a$ and $d=c$}
\State Let $m_1=\left\lfloor (a+b)/2 \right\rfloor$.
\State Add $\lbrace a,\ldots,m_1\rbrace\times\lbrace c\rbrace$ and $\lbrace m_1+1,\ldots,b\rbrace\times\lbrace c\rbrace$ to $\pi^{\dagger\dagger}_{p+1}$.
\ElsIf{$d>c$ and $b=a$}
\State Let $m_2=\left\lfloor (c+d)/2 \right\rfloor$.
\State Add $\lbrace a\rbrace\times\lbrace c,\ldots,m_2\rbrace$ and $\lbrace a\rbrace\times\lbrace m_2+1,\ldots,d\rbrace$ to $\pi^{\dagger\dagger}_{p+1}$.
\Else
\State Add $C$ to $\pi^{\dagger\dagger}_{p+1}$.
\EndIf
\EndFor
\State Set $p=p+1$.
\EndWhile
\State Set $q=p$.
\State \textbf{Output:} A sequence of nested partitions $\lbrace \pi^{\dagger\dagger}_p\rbrace_{p=1}^q$.
\end{algorithmic}
\label{AL:nestedpartition}    
\end{algorithm}

\subsection{Detailed computational strategy specific to A-Spat-RPM}
\label{SEC:AOS computationS}
 The parameters of interest are $(\pi^\dagger,\pi^\ast,\tmmathbf{\theta})$. Since we assign a conjugate prior to $\tmmathbf{\theta}$ in (\ref{EQ:priortheta}), the conditional posterior distribution $\mathbb{P}(\tmmathbf{\theta} \mid \mathcal{G}, \mathfrak{D},\pi^\dagger, \pi^\ast)$ has a closed form. Sampling $\tmmathbf{\theta}$ is therefore standard, and the details are omitted. Thus, the main computational task is to sample from $\mathbb{P}(\pi^\dagger,\pi^\ast\mid \mathcal{G},\mathfrak{D})$. Following Section~\ref{SEC:Bayesian computation}, at each MCMC iteration, we randomly choose to update either $\pi^{\ast}\mid \mathcal{G}, \mathfrak{D}, \pi^{\dagger}$ or $\pi^{\dagger}\mid \mathcal{G}, \mathfrak{D},\pi^{\ast}$, with probabilities $r^\ast$ and $r^\dagger$, respectively. The specific details of these two updates are given in Sections \ref{SEC:MCMC update piast} and \ref{SEC:MCMC update pidagger}, respectively.

\subsubsection[Bayesian computation]{Update $\pi^{\ast} \mid \mathcal{G}, \mathfrak{D}, \pi^{\dagger}$}
\label{SEC:MCMC update piast}

Conditional on $\pi^\dagger$, updating $\pi^\ast$ is equivalent to updating $\pi^\ast/\pi^\dagger$, the partition of clusters of $\pi^\dagger$. Recall that the spanning tree $\mathcal{T}$ appears in the prior construction of $\pi^\ast/\pi^\dagger$ (Section \ref{SEC:Ppiastconditional}). Although $\mathcal{T}$ is not itself a parameter of interest, we retain it explicitly as an auxiliary variable for computational convenience. Accordingly, conditional on $\pi^\dagger$, we jointly update $(\pi^\ast/\pi^\dagger,\mathcal{T})$ and the posterior samples of $\pi^{\ast}/\pi^\dagger \mid \mathcal{G}, \mathfrak{D}, \pi^{\dagger}$ (hence $\pi^{\ast} \mid \mathcal{G}, \mathfrak{D}, \pi^{\dagger}$) are then obtained by marginalizing out, or simply discarding, $\mathcal{T}$. 

We take the same four moves as the MCMC algorithm in Spat-RPM: birth, death, change, and hyper moves, with probabilities $r_b(k),r_d(k),r_c(k)$ and $r_h(k)$, respectively. In the birth move, one cluster in the current $\pi^\ast/\pi^\dagger$ is split into two clusters by randomly cutting an edge in $\mathcal{T}$ which connects two quotient vertices within the same cluster. We compute the M-H acceptance ratio as
\[ \min \left\{ 1, \frac{\lambda}{k+1} \times \frac{r_d (k + 1)}{r_b (k)}
   \times \frac{\mathbb{P}\{\tmmathbf{y} \mid \pi^\ast_{\tmop{new}}
   , \mathcal{S},\mathcal{X}\}}{\mathbb{P}\{\tmmathbf{y} \mid \pi ^\ast, 
   \mathcal{S},\mathcal{X}\}} \right\}, \]
where $\pi^\ast_{\tmop{new}}$ is the new induced partition after the edge cutting, and $\mathbb{P}\{\tmmathbf{y} \mid \pi ^\ast, \mathcal{S},\mathcal{X}\}$ is the integrated likelihood after marginalizing $\tmmathbf{\theta}$ out. The closed form of $\mathbb{P}\{\tmmathbf{y} \mid \pi ^\ast,  \mathcal{S},\mathcal{X}\}$ is provided in Equation (\ref{EQ:formlikelihood}) in Supplementary Material. In the death move, we randomly merge two adjacent clusters in $\pi^\ast/\pi^\dagger$. Specifically, we uniformly select an edge in $\mathcal{T}$ connecting two different clusters in $\pi^\ast/\pi^\dagger$ and then merge those two clusters into one. We compute the M-H acceptance ratio as
\[ \min \left\{ 1, \frac{k}{\lambda} \times \frac{r_b (k - 1)}{r_d (k)} \times
   \frac{\mathbb{P}\{\tmmathbf{y} \mid \pi^\ast_{\tmop{new}} ,
    \mathcal{S},\mathcal{X}\}}{\mathbb{P}\{\tmmathbf{y}
   \mid \pi^\ast ,  \mathcal{S},\mathcal{X}\}} \right\} .
\]

In the change move, we first perform a death move and then a birth move. The change move is to improve the mixing of the sampler, while keeping the number of clusters unchanged.

Finally, in the hyper move, we propose a new spanning tree that is compatible with the current partition $\pi^\ast/\pi^\dagger$. Conditional on $\mathcal{G}/\pi^\dagger$, let $(\mathcal{V}^\dagger,\mathcal{E}^\dagger)$ denote its quotient vertex and edge sets, respectively, and write $\pi^\ast/\pi^\dagger = \{\mathcal{V}^\dagger_j\}_{j=1}^k$, where $\mathcal{V}^\dagger_j\subseteq \mathcal{V}^\dagger$. For each $j=1,\ldots,k$, we first sample a UST using Wilson's algorithm \citep{wilson1996generating}, denoted by $\mathcal{T}_j$, from the subgraph of $\mathcal{G}/\pi^\dagger$ induced by $\mathcal{V}_j^\dagger$. To obtain a spanning tree of $\mathcal{G}/\pi^\dagger$, it remains to connect $\{\mathcal{T}_j\}_{j=1}^{k}$ using edges in $\mathcal{E}^\dagger$ without forming cycles. To this end, we construct the quotient multigraph of $\mathcal{G}/\pi^\dagger$ w.r.t. $\pi^\ast/\pi^\dagger$ and sample a UST from this multigraph. The original edges in $\mathcal{E}^\dagger$ corresponding to this sampled UST are then used to connect $\{\mathcal{T}_j\}_{j=1}^{k}$, yielding a spanning tree of $\mathcal{G}/\pi^\dagger$, denoted by $\mathcal{T}_{\tmop{new}}$. By construction, $\mathcal{T}_{\tmop{new}}$ is compatible with the current partition $\pi^\ast/\pi^\dagger$, and the M-H acceptance ratio is $1$.

\subsubsection[Bayesian computation]{Update $\pi^{\dagger}\mid \mathcal{G}, \mathfrak{D},\pi^{\ast}$} 
\label{SEC:MCMC update pidagger}

The procedure of proposing a new intermediate partition $\pi^\dagger_{\tmop{new}}$ is the same as that in Section \ref{SEC:Bayesian computation}. Following priors in Section \ref{SEC:Ppiastconditional}, the M-H acceptance ratio is computed as 
\[
\min \left\{ 1, \frac{| \Delta (\mathcal{G}/\pi^{\dagger}) | | \Delta
  (\pi^{\ast}/\pi^\dagger, \mathcal{G}/\pi^{\dagger}_{\tmop{new}}) |}{| \Delta (\pi^{\ast}/\pi^\dagger, \mathcal{G}/\pi^{\dagger})
  | | \Delta (\mathcal{G}/\pi^{\dagger}_{\tmop{new}}) |} \right\} ,
\]
where $\Delta (\pi^{\ast}/\pi^\dagger, \mathcal{G}/\pi^{\dagger})$ is the set of spanning trees of $\mathcal{G}/\pi^{\dagger}$ compatible with $\pi^{\ast}/\pi^\dagger$. 

In practice,  $|\Delta (\mathcal{G}/\pi^{\dagger})|$ is counted by Matrix Tree Theorem \citep{west2001introduction}, and $|\Delta (\pi^{\ast}/\pi^\dagger, \mathcal{G}/\pi^{\dagger})|$ is counted in a similar way as \cite{meila2018sample}. Since $\mathcal{G}$ is a planar mesh grid graph and $\pi^\dagger$ is contiguous w.r.t. $\mathcal{G}$, the quotient graph $\mathcal{G}/\pi^\dagger$ is also planar. Thus, generalized nested dissection \citep{lipton1979generalized} can be applied to the reduced Laplacian matrices involved in the Matrix-Tree Theorem. In particular, Theorem 4 of \citet{lipton1979generalized} implies that the required factorization has complexity $O((k^\dagger)^{3/2})$, and hence both $|\Delta (\mathcal{G}/\pi^{\dagger})|$ and $|\Delta (\pi^{\ast}/\pi^\dagger, \mathcal{G}/\pi^{\dagger})|$ can be computed within this complexity order.

\subsubsection{Data-informed initialization}
\label{SEC:initial}

We propose a data-informed initialization for the MCMC algorithm to reduce the burn-in period and facilitate convergence. The idea is to use the data to quickly identify candidate pairs of $\pi^\dagger$ and the corresponding induced $\pi^\ast$ that may closely approximate the true domain partition. To this end, at each resolution level $2 \leqslant p \leqslant q$, we search for an initialization candidate, denoted by $ (\pi^{\dagger}_{\tmop{ini}, p}, \pi^{\ast}_{\tmop{ini}, p})$, within the space induced by $\mathcal{G}/\pi^{\dagger \dagger}_p$. This yields a collection of initialization candidates $\{ (\pi^{\dagger}_{\tmop{ini}, p}, \pi^{\ast}_{\tmop{ini}, p}) \}_{p = 2}^q$. We then select the most promising candidate according to a specified criterion, such as the candidate with the largest posterior value. The details of constructing $(\pi^{\dagger}_{\tmop{ini}, p}, \pi^{\ast}_{\tmop{ini}, p})$ are provided as follows.

At resolution level $p$, let $\mathcal{E}_p$ denote the subset of edges in $\mathcal{T}^\dagger$ that connect distinct clusters of $\pi^{\dagger \dagger}_p$. Such edges exist because $\mathcal{T}^\dagger$ is compatible with $\pi^{\dagger \dagger}_p$. Each cluster of $\pi^{\dagger \dagger}_p$ corresponds to a union of blocks, and we denote the $m$-th union of blocks by $ B_{m, p}$. To construct $\pi^{\dagger}_{\tmop{ini}, p}$, we cut the $k^{\dagger} - 1$ edges in $\mathcal{E}_p$ that connect two unions of blocks exhibiting the largest differences in their estimated regression coefficients. Specifically, let $ \tmmathbf{x}_{m, p} $ and $
\tmmathbf{y}_{m, p} $ denote the corresponding covariate matrices and response vectors for the data within $ B_{m, p} $. Suppose
$\tmmathbf{x}^T_{m, p} \tmmathbf{x}_{m, p}$ is invertible, an event whose
probability has been shown to converge to $1$ in Section \ref{SEC:events}. For each edge $e \in
\mathcal{E}_p$ connecting two unions of blocks $B_{m,p}$ and $B_{m',p}$, we compute the weight
\begin{equation}
  w_{e, p} = \| V_{m, m', p}^{- 1 / 2} (\widehat{\tmmathbf{\theta}}_{m, p} -
  \widehat{\tmmathbf{\theta}}_{m', p}) \|, \label{EQ:wep}
\end{equation}
where $\widehat{\tmmathbf{\theta}}_{m, p}$ is the ordinary least squares
estimator of the regression coefficient based on the data within $B_{m, p}$,
i.e., $\widehat{\tmmathbf{\theta}}_{m, p} = (\tmmathbf{x}^T_{m, p} \tmmathbf{x}_{m,
p})^{- 1} \tmmathbf{x}^T_{m, p} \tmmathbf{y}_{m, p}$, and $V_{m, m', p}$ is
the covariance matrix of $\sigma_0^{- 1} (\widehat{\tmmathbf{\theta}}_{m, p} -
\widehat{\tmmathbf{\theta}}_{m', p})$, conditional on $\{ \tmmathbf{x}
(\tmmathbf{s}_i) \}_{i = 1}^n$, i.e.,
\[ \begin{array}{lll}
     V_{m, m', p} & = & \{ (\tmmathbf{x}_{m, p}^T \tmmathbf{x}_{m, p})^{- 1} +
     (\tmmathbf{x}_{m', p}^T \tmmathbf{x}_{m', p})^{- 1} \}
   \end{array}, \]
where $\sigma_0^2$ is the true variance of $\epsilon(\tmmathbf{s}_i)$. We then construct $\pi^{\dagger}_{\tmop{ini}, p}$ by cutting
the $k^{\dagger} - 1$ edges in $\mathcal{E}_p$ with the largest values of $\{w_{e, p}\}$.

The construction of (\ref{EQ:wep}) is motivated by testing whether two unions of 
blocks are fully contained in the same true cluster. If two unions of blocks
$B_{m, p}$ and $B_{m', p}$ are both fully contained in the same true cluster,
then $\sigma_0^{- 2} w_{e, p}^2$ follows a chi-square distribution with $d$
degrees of freedom. Thus, $\sigma_0^{- 2} w_{e, p}^2$ serves as a test
statistic for this hypothesis. A larger value of $w_{e, p}^2$ provides
stronger evidence against the hypothesis that the two unions of blocks belong to the same true cluster.
We therefore prioritize cutting the corresponding edge $e$ when constructing
$\pi^{\dagger}_{\tmop{ini}, p}$, with the aim of obtaining an intermediate
partition that is more likely to induce a domain partition closely
approximating the truth. In practice, if
$\tmmathbf{x}^T_{m, p} \tmmathbf{x}_{m, p}$ is not invertible, its pseudoinverse is used in place of the inverse when constructing
$\widehat{\tmmathbf{\theta}}_{m, p}$.

Conditional on $\pi^{\dagger}_{\tmop{ini}, p}$, we update $\pi^{\ast}$ using the method described in Section \ref{SEC:MCMC update piast} and obtain $\pi^{\ast}_{\tmop{ini}, p}$ as a post-burn-in sample. When $k^{\dagger}$ is small, a short burn-in period is typically sufficient for this conditional sampling because the space $\Xi^{\ast} (\pi^{\dagger}_{\tmop{ini}, p})$ is correspondingly small, allowing the MCMC chain to mix rapidly. Also note that, under the construction of $\lbrace \pi^{\dagger\dagger}_p\rbrace_{p=1}^q$ in Algorithm \ref{AL:nestedpartition}, the number of resolution levels satisfies $q=O(\log_2(K))$ and therefore grows slowly with $K$. Consequently, repeating this procedure over $2\leq p \leq q$ requires only modest computational effort.





\section{Simulation details}
\label{SEC:simulationdetail}
We adopt the data-generating process in Section 4.1 of \cite{huang2025consistent}. The true domain partition $\{\mathcal{D}_{l,0}\}_{l=1}^{k_0}$ is generated by the centroidal Voronoi tessellation using Lloyd's algorithm \citep{lloyd1982least}. See Figure \ref{FIG:piastanddagger}(c) in the main paper for the generated $\{\mathcal{D}_{l,0}\}_{l=1}^{k_0}$ with $k_0=19$. The observed locations $\{\tmmathbf{s}_i\}_{i=1}^n$ are $i.i.d.$ distributed within $\mathcal{D}$, and the covariate $\tmmathbf{x} (\tmmathbf{s}_i) = \{ x_1 (\tmmathbf{s}_i), x_2
	(\tmmathbf{s}_i) \}^T$ is a two dimensional vector, with $\{x_1 (\tmmathbf{s}_i)\}_{i=1}^{n}$ and $\{x_2 (\tmmathbf{s}_i)\}_{i=1}^{n}$ being $i.i.d.$ $\tmop{Unif}(-1,1)$. Given the domain partition, let $\tmmathbf{c}_{l,0}$ be the centroid of $\mathcal{D}_{l,0}$. We set $\tmmathbf{\theta}(\tmmathbf{s})=5\tmmathbf{c}_{l,0}$ if $\tmmathbf{s} \in \mathcal{D}_{l,0}$. The noise $\{\epsilon (\tmmathbf{s}_i)\}_{i=1}^{n}$  are $i.i.d.$ Gaussian with mean
	$0$ and variance $\sigma_0^2=0.01$.

Let $\lfloor a \rceil$ denote the nearest integer to $a$. For a given $n$, we set $k_0=\lfloor 5n^{1/8}\rceil$,
 which satisfies Equation (\ref{EQ:k0oarder}). Following Assumption \ref{AS:hyperpara}, we choose hyperparameters as $ K = \lfloor c_b n^{{1/2}}  \log^{-(1 + \alpha_b) / 2} (n)\rfloor$, $k^{\dagger} = \lfloor c^\dagger K \log^{\alpha_{\dagger}} (n)\rfloor$,  and $\log(\lambda^{-1}) = c_p n^{2/3}$, where $\alpha_b=1,\alpha^\dagger=0.1$, and constants $(c_b,c^\dagger,c_p)=(5,1,0.05)$. We set $\gamma$ in (\ref{EQ:priortheta}) to be $1$, and fix $\sigma^2=1$ throughout the simulation. Note that the value of $\sigma^2$ is different from the true variance $\sigma_0^2$. We will see from the simulation result that this misspecification doesn't affect the partition consistency of A-Spat-RPM, which aligns with our theoretical result. 
    
\section{Additional figures in real data analysis}
\label{SEC:Addfigure_data}

\begin{figure}[H]
		\centering
		\begin{tabular}{cc}
			{\includegraphics[width=0.35\linewidth,height=0.16\textheight]{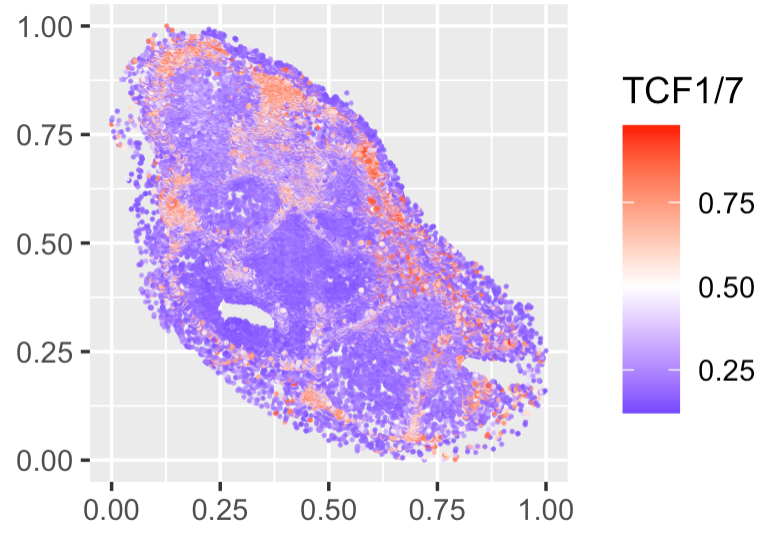}}&
			{\includegraphics[width=0.35\linewidth,height=0.16\textheight]{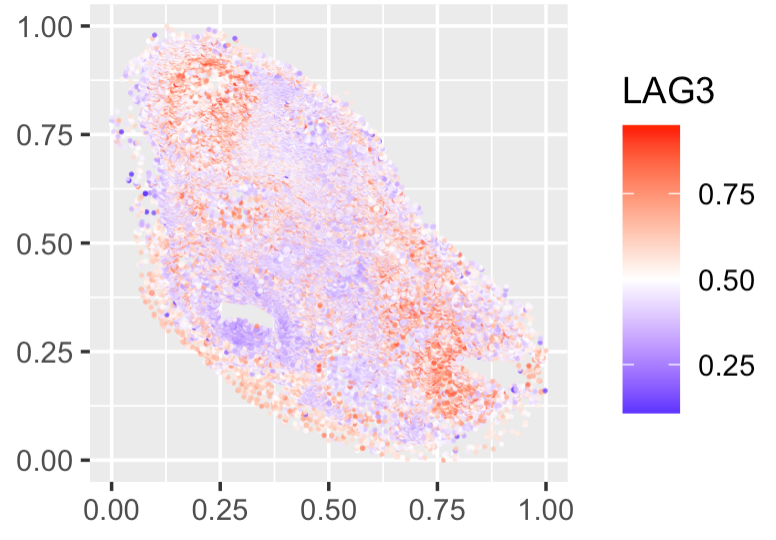}}\\
            {\small (a) } &{\small (b)} \\
            {\includegraphics[width=0.35\linewidth,height=0.16\textheight]{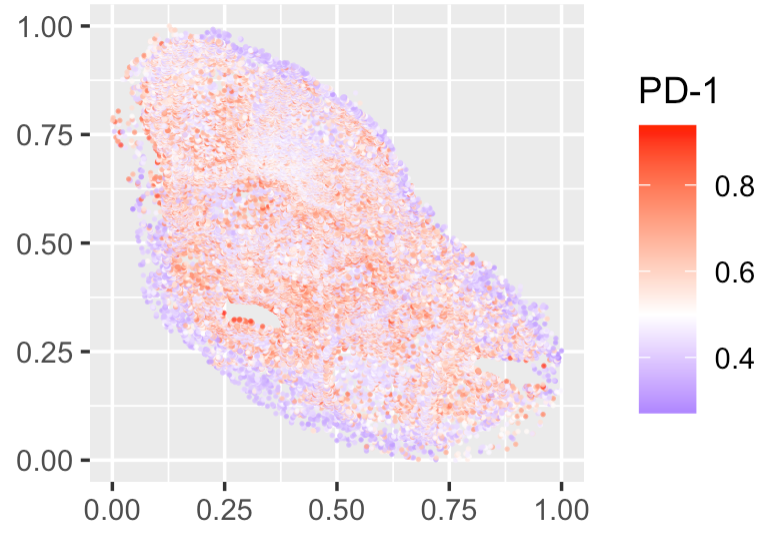}}&
            {\includegraphics[width=0.35\linewidth,height=0.16\textheight]{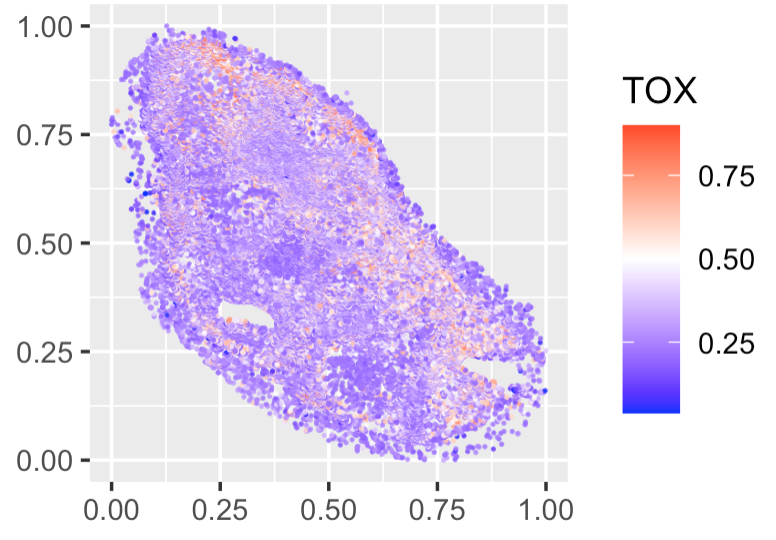}}
			\\
			  {\small (c)}  & {\small (d)}
		\end{tabular}
		\caption{(a - d) Heatmaps of expression levels of different T-cell functional markers.} 
		\label{FIG:addrealdataplot}
        \vspace{-10pt}
	\end{figure}

\section{Proofs}
\label{SEC:allproofs}

This section is organized as follows. Sections \ref{SEC:proofin2}, \ref{SEC:proofin42} and \ref{SEC:proofsforaosconsistency} provide proofs in Sections \ref{SEC:general graph}, \ref{SEC:existenceofgoodpartition} and \ref{SEC:pidaggerpiastconcertration} of the main paper, respectively. 

\subsection{Proofs in Section \ref{SEC:general graph}}
\label{SEC:proofin2}
\subsubsection{Proof of Lemma \ref{LM:verifyAS1_H}}
\label{SEC:proofoflemma7}
\begin{proof}


Suppose $\pi^{\ast} \in \Xi_{\tmop{direct}}^{\ast} (\mathcal{G})$ and
    $\pi^{\ast} \succeq \pi^{\dagger}$. For a cluster $C$ in $\pi^{\ast}$,
    write the set of its child clusters (i.e., clusters contained in $C$) in $\pi^{\dagger}$ as $S (C)$. Since $C$ is connected (because $\pi^{\ast}$ is a contiguous partition w.r.t. $\mathcal{G}$), the quotient vertices corresponding to $S(C)$ are also connected w.r.t. $\mathcal{G}/\pi^\dagger$. This means that $\pi^{\ast}/\pi^\dagger$ is a contiguous partition of quotient vertices w.r.t. $\mathcal{G} / \pi^{\dagger}$. Therefore, $\pi^{\ast}/\pi^\dagger $ is within the support of the imposed random contiguous partition model, which further leads to $\pi^{\ast} \in \Xi^{\ast} (\pi^{\dagger})$. Hence, we conclude
    $\Xi^{\ast} (\pi^{\dagger}) \supseteq \{ \pi^{\ast} \in \Xi_{\tmop{direct}}^{\ast}
    (\mathcal{G}) : \pi^{\ast} \succeq \pi^{\dagger} \}$ and the lemma is proved.

  \end{proof}

\subsubsection{Proof of Lemma \ref{PP:goodcondition}}
\label{ProofofPPgoodcondition}
  \begin{proof}
    Under Assumption \ref{AS:pidaggerspace}, it suffices to show that there
    exists a $\pi^{\dagger}_0\in \Xi^\dagger$, such that $\pi^{\ast}_0
    \succeq \pi^{\dagger}_0$. For the simplicity of notation, we set $p=p^\dagger$ in this proof, and Assumption \ref{AS:Tdaggercompatible} entails that $\mathcal{T}^{\dagger}$ is compatible with
    $\pi_p^{\dagger \dagger}$. We write $\mathcal{E} (\mathcal{T}^{\dagger},
    \pi_p^{\dagger \dagger})$ as the set of cutted edges of
    $\mathcal{T}^{\dagger}$ to induce $\pi_p^{\dagger \dagger}$. We construct a partition, say
  $\bar{\pi}^{\dagger}_0$, as the one induced by the spanning tree $\mathcal{T}^\dagger$ after cutting edges in
    $\overline{\mathcal{E}} (\mathcal{T}^{\dagger}, \pi_p^{\dagger \dagger})$,
    where
    \[ \overline{\mathcal{E}} (\mathcal{T}^{\dagger}, \pi_p^{\dagger \dagger})
       = \left\{ e \in \mathcal{E} (\mathcal{T}^{\dagger}, \pi_p^{\dagger
       \dagger}) : \text{the endpoints of $e$ belong to distinct clusters
       of $\pi_0^{\ast}$} \right\} . \]
    We next show that the constructed $\bar{\pi}^{\dagger}_0$ satisfies
    1) $\pi^{\ast}_0 \succeq \bar{\pi}^{\dagger}_0$, and 2) $|\bar{\pi}^{\dagger}_0|$, the number of clusters in $\bar{\pi}^{\dagger}_0$, is no larger than $k^{\dagger}$.

    To prove 1), note that Assumption \ref{AS:pi0existence} entails that any edge of $\mathcal{T}^{\dagger}$ that is not in $\mathcal{E}(\mathcal{T}^{\dagger}, \pi_p^{\dagger \dagger})$ has endpoints belonging to the same cluster of $\pi_0^{\ast}$. Hence, the
    constructed $\bar{\pi}^{\dagger}_0$ is the same as the partition induced by the spanning tree $\mathcal{T}^\dagger$ after cutting all edges whose endpoints belong to distinct clusters of $\pi_0^{\ast}$. Such a partition naturally
    nests in $\pi_0^{\ast}$. Thus, we have $ \pi^{\ast}_0 \succeq \bar{\pi}^{\dagger}_0$.
    
    To prove 2), it suffices to show that $ | \overline{\mathcal{E}} (\mathcal{T}^{\dagger}, \pi^{\dagger \dagger}_p)
    | \leqslant
    k^{\dagger} - 1$. Write $\pi_p^{\dagger \dagger} =\left\{C_j\right\}_{j=1}^{|\pi_p^{\dagger \dagger}|} $, where $C_j$ is the $j$-th cluster of $\pi_p^{\dagger \dagger}$. For an edge $e \in
    \overline{\mathcal{E}} (\mathcal{T}^{\dagger}, \pi_p^{\dagger \dagger})$,
    its endpoints belong to two distinct $C_j, C_{j'} \in \pi_p^{\dagger \dagger}$, whose corresponding quotient vertices
    also belong to distinct clusters under $\pi_0^{\ast} / \pi_p^{\dagger
    \dagger}$. On the other hand, since $C_j, C_{j'}$ are connected by $e$,
    their corresponding quotient vertices are also connected by an edge $e'$ under $\mathcal{G} /
    \pi_p^{\dagger \dagger}$. Hence, $e' \in \Omega_{\pi_0^{\ast}}
    (\pi^{\dagger \dagger}_p)$. Since different $e \in \overline{\mathcal{E}}
    (\mathcal{T}^{\dagger}, \pi_p^{\dagger \dagger})$ corresponds to different
    $e' \in \Omega_{\pi_0^{\ast}} (\pi^{\dagger \dagger}_p)$, we conclude that
    $| \overline{\mathcal{E}} (\mathcal{T}^{\dagger}, \pi^{\dagger \dagger}_p)
    | \leqslant | \Omega_{\pi_0^{\ast}} (\pi^{\dagger \dagger}_p) | \leqslant
    k^{\dagger} - 1$ under Assumption \ref{AS:Tdaggercompatible}. 
    
    Since $|\bar{\pi}^{\dagger}_0|\leq k^\dagger$, we can further cut edges of $\mathcal{T}^\dagger$, in addition to those in $\overline{\mathcal{E}}(\mathcal{T}^{\dagger}, \pi_p^{\dagger \dagger})$, to obtain a partition $\pi_0^\dagger \in \Xi^\dagger$ with $|\pi_0^\dagger|=k^\dagger$. The resulting partition $\pi_0^\dagger$ is naturally nested in $\bar{\pi}^{\dagger}_0$, and hence also in $\pi_0^\ast$. Combining the above arguments proves the lemma.
  \end{proof}

\subsubsection{Proof of Lemma \ref{PP:Tistree}}
\label{ProofofPPTistree}
\begin{proof}
  Recall that we write $\mathcal{G}=(\mathcal{V},\mathcal{E})$. For each $p = 2, \ldots, q$, let $\mathcal{G}_p=(\mathcal{V},\mathcal{E}_p)$ be the graph, where the edge set $\mathcal{E}_p$ contains original edges in $\mathcal{G}$ corresponding to the edges in
  \[ \{\mathcal{T}_{p'}^{\dagger} (C) : C \in \pi^{\dagger \dagger}_{p' - 1},
     2 \le p' \le p\} . \]
Recall the definition of quotient multigraph in Section \ref{SEC:graphbasic} of the main paper. Let $\tilde{\mathcal{G}}_p$ be the quotient multigraph of $\mathcal{G}_p$ w.r.t. $\pi^{\dagger \dagger}_p$. 
We claim that $\tilde{\mathcal{G}}_p$ is a spanning tree for every $p = 2, \ldots, q$.
  

  We prove the claim by induction. Note that $\pi_1^{\dagger\dagger}$ contains only one cluster $\mathcal{V}$. When $p = 2$, the graph $\tilde{\mathcal{G}}_2$ is exactly $\mathcal{T}_2^{\dagger} (\mathcal{V})$, which is constructed as a spanning tree on the child clusters of $\mathcal{V}$. Hence, $\tilde{\mathcal{G}}_2$ is a spanning tree. 
  
  Now suppose that $\tilde{\mathcal{G}}_{p}$ is a spanning tree, for $p\geq 2$. When passing from $\tilde{\mathcal{G}}_{p }$ to $\tilde{\mathcal{G}}_{p+1}$, each parent
  cluster $C \in \pi^{\dagger \dagger}_{p}$ is divided into the child
  clusters in $\mathcal{S}_{p+1} (C)$. By construction, the edges in
  $\mathcal{T}_{p+1}^{\dagger} (C)$ form a spanning tree on these child clusters.
  Thus, passing from $\tilde{\mathcal{G}}_{p }$ to $\tilde{\mathcal{G}}_{p+1}$
  replaces each vertex corresponding to a parent cluster $C$ by a tree on its
  child clusters. Since $\tilde{\mathcal{G}}_{p }$ is a spanning tree,
  replacing each of its vertices by a tree preserves connectedness and does
  not create cycles. Therefore, $\tilde{\mathcal{G}}_{p+1}$ is also a spanning
  tree. This proves the claim by
  induction.
  
  Finally, since $\pi^{\dagger \dagger}_q$ has each vertex in $\mathcal{V}$ as a cluster, contracting the clusters in $\pi^{\dagger \dagger}_q$ does
  not change the graph. Hence $\tilde{\mathcal{G}}_q =\mathcal{G}_q
  =\mathcal{T}^{\dagger}$. Therefore, $\mathcal{T}^{\dagger}$ is a spanning
  tree of $\mathcal{G}$ and the lemma is proved.
\end{proof}

\subsubsection{Proof of Lemma \ref{LM:Tdaggercompatible}}
\label{proofofTdaggercompatible}
  \begin{proof}
As shown in the proof of Lemma~\ref{PP:Tistree}, $\tilde{\mathcal{G}}_p$ is a spanning tree whose vertices are the clusters of $\pi^{\dagger \dagger}_p$. By construction of the quotient multigraph, the edges of $\tilde{\mathcal{G}}_p$ record exactly those edges of $\mathcal{T}^{\dagger}$ whose endpoints lie in different clusters of $\pi^{\dagger \dagger}_p$. Therefore, deleting these edges from $\mathcal{T}^{\dagger}$ separates different clusters of $\pi^{\dagger \dagger}_p$. The remaining edges of $\mathcal{T}^\dagger$ are those generated within the clusters at finer resolution levels, so the resulting connected components are exactly the clusters of $\pi^{\dagger \dagger}_p$. Therefore, $\mathcal{T}^{\dagger}$ is compatible with $\pi^{\dagger \dagger}_p$, and the lemma is proved.

  \end{proof}

\subsubsection{Proof of Lemma \ref{LM:pidaggerconsistency_general}}
\label{ProofofLMpidaggerconsistency_general}
\begin{proof}
    First, for any $\pi^{\dagger} \in  \Xi^{\dagger} \cap
    (\Pi^{\dagger}_0)^c $, we compute
    \begin{eqnarray*}
      \frac{\mathbb{P} (\pi^{\dagger} \mid \mathcal{G},
      \mathfrak{D})}{\mathbb{P} (\pi^{\dagger}_0 \mid \mathcal{G},
      \mathfrak{D})} & = & \frac{\sum_{\pi^{\ast} \in \Xi^{\ast}
      (\pi^{\dagger})} \mathbb{P} (\pi^{\dagger}, \pi^{\ast} \mid \mathcal{G},
      \mathfrak{D})}{\sum_{\pi^{\ast} \in \Xi^{\ast} (\pi^{\dagger}_0)}
      \mathbb{P} (\pi^{\dagger}_0, \pi^{\ast} \mid \mathcal{G},
      \mathfrak{D})}\\
      & \leqslant & \sum_{\pi^{\ast} \in \Xi^{\ast} (\pi^{\dagger})}
      \frac{\mathbb{P} (\pi^{\dagger}, \pi^{\ast} \mid \mathcal{G},
      \mathfrak{D})}{\mathbb{P} (\pi^{\dagger}_0, \pi^{\ast}_0 \mid
      \mathcal{G}, \mathfrak{D})} = \sum_{\pi^{\ast} \in \Xi^{\ast}
      (\pi^{\dagger})} \frac{\mathbb{P} (\pi^{\dagger} \mid \mathcal{G})
      \mathbb{P} (\pi^{\ast} \mid \mathcal{G}, \pi^{\dagger}) \mathbb{P}
      (\mathfrak{D} \mid \pi^{\ast})}{\mathbb{P} (\pi^{\dagger}_0 \mid
      \mathcal{G}) \mathbb{P} (\pi^{\ast}_0 \mid \mathcal{G}, \pi^{\dagger}_0)
      \mathbb{P} (\mathfrak{D} \mid \pi^{\ast}_0)}\\
      & = & \sum_{\pi^{\ast} \in \Xi^{\ast} (\pi^{\dagger})} \frac{\mathbb{P}
      (\pi^{\ast} \mid \mathcal{G}, \pi^{\dagger}) \mathbb{P} (\mathfrak{D}
      \mid \pi^{\ast})}{\mathbb{P} (\pi^{\ast}_0 \mid \mathcal{G},
      \pi^{\dagger}_0) \mathbb{P} (\mathfrak{D} \mid \pi^{\ast}_0)} =
      \sum_{\pi^{\ast} \in  \Xi^{\ast} (\pi^{\dagger}) \cap (\Pi_0^{\ast})^c
      } \frac{\mathbb{P} (\pi^{\ast} \mid \mathcal{G}, \pi^{\dagger})
      \mathbb{P} (\mathfrak{D} \mid \pi^{\ast})}{\mathbb{P} (\pi^{\ast}_0 \mid
      \mathcal{G}, \pi^{\dagger}_0) \mathbb{P} (\mathfrak{D} \mid
      \pi^{\ast}_0)},
    \end{eqnarray*}
    where the last equality uses the fact that any $\pi^{\ast} \in
    \Xi^{\ast} (\pi^{\dagger})$ is within $(\Pi_0^{\ast})^c$ since
    $\pi^{\dagger} \nin \Pi^{\dagger}_0$. Hence, we conclude that with
    probability tending to 1,
    \begin{eqnarray}
      \mathbb{P} (\pi^{\dagger} \nin \Pi^{\dagger}_0 \mid \mathcal{G},
      \mathfrak{D}) & = & \sum_{\pi^{\dagger} \in  \Xi^{\dagger} \cap
      (\Pi^{\dagger}_0)^c } \mathbb{P} (\pi^{\dagger} \mid \mathcal{G},
      \mathfrak{D})  \nonumber\\
      &\leqslant& \sum_{\pi^{\dagger} \in  \Xi^{\dagger} \cap
      (\Pi^{\dagger}_0)^c } \sum_{\pi^{\ast} \in  \Xi^{\ast}
      (\pi^{\dagger}) \cap (\Pi_0^{\ast})^c } \frac{\mathbb{P} (\pi^{\ast}
      \mid \mathcal{G}, \pi^{\dagger}) \mathbb{P} (\mathfrak{D} \mid
      \pi^{\ast})}{\mathbb{P} (\pi^{\ast}_0 \mid \mathcal{G}, \pi^{\dagger}_0)
      \mathbb{P} (\mathfrak{D} \mid \pi^{\ast}_0)} \nonumber \\
      & \leqslant & \sum_{\pi^{\dagger} \in \Xi^{\dagger}} \sum_{\pi^{\ast}
      \in  \Xi^{\ast} (\pi^{\dagger}) \cap (\Pi_0^{\ast})^c }
      \frac{\mathbb{P} (\pi^{\ast} \mid \mathcal{G}, \pi^{\dagger}) \mathbb{P}
      (\mathfrak{D} \mid \pi^{\ast})}{\mathbb{P} (\pi^{\ast}_0 \mid
      \mathcal{G}, \pi^{\dagger}_0) \mathbb{P} (\mathfrak{D} \mid
      \pi^{\ast}_0)} \overset{\text{Assumption \ref{AS:ansum}}}{\rightarrow}
      0. \label{EQ:pidaggerconsistency} 
    \end{eqnarray}
    Second, for $\pi^{\ast} \nin \Pi^{\ast}_0$, we compute
    \begin{eqnarray*}
      \frac{\mathbb{P} (\pi^{\ast} \mid \mathcal{G}, \mathfrak{D})}{\mathbb{P}
      (\pi^{\ast}_0 \mid \mathcal{G}, \mathfrak{D})} & = &
      \frac{\sum_{\pi^{\dagger} \in \Xi^{\dagger}} \mathbb{P} (\pi^{\ast},
      \pi^{\dagger} \mid \mathcal{G}, \mathcal{}
      \mathfrak{D})}{\sum_{\pi^{\dagger} \in \Xi^{\dagger}} \mathbb{P}
      (\pi^{\ast}_0, \pi^{\dagger} \mid \mathcal{} \mathcal{G}, \mathfrak{D})}
      = \frac{\sum_{\pi^{\dagger} \in \Xi^{\dagger}} \mathbb{P} (\pi^{\dagger}
      \mid \mathcal{G}) \mathbb{P} (\pi^{\ast} \mid \mathcal{G},
      \pi^{\dagger}) \mathbb{P} (\mathfrak{D} \mid
      \pi^{\ast})}{\sum_{\pi^{\dagger} \in \Xi^{\dagger}} \mathbb{P}
      (\pi^{\dagger} \mid \mathcal{G}) \mathbb{P} (\pi^{\ast}_0 \mid
      \mathcal{G}, \pi^{\dagger}) \mathbb{P} (\mathfrak{D} \mid
      \pi^{\ast}_0)}\\
      & \leqslant & \sum_{\pi^{\dagger} \in \Xi^{\dagger}} \frac{\mathbb{P}
      (\pi^{\dagger} \mid \mathcal{G}) \mathbb{P} (\pi^{\ast} \mid
      \mathcal{G}, \pi^{\dagger}) \mathbb{P} (\mathfrak{D} \mid
      \pi^{\ast})}{\mathbb{P} (\pi^{\dagger}_0 \mid \mathcal{G}) \mathbb{P}
      (\pi^{\ast}_0 \mid \mathcal{G}, \pi^{\dagger}_0) \mathbb{P}
      (\mathfrak{D} \mid \pi^{\ast}_0)} = \sum_{\pi^{\dagger} \in
      \Xi^{\dagger}} \frac{\mathbb{P} (\pi^{\ast} \mid \mathcal{G},
      \pi^{\dagger}) \mathbb{P} (\mathfrak{D} \mid \pi^{\ast})}{\mathbb{P}
      (\pi^{\ast}_0 \mid \mathcal{G}, \pi^{\dagger}_0) \mathbb{P}
      (\mathfrak{D} \mid \pi^{\ast}_0)} .
    \end{eqnarray*}
    Hence, we conclude that with probability tending to 1,
    \begin{eqnarray}
      \mathbb{P} (\pi^{\ast} \nin \Pi^{\ast}_0 \mid \mathcal{G}, \mathfrak{D})
      & = & \sum_{\pi^{\ast} \in (\Pi^{\ast}_0)^c} \mathbb{P} (\pi^{\ast} \mid
      \mathcal{G}, \mathfrak{D}) \leqslant \sum_{\pi^{\ast} \in
      (\Pi^{\ast}_0)^c} \sum_{\pi^{\dagger} \in \Xi^{\dagger}}
      \frac{\mathbb{P} (\pi^{\ast} \mid \mathcal{G}, \pi^{\dagger}) \mathbb{P}
      (\mathfrak{D} \mid \pi^{\ast})}{\mathbb{P} (\pi^{\ast}_0 \mid
      \mathcal{G}, \pi^{\dagger}_0) \mathbb{P} (\mathfrak{D} \mid
      \pi^{\ast}_0)} \nonumber\\
      & = & \sum_{\pi^{\dagger} \in \Xi^{\dagger}} \sum_{\pi^{\ast} \in 
      \Xi^{\ast} (\pi^{\dagger}) \cap (\Pi_0^{\ast})^c } \frac{\mathbb{P}
      (\pi^{\ast} \mid \mathcal{G}, \pi^{\dagger}) \mathbb{P} (\mathfrak{D}
      \mid \pi^{\ast})}{\mathbb{P} (\pi^{\ast}_0 \mid \mathcal{G},
      \pi^{\dagger}_0) \mathbb{P} (\mathfrak{D} \mid \pi^{\ast}_0)}
      \overset{\text{Assumption \ref{AS:ansum}}}{\rightarrow} 0.
      \label{EQ:piastconsistency} 
    \end{eqnarray}
    
    The lemma is proved by combining Equations (\ref{EQ:pidaggerconsistency})
    - (\ref{EQ:piastconsistency})
  \end{proof}

\subsubsection{Proof of Lemma \ref{LM:piastconditionalpidagger}}
\label{ProofofLMpiastconditionalpidagger}

  \begin{proof}
    We compute
    \begin{eqnarray*}
      \frac{\mathbb{P} (\pi^{\ast} \mid \mathcal{G}, \mathfrak{D},
      \pi^{\dagger})}{\mathbb{P} (\pi^{\ast}_0 (\pi^{\dagger}) \mid
      \mathcal{G}, \mathfrak{D}, \pi^{\dagger})} & = & \frac{\mathbb{P}
      (\pi^{\ast} \mid \mathcal{G}, \pi^{\dagger}) \mathbb{P} (\mathfrak{D}
      \mid \pi^{\ast})}{\mathbb{P} (\pi^{\ast}_0 (\pi^{\dagger}) \mid
      \mathcal{G}, \pi^{\dagger}) \mathbb{P} (\mathfrak{D} \mid \pi^{\ast}_0
      (\pi^{\dagger}))}\\
      & = & \frac{\mathbb{P} (\pi^{\ast} \mid \mathcal{G},
      \pi^{\dagger})}{\mathbb{P} (\pi^{\ast}_0 (\pi^{\dagger}) \mid
      \mathcal{G}, \pi^{\dagger})} \times \frac{\mathbb{P} (\pi^{\ast}_0 \mid
      \mathcal{G}, \pi^{\dagger}_0)}{\mathbb{P} (\pi^{\ast}_0 \mid
      \mathcal{G}, \pi^{\dagger}_0)} \times \frac{\mathbb{P} (\mathfrak{D}
      \mid \pi^{\ast}_0)}{\mathbb{P} (\mathfrak{D} \mid \pi^{\ast}_0)} \times
      \frac{\mathbb{P} (\mathfrak{D} \mid \pi^{\ast})}{\mathbb{P}
      (\mathfrak{D} \mid \pi^{\ast}_0 (\pi^{\dagger}))}\\
      & = & \frac{\mathbb{P} (\pi^{\ast}_0 \mid \mathcal{G},
      \pi^{\dagger}_0)}{\mathbb{P} (\pi^{\ast}_0 (\pi^{\dagger}) \mid
      \mathcal{G}, \pi^{\dagger})} \times \frac{\mathbb{P} (\mathfrak{D} \mid
      \pi^{\ast}_0)}{\mathbb{P} (\mathfrak{D} \mid \pi^{\ast}_0
      (\pi^{\dagger}))} \times \frac{\mathbb{P} (\pi^{\ast} \mid \mathcal{G},
      \pi^{\dagger})}{\mathbb{P} (\pi^{\ast}_0 \mid \mathcal{G},
      \pi^{\dagger}_0)} \times \frac{\mathbb{P} (\mathfrak{D} \mid
      \pi^{\ast})}{\mathbb{P} (\mathfrak{D} \mid \pi^{\ast}_0)} .
    \end{eqnarray*}
    Thus, we conclude that with probability tending to 1,
    \begin{eqnarray*}
      &  & {\sup_{\pi^{\dagger} \in \Xi^{\dagger} \cap \Pi^{\dagger}_0}} 
      \mathbb{P} (\pi^{\ast} \nin \Pi^{\ast}_0 \mid \mathcal{G}, \mathfrak{D},
      \pi^{\dagger})\\
      & = & {\sup_{\pi^{\dagger} \in \Xi^{\dagger} \cap \Pi^{\dagger}_0}} 
      \sum_{\pi^{\ast} \in  \Xi^{\ast} (\pi^{\dagger}) \cap (\Pi_0^{\ast})^c
      } \mathbb{P} (\pi^{\ast} \mid \mathcal{G}, \mathfrak{D},
      \pi^{\dagger})\\
      & \leqslant & {\sup_{\pi^{\dagger} \in \Xi^{\dagger} \cap
      \Pi^{\dagger}_0}}  \left\{ \frac{\mathbb{P} (\pi^{\ast}_0 \mid
      \mathcal{G}, \pi^{\dagger}_0)}{\mathbb{P} (\pi^{\ast}_0 (\pi^{\dagger})
      \mid \mathcal{G}, \pi^{\dagger})} \times \frac{\mathbb{P} (\mathfrak{D}
      \mid \pi^{\ast}_0)}{\mathbb{P} (\mathfrak{D} \mid \pi^{\ast}_0
      (\pi^{\dagger}))} \right. \left. \times \sum_{\pi^{\ast} \in 
      \Xi^{\ast} (\pi^{\dagger}) \cap (\Pi_0^{\ast})^c } \frac{\mathbb{P}
      (\pi^{\ast} \mid \mathcal{G}, \pi^{\dagger})}{\mathbb{P} (\pi^{\ast}_0
      \mid \mathcal{G}, \pi^{\dagger}_0)} \times \frac{\mathbb{P}
      (\mathfrak{D} \mid \pi^{\ast})}{\mathbb{P} (\mathfrak{D} \mid
      \pi^{\ast}_0)} \right\}\\
      &  & \overset{\text{Assumption \ref{AS:anfurther}}}{\rightarrow} 0.
    \end{eqnarray*}
    The lemma is proved.
  \end{proof}

\subsection{Proofs in Section \ref{SEC:existenceofgoodpartition}}
\label{SEC:proofin42}
\subsubsection{Proof of Lemma \ref{LM:goodpartition}}
\label{SEC:proofoflemma6}
  \begin{proof}
  Note that the quotient graph $\mathcal{G} / \pi^{\dagger \dagger}_{p^{\dagger}}$ is a $2^{p^{\dagger}-1}$ by $2^{p^{\dagger}-1}$ mesh grid, where each quotient vertex represents a union of $2^{q-p^{\dagger}}\times 2^{q-p^{\dagger}}$ blocks in $\{B_m\}_{m=1}^{K^2}$. Let $B_{m, p^{\dagger}}$ be the $m$-th union of blocks corresponding to quotient vertices of the graph $\mathcal{G} / \pi_{p^{\dagger}}^{\dagger \dagger}$. We can see that each $B_{m, p^{\dagger}}$ has a side length $2^{- (p^{\dagger} - 1)}$.
  
According to Proposition 1
    of \cite{huang2025consistent}, there exists a contiguous partition of quotient vertices w.r.t. $\mathcal{G} / \pi_{p^{\dagger}}^{\dagger \dagger}$, and an induced domain partition $\pi^\ast_0 = \{\mathcal{D}_{1, 0}^{\ast}
   , \ldots \mathcal{D}_{k_0, 0}^{\ast}  \}$, such that
   \begin{equation}
       \sum_{j = 1}^{k_0} | \{B_{m, p^{\dagger}} : B_{m, p^{\dagger}}
       \subseteq \mathcal{D}_{j, 0}^{\ast} , B_{m, p^{\dagger}}
       \subsetneq \mathcal{D}_{j, 0} \} | \leqslant c k_0^{1 / 2}
       2^{p^{\dagger} - 1}
       \label{EQ:WBmp}
   \end{equation}
       for some constant $c$. Since $\pi_0^\ast$ is induced by $\mathcal{G}/\pi_{p^{\dagger}}^{\dagger \dagger}$, $ \pi_{p^{\dagger}}^{\dagger \dagger}$ is naturally nested in $\pi_0^\ast$. Besides, since each $B_{m, p^{\dagger}}$ contains $2^{2(q-p^{\dagger})}$
    blocks in $\{B_m\}_{m=1}^{K^2}$, we conclude that 
    \[ \sum_{j = 1}^{k_0} | \{B_m : B_m \subseteq \mathcal{D}_{j, 0}^{\ast},
       B_m \subsetneq \mathcal{D}_{j, 0} \} | \leqslant c k_0^{1 / 2}
       2^{p^{\dagger} - 1} \times 2^{2 (q-p^{\dagger})} \leqslant c
       k_0 K. \]
       The lemma is proved.
  \end{proof}

\subsubsection{Proof of Lemma \ref{LM:goodpidagger}}
\label{SEC:proofoflemma8}
\begin{proof}
  We retain notation used in Lemma \ref{LM:goodpartition}. We consider the quotient graph $\mathcal{G}/\pi^{\dagger \dagger}_{p^{\dagger}}$, and recall $\{B_{m,p^\dagger}\}$ defined in the proof of Lemma \ref{LM:goodpartition}. Write $\mathcal{W}_{\pi_0^{\ast} / \pi_{p^{\dagger}}^{\dagger \dagger}} =
    \cup_{j = 1}^{k_0} \{B_{m, p^{\dagger}} : B_{m, p^{\dagger}} \subseteq
    \mathcal{D}_{j, 0}^{\ast} , B_{m, p^{\dagger}} \subsetneq
    \mathcal{D}_{j, 0} \}$. We claim that, if an edge $e$ in $\mathcal{G}/\pi^{\dagger \dagger}_{p^{\dagger}}$ connects two quotient vertices (i.e., unions of blocks) from two distinct clusters of $\pi_0^{\ast} / \pi_{p^{\dagger}}^{\dagger \dagger}$, say
    $B_{m, p^{\dagger}} \subseteq \mathcal{D}_{j, 0}^{\ast}$ and
    $B_{m, p^{\dagger}} \subseteq \mathcal{D}_{j', 0}^{\ast}$
    for $j \neq j'$, one of $B_{m, p^{\dagger}}$ and $B_{m',
    p^{\dagger}}$ must be within $\mathcal{W}_{\pi_0^{\ast} /
    \pi_{p^{\dagger}}^{\dagger \dagger}} \cup \mathbb{B}_p$, where
    $\mathbb{B}_p$ is the blocks in $\{ B_{m, p^{\dagger}} \}$ that intersect the boundary of the true domain partition. 
    
    Suppose neither of $B_{m, p^{\dagger}}$ and $B_{m', p^{\dagger}}$ is
    within $\mathcal{W}_{\pi_0^{\ast} / \pi_{p^{\dagger}}^{\dagger \dagger}}
    \cup \mathbb{B}_p$. Then we have $B_{m, p^{\dagger}} \subseteq
    \mathcal{D}_{j, 0}$ and $B_{m', p^{\dagger}} \subseteq \mathcal{D}_{j',
    0}$. On the other hand, since $B_{m, p^{\dagger}}$ and $B_{m',
    p^{\dagger}}$ are connected by an edge, one of them must intersect the boundary. By contradiction, the claim is proved.
    
    By Equation (\ref{EQ:WBmp}) and Assumption \ref{AS:lengthofboundary}, we have $\left| \mathcal{W}_{\pi_0^{\ast} / \pi_{p^{\dagger}}^{\dagger
    \dagger}} \cup \mathbb{B}_p \right| \leqslant c k_0^{1 / 2}
    2^{p^{\dagger} - 1}$. On the other hand, since each $B_{m, p^{\dagger}}$ has at most 4 edges
    connected to it under $\mathcal{G}/\pi^{\dagger \dagger}_{p^\dagger}$, we conclude that $| \Omega_{\pi_0^{\ast}}
    (\pi^{\dagger \dagger}_{p^{\dagger}}) | \leqslant c k_0^{1 / 2}
    2^{p^{\dagger} - 1} \leqslant c K$. Thus, according to Assumption \ref{AS:hyperpara}, $k^{\dagger} \sim K
    \log^{\alpha_{\dagger}} (n) \gg | \Omega_{\pi_0^{\ast}}
    (\pi^{\dagger \dagger}_{p^{\dagger}}) |$ and the lemma is proved.
  \end{proof}

\subsection{Proofs in Section \ref{SEC:pidaggerpiastconcertration}}
\label{SEC:proofsforaosconsistency}
\subsubsection{Proof framework}
\label{SEC:proofframework}
Since $\mathcal{G}$ is fixed throughout our model, we suppress conditioning on $\mathcal{G}$ in the proofs of this section. For example, we write $\mathbb{P}(\pi^\ast\mid\pi^\dagger)$ in place of $\mathbb{P}(\pi^\ast \mid \mathcal{G},\pi^\dagger)$. We refer to Section \ref{SEC:notation2} for notation used throughout the proof, where we recall the decomposition $\epsilon(\cdot,\cdot) = \epsilon_1(\cdot,\cdot) +\epsilon_2(\cdot,\cdot)$, and the definition of $\mathcal{M}(\mathcal{D}^{\ast}_j)$ and $\Xi^\ast$. We divide $\Xi^{\ast}$ into five
categories:
\[ \Pi_1^{\ast} = \{\pi^{\ast} \in \Xi^{\ast} : | \pi^{\ast} | < k_0 \},
   \Pi_2^{\ast} = \{\pi^{\ast} \in \Xi^{\ast} : | \pi^{\ast} | > k_0 \}, \]
\[ \Pi_3^{\ast} = \left\{ \pi^{\ast} \in \Xi^{\ast} : | \pi^{\ast} | = k_0 
   \text{ and } \{\mathcal{M}(\mathcal{D}^{\ast}_j)\}_{j = 1}^{| \pi^{\ast} |}
   \subsetneq \{1, \ldots, k_0 \} \right\}, \]
\[ \Pi_4^{\ast} = \left\{ \pi^{\ast} \in \Xi^{\ast} : | \pi^{\ast} | = k_0,
   \{\mathcal{M}(\mathcal{D}^{\ast}_j)\}_{j = 1}^{| \pi^{\ast} |} =\{1,
   \ldots, k_0 \} \infixand \epsilon_1 (\pi^{\ast}, \pi_0^{\ast}) {> k_0^{1 +
   2 / d}}  n^{-1/2}\log^{\alpha_0+(1+\alpha_b)/2}(n) \right\}, \]
and
\[ \Pi_5^{\ast} = \left\{ \pi^{\ast} \in \Xi^{\ast} : | \pi^{\ast} | = k_0,
   \{\mathcal{M}(\mathcal{D}^{\ast}_j)\}_{j = 1}^{| \pi^{\ast} |} =\{1,
   \ldots, k_0 \} \infixand \epsilon_1 (\pi^{\ast}, \pi_0^{\ast}) {\leqslant
   k_0^{1 + 2 / d}}  n^{-1/2}\log^{\alpha_0+(1+\alpha_b)/2}(n) \right\} . \]
Both $\Pi_1^{\ast}$ and $\Pi_3^{\ast}$ are sets of ``underfitted" partitions,
in the sense that $\lvert \{\mathcal{M}(\mathcal{D}^{\ast}_j)\}_{j = 1}^{|
\pi^{\ast} |} \rvert < k_0$ for $\pi^{\ast} \in \Pi_1^{\ast} \cup
\Pi_3^{\ast}$. $\Pi_2^{\ast}$ is the set of ``overfitted" partitions, in the
sense that $| \pi^{\ast} | > k_0$ for $\pi^{\ast} \in \Pi_2^{\ast}$.
$\Pi_4^{\ast}$ is the set of partitions with relatively large $\epsilon_1
(\pi^{\ast}, \pi_0^{\ast})$, although having the correct number of clusters.
$\Pi_5^{\ast}$ is the set of ``good" partitions, in the sense that the number
of clusters is correct and $\epsilon_1 (\pi^{\ast}, \pi_0^{\ast})$ is small.

$\Pi_5^\ast$ is closely related to $\Pi_0^\ast$. Specifically, following similar arguments of Lemma S.4 in \cite{huang2025consistent}, it can be shown that under Assumptions \ref{AS:lengthofboundary} and \ref{AS:hyperpara}, if
$\pi^{\ast} \in \Pi_5^{\ast}$, we have $\epsilon_1 (\pi^{\ast}, \pi_0^{\ast}) =
\epsilon_2 (\pi^{\ast}, \pi_0^{\ast})$ when $n$ is large. Thus, together with the definition of
$\Pi_5^{\ast}$ and $\Pi_0^{\ast}$, we conclude that $\Pi_5^{\ast} \subseteq
\Pi_0^{\ast}$ when $n$ is large. Consequently, to establish Theorems \ref{TH:piast} - \ref{TH:piastpidagger}, it suffices to 
verify Assumptions \ref{AS:ansum} - \ref{AS:anfurther}
on the set $\Pi_5^\ast$ (i.e., substituting $\Pi_0^\ast$ in  Assumptions \ref{AS:ansum} - \ref{AS:anfurther} with $\Pi_5^\ast$).

A key component in Assumptions \ref{AS:ansum} - \ref{AS:anfurther} is $\frac{\mathbb{P}
(\mathfrak{D} \mid \pi^{\ast})}{\mathbb{P} (\mathfrak{D} \mid \pi^{\ast}_0)} =
\frac{\mathbb{P} (\tmmathbf{y} \mid \pi^{\ast}, \mathcal{X},
\mathcal{S})}{\mathbb{P} (\tmmathbf{y} \mid \pi^{\ast}_0, \mathcal{X},
\mathcal{S})}$. To study it, we first define two events, say $\mathcal{A}_n$ and
$\mathcal{E}_n$, and show that $\mathbb{P} (\mathcal{A}_n
\cap \mathcal{E}_n) \rightarrow 1$. The definitions of $\mathcal{A}_n$ and
$\mathcal{E}_n$ are deferred to Section \ref{SEC:events}. Under $\mathcal{A}_n \cap \mathcal{E}_n$, we have the following result.

\begin{proposition}
  \label{PP:likelihoodratio}Under events $\mathcal{A}_n \cap \mathcal{E}_n$
  and Assumptions \ref{AS:distributionloc} - \ref{AS:hyperpara}, there exists a constant $c$, such that
  \[ \frac{\mathbb{P} (\tmmathbf{y} \mid \pi^{\ast}, \mathcal{X}, \mathcal{S})
     \lambda^{| \pi^{\ast} |}}{\mathbb{P} (\tmmathbf{y} \mid \pi^{\ast}_0,
     \mathcal{X}, \mathcal{S}) \lambda^{| \pi_0^{\ast} |}} \leqslant \left\{
     \begin{array}{l}
       \exp (- c k_0^{- 1 - 2 / d} n), \text{if } \pi^{\ast} \in
       \Pi_1^{\ast}\\
       \exp \{- c \log (\lambda^{- 1}) \}, \text{if } \pi^{\ast} \in
       \Pi_2^{\ast}\\
       \exp (- ck_0^{- 1 - 2 / d} n), \text{if } \pi^{\ast} \in \Pi_3^{\ast}\\
       \exp \{- c n k_0 K^{- 1} \log^{\alpha_0} (n) \}, \text{if } \pi^{\ast}
       \in \Pi_4^{\ast}
     \end{array} \right. . \]
  \begin{proof}
    The proof is deferred to Section \ref{SEC:proofofPP5}.
  \end{proof}
\end{proposition}


Proposition \ref{PP:likelihoodratio} bounds $\frac{\mathbb{P} (\tmmathbf{y} \mid \pi^{\ast}, \mathcal{X}, \mathcal{S})
     \lambda^{| \pi^{\ast} |}}{\mathbb{P} (\tmmathbf{y} \mid \pi^{\ast}_0,
     \mathcal{X}, \mathcal{S}) \lambda^{| \pi_0^{\ast} |}}$ for $\pi^\ast \in (\Pi_5^\ast)^c$, which is useful for verifying Assumptions \ref{AS:ansum}--\ref{AS:anfurther}. Note that Proposition \ref{PP:likelihoodratio} appears similar to Proposition S.1 in \cite{huang2025consistent}. However, we emphasize that the proof techniques are different, while following a similar overall framework. Specifically, our hyperparameter condition in Assumption \ref{AS:hyperpara} is developed under the FDG regime, whereas the hyperparameter condition in Assumption 5 of \cite{huang2025consistent} is developed under the VDG regime with a known $k_0$ order. We obtain bounds of the same order in Proposition \ref{PP:likelihoodratio} because the reduced partition space of A-Spat-RPM mitigates the noise effect from $\epsilon(\cdot)$; see the definition of $\mathcal{E}_n$ in Section \ref{SEC:events} for details.

     The remaining proof in this section is organized as follows. In Section \ref{SEC:prelemmas},
we provide some basic lemmas regarding probabilities, distributions, and
inequalities, which are useful for later proofs. In Section
\ref{SEC:events}, we define two events $\mathcal{A}_n$ and $\mathcal{E}_n$,
and show that $\mathbb{P} (\mathcal{A}_n \cap \mathcal{E}_n) \rightarrow 1$.
In Section \ref{SEC:proofoftheorem12}, we prove Theorems \ref{TH:piast} -
\ref{TH:piastpidagger} by verifying Assumptions \ref{AS:ansum} - \ref{AS:anfurther}, making use of Proposition \ref{PP:likelihoodratio}. In
Section \ref{SEC:proofofPP5}, we prove Proposition \ref{PP:likelihoodratio}.

\subsubsection{Preliminary lemmas}\label{SEC:prelemmas}


\begin{lemma}
  \label{LM:chisquarebound}(Lemma 1 of {\cite{laurent2000adaptive}}) Let
  $\chi^2_r$ be a chi-square distribution with a degree of freedom $r$. The
  following concentration inequalities hold for any $x > 0$:
  \[ \mathbb{P} \left( \chi^2_r > r + 2 x + 2 \sqrt{rx} \right) \leq \exp (-
     x), \]
  and
  \[ \mathbb{P} \left( \chi^2_r < r - 2 \sqrt{rx} \right) \leq \exp (- x) . \]
\end{lemma}

From Lemma \ref{LM:chisquarebound}, we can see that for $x > r$, we have
\begin{equation}
  \mathbb{P} (\chi^2_r > 5 x) \leqslant \mathbb{P} \left( \chi^2_r > r + 2 x +
  2 \sqrt{rx} \right) \leq \exp (- x) . \label{EQ:chisqbound}
\end{equation}
\begin{lemma}
  \label{LM:binomialcoe}The binomial coefficient satisfies
  \[ \left( \frac{n}{k} \right)^k \leqslant \left( \begin{array}{c}
       n\\
       k
     \end{array} \right) \leqslant \left( \frac{en}{k} \right)^k \]
  and
  \[ \sum_{i = 0}^k \left( \begin{array}{c}
       n\\
       i
     \end{array} \right) \leqslant (n + 1)^k \]
  for k=0,...,n.
  
  \begin{proof}
    The proof is trivial and omitted.
  \end{proof}
\end{lemma}

\begin{lemma}
  \label{LM:chisquaredis}For a real symmetric matrix $\tmmathbf{A}$ satisfying
  $\tmmathbf{A}^2 =\tmmathbf{A}$ and a standard Gaussian random vector
  $\tmmathbf{e}$, we have
  \[ \tmmathbf{e}^T \tmmathbf{A}\tmmathbf{e} \sim \chi_r^2, \]
  where $r$ is the number of positive eigenvalues of $\tmmathbf{A}$.
  
  \begin{proof}
    The proof is trivial by performing eigen-decomposition of $\tmmathbf{A}$
    and noting that the eigenvalues of $\tmmathbf{A}$ equal either $0$ or $1$.
  \end{proof}
\end{lemma}

\begin{lemma}
  \label{LM:spacesize}Under our model, we have
  \[ | \Xi^{\dagger} | \leqslant \left( \begin{array}{l}
       K^2 - 1\\
       k^{\dagger} - 1
     \end{array} \right) \leqslant \exp \{ 2 k^{\dagger} \log (K) \},
     \text{and } | \Xi^{\ast} (\pi^{\dagger}) | \leqslant | \pi^{\dagger} |^{|
     \pi^{\dagger} |} = \exp \{ k^{\dagger} \log (k^{\dagger}) \} . \]
  Consequently,
  \[ | \Xi^{\ast} | \leqslant | \Xi^{\dagger} | \sup_{\pi^{\dagger} \in
     \Xi^{\dagger}} | \Xi^{\ast} (\pi^{\dagger}) | \leqslant \exp \{ 2
     k^{\dagger} \log (K) + k^{\dagger} \log (k^{\dagger}) \} . \]
\end{lemma}

\subsubsection{Definition of events}\label{SEC:events}

We define $\mathcal{A}_n =\mathcal{A}_{1 n} \cap \mathcal{A}_{2 n}$ as events
with respect to $\mathcal{S}$ and $\mathcal{X}$, where
\begin{equation}
  \mathcal{A}_{1 n} = \left\{ c < \frac{\min_{1 \leqslant m \leqslant K^2}
  \|B_m \|}{nK^{- 2}} \leqslant \frac{\max_{1 \leqslant m \leqslant K^2} \|B_m
  \|}{nK^{- 2}} < C \right\}, \text{and} \label{EQ:A1n}
\end{equation}
\begin{equation}
  \mathcal{A}_{2 n} = \left\{ 0 < c <\|B_m \|^{- 1}  \sum_{\tmmathbf{s}_i \in
  B_m} \tmmathbf{x}(\tmmathbf{s}_i)\tmmathbf{x}^T (\tmmathbf{s}_i) < C <
  \infty, m = 1, 2, \ldots, K^2 \right\} \label{EQ:A2n}
\end{equation}
for some constants $c, C > 0$ to be determined later. The following result is from Lemma S.1 in \cite{huang2025consistent}.

\begin{lemma}
  \label{LM:An}Under Assumptions \ref{AS:distributionloc}, \ref{AS:covariates}
  and \ref{AS:hyperpara}, there exist constants $c, C > 0$, such that
  $\mathbb{P} (\mathcal{A}_n) \rightarrow 1$.
\end{lemma}

In what follows, we assume that the constants $c$ and $C$ in (\ref{EQ:A1n}) -
(\ref{EQ:A2n}) are chosen such that $\mathbb{P} (\mathcal{A}_n) \rightarrow
1$. Under $\mathcal{A}_n$, we can replace the pseudoinverse notation in
(\ref{EQ:priortheta}) with inverse. For the simplicity of proof, we take the
value of $\gamma$ in (\ref{EQ:priortheta}) to be $1$, and it is trivial to
extend the proof to the case for any fixed $\gamma > 0$. For a given
$\pi^{\ast}$, following (\ref{EQ:priortheta}) - (\ref{EQ:priory}), and after
replacing the pseudoinverse notation with inverse, we have
\begin{equation}
  \tmmathbf{\theta} \mid \pi^{\ast}, k, \mathcal{T}, \mathcal{S}, \mathcal{X}
  \sim \prod_{j = 1}^k \mathbb{P}_{\tmop{Gaussian}} \{\tmmathbf{\theta}_j ;
  \tmmathbf{0}, n \sigma^2 (\tmmathbf{x}_j^T \tmmathbf{x}_j)^{- 1} \},
  \text{and} \label{EQ:thetajdis}
\end{equation}
\begin{equation}
  \tmmathbf{y} \mid \pi^{\ast}, k, \mathcal{T}, \mathcal{S}, \mathcal{X},
  \tmmathbf{\theta} \sim \prod_{j = 1}^k \mathbb{P}_{\tmop{Gaussian}}
  (\tmmathbf{y}_j ; \tmmathbf{x}_j \tmmathbf{\theta}_j, \sigma^2
  \tmmathbf{I}_{\|\mathcal{D}_j^{\ast} \|}), \label{EQ:yjmodel}
\end{equation}
where $\tmmathbf{x}_j$ is a $\|\mathcal{D}_j^{\ast} \| \times d$ design matrix
of covariates in $\mathcal{D}_j^{\ast}$, and $\tmmathbf{y}_j = \{y
(\tmmathbf{s}_i) : \tmmathbf{s}_i \in \mathcal{D}_j^{\ast} \}$. For a given
$\pi^{\ast}$, following Equations (\ref{EQ:thetajdis}) - (\ref{EQ:yjmodel}),
$\tmmathbf{y}_j \mid \pi^{\ast}, k, \mathcal{T}, \mathcal{S}, \mathcal{X} \sim
\mathbb{P}_{\tmop{Gaussian}} [\tmmathbf{y}_j ; \tmmathbf{0}, \sigma^2
\{n\tmmathbf{x}_j (\tmmathbf{x}_j^T \tmmathbf{x}_j)^{- 1} \tmmathbf{x}_j^T
+\tmmathbf{I}_{\|\mathcal{D}_j \|} \}]$. We can thus write
\begin{eqnarray}
  \mathbb{P} (\tmmathbf{y} \mid \pi^{\ast}, \mathcal{X}, \mathcal{S}) & = &
  \mathbb{P} (\tmmathbf{y} \mid \pi^{\ast}, k, \mathcal{T}, \mathcal{X},
  \mathcal{S}) = \prod_{j = 1}^k \mathbb{P}_{\tmop{Gaussian}} [\tmmathbf{y}_j
  ; \tmmathbf{0}, \sigma^2 \{n\tmmathbf{\phi}_j +\tmmathbf{I}_{\|\mathcal{D}_j
  \|} \}] \nonumber\\
  & = & (2 \pi \sigma^2)^{- n / 2} \exp \left( - \frac{\tmmathbf{y}^T
  \tmmathbf{y}}{2 \sigma^2} \right)  (n + 1)^{- kd / 2} \exp \left\{
  \frac{n}{2 \sigma^2 (n + 1)} \tmmathbf{y}^T \tmmathbf{\phi}_{\pi^{\ast}}
  \tmmathbf{y} \right\},  \label{EQ:formlikelihood}
\end{eqnarray}
where $\tmmathbf{\phi}_j =\tmmathbf{x}_j  (\tmmathbf{x}_j^T \tmmathbf{x}_j)^{-
1} \tmmathbf{x}_j^T$ is a projection matrix, $\tmmathbf{\phi}_{\pi^{\ast}}
=\tmmathbf{P}_{\pi^{\ast}}^T \tmop{diag} (\tmmathbf{\phi}_1, \ldots
\tmmathbf{\phi}_k) \tmmathbf{P}_{\pi^{\ast}}$, and $\tmmathbf{P}_{\pi^{\ast}}$
ia a permutation matrix such that $\tmmathbf{P}_{\pi^{\ast}} \tmmathbf{y}=
(\tmmathbf{y}^T_1, \ldots, \tmmathbf{y}^T_k)^T$. We will use Equation
(\ref{EQ:formlikelihood}) in the proofs of Theorems
\ref{TH:piast} - \ref{TH:piastpidagger} later.

Let $\tmmathbf{\epsilon}= \{\epsilon (\tmmathbf{s}_i)\}_{i = 1}^n$ for $\epsilon(\tmmathbf{s}_i)$ in (\ref{EQ:model}). We define
$\mathcal{E}_n$ as
\begin{equation}
  \mathcal{E}_n = \{\tmmathbf{\epsilon}: \tmmathbf{\epsilon}^T
  (\tmmathbf{\phi}_{\pi^{\ast} \cap \pi^{\ast}_0}
  +\tmmathbf{\phi}_{\pi^{\ast}})\tmmathbf{\epsilon} \leqslant Ck^{\dagger}
  \log (n), \forall \pi^{\ast} \in \Xi^{\ast} \} \label{EQ:epsilon1n},
\end{equation}
for some constant $C$ to be determined later. We have the following lemma.

\begin{lemma}
  \label{LM:EpsilonT}Under our model and Assumption
  \ref{AS:hyperpara}, there exists a constant $C$, such that for
  $\mathcal{E}_n$ defined in (\ref{EQ:epsilon1n}), we have
  \[ \mathbb{P} (\mathcal{E}_n) \rightarrow 1. \]
  \begin{proof}
    Recall that $\sigma_0^2$ is the true variance of  $\epsilon (\tmmathbf{s}_i)$. We set the constant $C$ in (\ref{EQ:epsilon1n}) to be $40 d \sigma_0^2$.
    To show $\mathbb{P} (\mathcal{E}_n) \rightarrow 1$, it suffices to show
    that
    \begin{equation}
      \mathbb{P} [\tmmathbf{\epsilon}^T \tmmathbf{\phi}_{\pi^{\ast} \cap
      \pi^{\ast}_0} \tmmathbf{\epsilon}> 20 k^{\dagger} d \sigma_0^2 \log (n),
      \exists \pi^{\ast} \in \Xi^{\ast}] \rightarrow 0, \label{EQ:epsilon11}
    \end{equation}
    and
    \begin{equation}
      \mathbb{P} [\tmmathbf{\epsilon}^T \tmmathbf{\phi}_{\pi^{\ast}}
      \tmmathbf{\epsilon}> 20 k^{\dagger} d \sigma_0^2 \log (n), \exists
      \pi^{\ast} \in \Xi^{\ast}] \rightarrow 0. \label{EQ:epsilon12}
    \end{equation}
    
    We provide proof of Equation (\ref{EQ:epsilon11}), and the proof of
    Equation (\ref{EQ:epsilon12}) is similar. First of all, it is easy to see that
    $\tmmathbf{\phi}_{\pi^{\ast} \cap \pi^{\ast}_0}$ is a real symmetric
    matrix with $\tmmathbf{\phi}_{\pi^{\ast} \cap \pi^{\ast}_0}^2
    =\tmmathbf{\phi}_{\pi^{\ast} \cap \pi^{\ast}_0}$. Second of all,
    suppose $\pi^{\ast}\in \Xi^\ast(\pi^\dagger)$ for some $\pi^\dagger\in \Xi^\dagger$. Under our model, we have $\pi^{\dagger} \preceq \pi^{\ast}$, and $\pi_0^{\dagger} \preceq
    \pi^{\ast}_0$, leading to $(\pi^{\dagger} \cap \pi_0^{\dagger}) \preceq
    (\pi^{\ast} \cap \pi^{\ast}_0)$. Noting that $\pi^{\dagger}$
    and $\pi_0^{\dagger}$ are induced from the same spanning tree
    $\mathcal{T}^{\dagger}$, we conclude that $| \pi^{\ast} \cap \pi_0^{\ast}
    | \leqslant | \pi^{\dagger} \cap \pi_0^{\dagger} | \leqslant 2
    k^{\dagger}$. Hence, the number of positive eigenvalues of
    $\tmmathbf{\phi}_{\pi^{\ast} \cap \pi^{\ast}_0}$ is smaller than $2
    k^{\dagger} d$. Combining the above arguments and applying Lemma \ref{LM:chisquaredis}, $\sigma_0^{- 2}
    \tmmathbf{\epsilon}^T \tmmathbf{\phi}_{\pi^{\ast} \cap \pi^{\ast}_0}
    \tmmathbf{\epsilon}$ follows a chi-square distribution with degree of
    freedom no larger than $2 k^{\dagger} d$. Applying Equation
    (\ref{EQ:chisqbound}), we have
    \begin{eqnarray}
      &  & \mathbb{P} \{\tmmathbf{\epsilon}^T \tmmathbf{\phi}_{\pi^{\ast}
      \cap \pi^{\ast}_0} \tmmathbf{\epsilon}> 20 k^{\dagger} d \sigma_0^2 \log
      (n), \exists \pi^{\ast} \in \Xi^{\ast} \} \nonumber\\
      & \leqslant & \sum_{\pi^{\ast} \in \Xi^{\ast}} \mathbb{P} \{\sigma_0^{-
      2} \tmmathbf{\epsilon}^T \tmmathbf{\phi}_{\pi^{\ast} \cap \pi^{\ast}_0}
      \tmmathbf{\epsilon}> 20 k^{\dagger} d \log (n)\} \nonumber\\
      & \overset{\text{Equation } \left( \ref{EQ:chisqbound}
      \right)}{\leqslant} & \sum_{\pi^{\ast} \in \Xi^{\ast}} \exp \{- 4
      k^{\dagger} d \log (n)\} \nonumber\\
      & \overset{\text{Lemma \ref{LM:spacesize}}}{\leqslant} & \exp \{- 4
      k^{\dagger} d \log (n) + 2 k^{\dagger} \log (n) + k^{\dagger} \log (n)\}
      \nonumber\\
      & \leqslant & \exp \{- k^{\dagger} \log (n)\} \rightarrow 0. \nonumber
    \end{eqnarray}
    The lemma is proved.
  \end{proof}
\end{lemma}

In what follows, we assume that the constant $C$ in (\ref{EQ:epsilon1n}) is
chosen such that $\mathbb{P} (\mathcal{E}_n) \rightarrow 1$. For the remaining
proofs provided in the following sections, we will consider the probability
conditional on events $\mathcal{A}_n \cap \mathcal{E}_n$.

\subsubsection{Proof of Theorems \ref{TH:piast} -\ref{TH:piastpidagger}}\label{SEC:proofoftheorem12}

\begin{lemma}
  \label{LM:pidaggerastratio}Under our model and Assumption
  \ref{AS:hyperpara}, for $\forall \pi^{\dagger} \in \Xi^{\dagger}$ and
  $\pi^{\ast} \in \Xi^{\ast} (\pi^{\dagger})$, we have
  \[ \frac{\mathbb{P} (\tmmathbf{} \pi^{\ast} \mid \pi^{\dagger})}{\mathbb{P}
     (\tmmathbf{} \pi^{\ast}_0 \mid \pi^{\dagger}_0)} \leqslant c
     \frac{\lambda^k}{\lambda^{k_0}} \exp \{ c k^{\dagger} \log (n) \} \]
  for some constant $c$.
  
  \begin{proof}
    We compute
    \begin{eqnarray*}
      \frac{\mathbb{P} (\tmmathbf{} \pi^{\ast} \mid \pi^{\dagger})}{\mathbb{P}
      (\tmmathbf{} \pi^{\ast}_0 \mid \pi^{\dagger}_0)} & = & \frac{\mathbb{P}
      (| \pi^{\ast} |) \mathbb{P} (\text{$\mathcal{T}$ {induced} {by}
      $\mathcal{G}/\pi^{\dagger}$ {is compatible with} $\pi^{\ast}$}) \left(
      \begin{array}{l}
        k^{\dagger} - 1\\
        | \pi^{\ast} | - 1
      \end{array} \right)^{- 1}}{\mathbb{P} (| \pi^{\ast}_0 |) \mathbb{P}
      (\text{$\mathcal{T}$ {induced} {by}
      $\mathcal{G}/\pi^{\dagger}_0$ {is compatible with} $\pi^{\ast}_0$}) \left( \begin{array}{l}
        k^{\dagger} - 1\\
        | \pi^{\ast}_0 | - 1
      \end{array} \right)^{- 1}}\\
      & \overset{(i)}{\leqslant} & \frac{\lambda^k}{\lambda^{k_0}} \frac{k_0
      !}{k!} \times \exp \{ k^{\dagger} \log (n) \} \times \left(
      \begin{array}{l}
        k^{\dagger} - 1\\
        k_0 - 1
      \end{array} \right)\\
      & \leqslant & c \frac{\lambda^k}{\lambda^{k_0}} \exp \{ c k^{\dagger}
      \log (n) \} .
    \end{eqnarray*}
    where $(i)$ uses the fact that the number of spanning trees induced by
    $\mathcal{G}/\pi^{\dagger}_0$ is upper bounded by $| \pi^{\dagger}_0 |^{|
    \pi^{\dagger}_0 | - 2}$ \citep{Cayley}. The lemma is proved. 
  \end{proof}
\end{lemma}

We next verify Assumption \ref{AS:ansum} as follows, with $\Pi_0^\ast$ substituted with $\Pi_5^\ast$ following the argument in Section \ref{SEC:proofframework}.

\begin{proof}
  Following Lemma \ref{LM:pidaggerastratio}, we have
  \begin{eqnarray*}
    \frac{\mathbb{P} (\pi^{\ast} \mid \pi^{\dagger}) \mathbb{P} (\mathfrak{D}
    \mid \pi^{\ast})}{\mathbb{P} (\pi^{\ast}_0 \mid \pi^{\dagger}_0)
    \mathbb{P} (\mathfrak{D} \mid \pi^{\ast}_0)} = \frac{\mathbb{P}
    (\pi^{\ast} \mid \pi^{\dagger}) \mathbb{P} (\tmmathbf{y} \mid \pi^{\ast},
    \mathcal{X}, \mathcal{S})}{\mathbb{P} (\pi^{\ast}_0 \mid \pi^{\dagger}_0)
    \mathbb{P} (\tmmathbf{y} \mid \pi^{\ast}_0, \mathcal{X}, \mathcal{S})} &
    \leqslant & c \exp \{ c k^{\dagger} \log (n) \} \frac{\mathbb{P}
    (\tmmathbf{y} \mid \pi^{\ast}, \mathcal{X}, \mathcal{S})
    \lambda^k}{\mathbb{P} (\tmmathbf{y} \mid \pi^{\ast}_0, \mathcal{X},
    \mathcal{S}) \lambda^{k_0}} .
  \end{eqnarray*}
  On the other hand, from the result in Proposition \ref{PP:likelihoodratio},
  we have for $\pi^{\ast} \in (\Pi_5^{\ast})^c$, $\frac{\mathbb{P}
  (\tmmathbf{y} \mid \pi^{\ast}, \mathcal{X}, \mathcal{S})
  \lambda^k}{\mathbb{P} (\tmmathbf{y} \mid \pi^{\ast}_0, \mathcal{X},
  \mathcal{S}) \lambda^{k_0}} \leqslant \exp \{- c n k_0 K^{- 1}
  \log^{\alpha_0} (n) \}$. Thus, we compute
  \begin{eqnarray*}
    & &\sum_{\pi^{\dagger} \in \Xi^{\dagger}} \sum_{\pi^{\ast} \in \{ \Xi^{\ast}
    (\pi^{\dagger}) \cap (\Pi_5^{\ast})^c \}} \frac{\mathbb{P} (\pi^{\ast}
    \mid \pi^{\dagger}) \mathbb{P} (\mathfrak{D} \mid \pi^{\ast})}{\mathbb{P}
    (\pi^{\ast}_0 \mid \pi^{\dagger}_0) \mathbb{P} (\mathfrak{D} \mid
    \pi^{\ast}_0)} \\
    & \leqslant & \sum_{\pi^{\dagger} \in \Xi^{\dagger}}
    \sum_{\pi^{\ast} \in \{ \Xi^{\ast} (\pi^{\dagger}) \cap (\Pi_5^{\ast})^c
    \}} c \exp \{ c k^{\dagger} \log (n) - c n k_0 K^{- 1} \log^{\alpha_0} (n)
    \}\\
    & \overset{\text{Lemma \ref{LM:spacesize}}}{\leqslant} & \exp \{ c
    k^{\dagger} \log (n) - c n k_0 K^{- 1} \log^{\alpha_0} (n) \}
    \overset{\text{Assumption \ref{AS:hyperpara}}}{\rightarrow} 0.
  \end{eqnarray*}
  Assumption \ref{AS:ansum} is verified.
\end{proof}

Theorem \ref{TH:piast} is thus proved after applying Lemma \ref{LM:pidaggerconsistency_general} and using the fact that $\Pi_5^\ast \subseteq \Pi_0^\ast$. We furthermore verify Assumption \ref{AS:anfurther} as follows, with $\Pi_0^\ast$ substituted with $\Pi_5^\ast$.

\begin{proof}
We compute
\begin{eqnarray*}
  & \sup_{\pi^{\dagger} \in \Xi^\dagger \cap \Pi^{\dagger}_0} \left\{ \frac{\mathbb{P}
  (\pi^{\ast}_0 \mid \pi^{\dagger}_0)}{\mathbb{P} (\pi^{\ast}_0
  (\pi^{\dagger}) \mid \pi^{\dagger})} \frac{\mathbb{P} (\mathfrak{D} \mid
  \pi^{\ast}_0)}{\mathbb{P} (\mathfrak{D} \mid \pi^{\ast}_0 (\pi^{\dagger}))}
  \times \sum_{\pi^{\ast} \in  \Xi^{\ast} (\pi^{\dagger}) \cap
  (\Pi_5^{\ast})^c } \frac{\mathbb{P} (\pi^{\ast} \mid \pi^{\dagger})
  \mathbb{P} (\mathfrak{D} \mid \pi^{\ast})}{\mathbb{P} (\pi^{\ast}_0 \mid
  \pi^{\dagger}_0) \mathbb{P} (\mathfrak{D} \mid \pi^{\ast}_0)} \right\} & \\
  \leqslant & \sup_{\pi^{\dagger} \in \Xi^\dagger \cap \Pi^{\dagger}_0} \left\{
  \frac{\mathbb{P} (\pi^{\ast}_0 \mid \pi^{\dagger}_0)}{\mathbb{P}
  (\pi^{\ast}_0 (\pi^{\dagger}) \mid \pi^{\dagger})} \frac{\mathbb{P}
  (\mathfrak{D} \mid \pi^{\ast}_0)}{\mathbb{P} (\mathfrak{D} \mid \pi^{\ast}_0
  (\pi^{\dagger}))} \times c \exp \{ c k^{\dagger} \log (n) - c n k_0 K^{- 1}
  \log^{\alpha_0} (n) \} \right\} & \\
  \leqslant & \sup_{\pi^{\dagger} \in \Xi^\dagger \cap \Pi^{\dagger}_0} \left[ \frac{\mathbb{P}
  \text{($\mathcal{T}$ {induced} {by} $\mathcal{G}/\pi^{\dagger}_0$ {is compatible with} $\pi^{\ast}_0$)}}{\mathbb{P} \text{($\mathcal{T}$ {induced} {by} $\mathcal{G}/\pi^{\dagger}$ {is compatible with} $\pi^{\ast}_0(\pi^\dagger)$)}} \times \frac{\mathbb{P}(\tmmathbf{y} \mid \pi^{\ast}_0,
  \mathcal{X}, \mathcal{S})}{\mathbb{P} \{ \tmmathbf{y} \mid \pi^{\ast}_0
  (\pi^{\dagger}), \mathcal{X}, \mathcal{S} \}} \right.\\
  &\quad\quad\quad\quad\quad\quad\quad\quad\quad\quad\quad\quad\quad\quad\quad \left.\times c \exp \{ c k^{\dagger}
  \log (n) - c n k_0 K^{- 1} \log^{\alpha_0} (n) \} \right]&  \\
  \overset{(i)}{\leqslant} & \sup_{\pi^{\dagger} \in \Xi^\dagger \cap\Pi^{\dagger}_0} \left[
  \frac{\mathbb{P}(\tmmathbf{y} \mid \pi^{\ast}_0, \mathcal{X},
  \mathcal{S})}{\mathbb{P} \{ \tmmathbf{y} \mid \pi^{\ast}_0 (\pi^{\dagger}),
  \mathcal{X}, \mathcal{S} \}} \times c \exp \{ c k^{\dagger} \log (n) - c n
  k_0 K^{- 1} \log^{\alpha_0} (n) \} \right] & \\
  \overset{(i i)}{\leqslant} & c \exp \{ cn k_0 K^{- 1} + c k^{\dagger} \log
  (n) - c n k_0 K^{- 1} \log^{\alpha_0} (n) \}  \overset{\text{Assumption \ref{AS:hyperpara}}}{\rightarrow} 0, & 
\end{eqnarray*}
where $(i)$ uses the fact that the number of spanning trees induced by
$\mathcal{G}/\pi^{\dagger}$ is upper bounded by $| \pi^{\dagger} |^{| \pi^{\dagger} | -
2}$ \citep{Cayley}, and $(i i)$ uses Equations
(\ref{EQ:likelihoodratioexpre}) - (\ref{EQ:Rsquaredif}).
\end{proof}
Theorem \ref{TH:piastpidagger} then follows by applying Lemma
\ref{LM:piastconditionalpidagger}  and using the fact that $\Pi_5^\ast \subseteq \Pi_0^\ast$. 

\subsubsection{Proof of Proposition \ref{PP:likelihoodratio}}\label{SEC:proofofPP5}

We first present two lemmas analogous to those in \cite{huang2025consistent}. Let
$\tmmathbf{\mu}_0$ be the $n$ by $1$ true regression mean vector.

\begin{lemma}
  \label{LM:bestapp}Under the event $\mathcal{A}_n$ and Assumptions
  \ref{AS:lengthofboundary}, \ref{AS:covariates} and \ref{AS:hyperpara}, there
  exists a constant $c$, such that
  \[ \|(\tmmathbf{\phi}_{\pi^{\ast}_0} -\tmmathbf{I}_n)\tmmathbf{\mu}_0 \|^2_2
     \leqslant cn k_0 K^{- 1} . \]
  \begin{proof}
    The proof is similar to the proof of Lemma S.20 in \cite{huang2025consistent} and is thus
    omitted.
  \end{proof}
\end{lemma}

\begin{lemma}
  \label{LM:underfitted}Under the event $\mathcal{A}_n$ and Assumption
  \ref{AS:thetagap}, there exists a uniform constant $c > 0$, such that
  \begin{equation}
    \|(\tmmathbf{I}_n -\tmmathbf{\phi}_{\pi^{\ast}})\tmmathbf{\mu}_0 \|^2_2
    \geqslant c k_0^{- 2 / d} n \epsilon_1 (\pi^{\ast}, \pi_0^{\ast})
    \label{EQ:underfittederror}
  \end{equation}
  for any $\pi^{\ast} \in \Xi^{\ast}$ with $\epsilon_1 (\pi^{\ast},
  \pi_0^{\ast}) \geqslant k_0 K^{- 1} \log^{\alpha_0/2} (n)$.
  \begin{proof}
    The proof is similar to the proof of Lemma S.21 of \cite{huang2025consistent} and is thus
    omitted.
  \end{proof}
\end{lemma}

With the two technical lemmas above, we next prove Proposition
\ref{PP:likelihoodratio}.

\begin{proof}
  Following (\ref{EQ:formlikelihood}), the ratio in Proposition
  \ref{PP:likelihoodratio} is computed as
  \begin{equation}
    \frac{\mathbb{P} (\tmmathbf{y} \mid \pi^{\ast}, \mathcal{X}, \mathcal{S})
    \lambda^{| \pi^{\ast} |}}{\mathbb{P} (\tmmathbf{y} \mid \pi^{\ast}_0,
    \mathcal{X}, \mathcal{S}) \lambda^{| \pi^{\ast}_0 |}} = \exp \left[ (k_0 -
    | \pi^{\ast} |) \left\{ \log (\lambda^{- 1}) + \frac{d \log (n +
    1^{\nosymbol})}{2} \right\} + \frac{n\tmmathbf{y}^T
    (\tmmathbf{\phi}_{\pi^{\ast}}
    -\tmmathbf{\phi}_{\pi^{\ast}_0})\tmmathbf{y}}{2 \sigma^2 (n + 1)} \right]
    . \label{EQ:likelihoodratioexpre}
  \end{equation}
  For a given $\pi^{\ast} \in \Xi^{\ast}$, write $\tilde{\pi}^{\ast} =
  \pi^{\ast} \cap \pi^{\ast}_0$. We bound $\tmmathbf{y}^T 
  (\tmmathbf{\phi}_{\pi^{\ast}} -\tmmathbf{\phi}_{\pi_0^{\ast}}) \tmmathbf{y}$
  in (\ref{EQ:likelihoodratioexpre}) by
  \begin{eqnarray}
    \tmmathbf{y}^T  (\tmmathbf{\phi}_{\pi^{\ast}}
    -\tmmathbf{\phi}_{\pi_0^{\ast}}) \tmmathbf{y} & = & \tmmathbf{y}^T 
    (\tmmathbf{\phi}_{\tilde{\pi}^{\ast}} -\tmmathbf{\phi}_{\pi^{\ast}_0})
    \tmmathbf{y}+\tmmathbf{y}^T  (\tmmathbf{\phi}_{\pi^{\ast}}
    -\tmmathbf{\phi}_{\tilde{\pi}^{\ast}}) \tmmathbf{y} \nonumber\\
    & \overset{(i)}{=} & \|(\tmmathbf{\phi}_{\tilde{\pi}^{\ast}}
    -\tmmathbf{\phi}_{\pi^{\ast}_0})\tmmathbf{y}\|^2_2 -
    \|(\tmmathbf{\phi}_{\tilde{\pi}^{\ast}}
    -\tmmathbf{\phi}_{\pi^{\ast}})\tmmathbf{y}\|^2_2 \nonumber\\
    & \leqslant & 2\| (\tmmathbf{\phi}_{\tilde{\pi}^{\ast}}
    -\tmmathbf{\phi}_{\pi^{\ast}_0}) \tmmathbf{\mu}_0  \|^2_2 + 2\| 
    (\tmmathbf{\phi}_{\tilde{\pi}^{\ast}} -\tmmathbf{\phi}_{\pi^{\ast}_0})
    \tmmathbf{\epsilon}\|^2_2 - \|(\tmmathbf{\phi}_{\tilde{\pi}^{\ast}}
    -\tmmathbf{\phi}_{\pi^{\ast}})\tmmathbf{y}\|^2_2 \nonumber\\
    & \overset{(ii)}{\leqslant} & 8\| (\tmmathbf{I}_n
    -\tmmathbf{\phi}_{\pi^{\ast}_0}) \tmmathbf{\mu}_0  \|^2_2 + 4\|
    \tmmathbf{\phi}_{\tilde{\pi}^{\ast}} \tmmathbf{\epsilon} \|^2 + 4\|
    \tmmathbf{\phi}_{\pi^{\ast}_0} \tmmathbf{\epsilon}\|^2_2 -
    \|(\tmmathbf{\phi}_{\tilde{\pi}^{\ast}}
    -\tmmathbf{\phi}_{\pi^{\ast}})\tmmathbf{y}\|^2_2 \nonumber\\
    & \overset{(iii)}{\leqslant} & cn k_0 K^{- 1} + Ck^{\dagger} \log (n) -
    \|(\tmmathbf{\phi}_{\tilde{\pi}^{\ast}}
    -\tmmathbf{\phi}_{\pi^{\ast}})\tmmathbf{y}\|^2_2,  \label{EQ:Rsquaredif}
  \end{eqnarray}
  where $(i)$ uses the fact that $(\tmmathbf{\phi}_{\tilde{\pi}^{\ast}}
  -\tmmathbf{\phi}_{\pi^{\ast}_0})^2 =\tmmathbf{\phi}_{\tilde{\pi}^{\ast}}
  -\tmmathbf{\phi}_{\pi_0^{\ast}}$ and $(\tmmathbf{\phi}_{\tilde{\pi}^{\ast}}
  -\tmmathbf{\phi}_{\pi^{\ast}})^2 =\tmmathbf{\phi}_{\tilde{\pi}^{\ast}}
  -\tmmathbf{\phi}_{\pi^{\ast}}$ since $\tilde{\pi}^{\ast}$ is nested in
  $\pi^{\ast}$ and $\pi^{\ast}_0$, $(i i)$ is because
  \[ \|(\tmmathbf{\phi}_{\tilde{\pi}^{\ast}}
     -\tmmathbf{\phi}_{\pi^{\ast}_0})\tmmathbf{\mu}_0 \|^2_2 \leqslant 2
     \|(\tmmathbf{\phi}_{\tilde{\pi}^{\ast}} -\tmmathbf{I}_n)\tmmathbf{\mu}_0
     \|^2_2 + 2\|(\tmmathbf{I}_n
     -\tmmathbf{\phi}_{\pi^{\ast}_0})\tmmathbf{\mu}_0 \|^2_2 \leqslant 4
     \|(\tmmathbf{I}_n -\tmmathbf{\phi}_{\pi^{\ast}_0})\tmmathbf{\mu}_0
     \|^2_2, \]
  and $(iii)$ uses results from Lemma \ref{LM:bestapp} and $\mathcal{E}_n$.
  
  We next discuss cases when $\pi^{\ast} \in \Pi_1^{\ast}, \Pi_2^{\ast},
  \Pi_3^{\ast}  \text{ and } \Pi_4^{\ast}$, respectively.
  
  \noindent\tmtextbf{Cases when} $\pi^{\ast} \in \Pi_1^{\ast}$:
  
  Since each block has an area $K^{- 2}$, together with Assumption
  \ref{AS:lengthofboundary}, we can see for each $D_{l, 0}, l = 1, \ldots,
  k_0$, the number of blocks intersecting with it is of the order $ck_0^{- 1}
  K^2$. Combining with Lemma \ref{LM:goodpartition}, we
  further conclude that the number of blocks within each $D^{\ast}_{l, 0}, l =
  1, \ldots, k_0$ is larger than $c k_0^{- 1} K^2$. Similarly, it is easy to
  derive that the number of blocks within each $D^{\ast}_{l, 0}, l = 1,
  \ldots, k_0$ is smaller than $Ck_0^{- 1} K^2$ for another constant $C$.

  Based on the above result, we can see that there exists a constant $c$, such
  that $\epsilon_1 (\pi^{\ast}, \pi_0^{\ast}) K^2 > c (k_0 - k) k_0^{- 1} K^2$
  holds for $\pi^{\ast} \in \Pi_1^{\ast}$. Since
  {Assumption \ref{AS:hyperpara}} guarantees that $k_0^{- 1} \gg k_0
  K^{- 1} \log^{\alpha_0} (n)$, we can apply Lemma \ref{LM:underfitted} and
  obtain
  \begin{equation}
    \|(\tmmathbf{I}_n -\tmmathbf{\phi}_{\pi^{\ast}})\tmmathbf{\mu}_0 \|^2_2
    \geqslant c n (k_0 - k) k_0^{- 1 - 2 / d}, \forall \pi^{\ast} \in
    \Pi_1^{\ast} . \label{EQ:pi1underfitted}
  \end{equation}
  Next, applying Equation (\ref{EQ:Rsquaredif}), we have
  \begin{eqnarray*}
    \tmmathbf{y}^T  (\tmmathbf{\phi}_{\pi^{\ast}}
    -\tmmathbf{\phi}_{\pi_0^{\ast}}) \tmmathbf{y} & \leqslant & cn k_0 K^{- 1}
    + Ck^{\dagger} \log (n) - \|(\tmmathbf{\phi}_{\tilde{\pi}^{\ast}}
    -\tmmathbf{\phi}_{\pi^{\ast}})\tmmathbf{y}\|^2_2\\
    & \leqslant & cn k_0 K^{- 1} + Ck^{\dagger} \log (n) -
    \{\|(\tmmathbf{\phi}_{\tilde{\pi}^{\ast}}
    -\tmmathbf{\phi}_{\pi^{\ast}})\tmmathbf{\mu}_0 \|_2
    -\|(\tmmathbf{\phi}_{\tilde{\pi}^{\ast}}
    -\tmmathbf{\phi}_{\pi^{\ast}})\tmmathbf{\epsilon}\|_2 \}^2 .
  \end{eqnarray*}
  Note that
  \begin{eqnarray}
    \|(\tmmathbf{\phi}_{\tilde{\pi}^{\ast}}
    -\tmmathbf{\phi}_{\pi^{\ast}})\tmmathbf{\mu}_0 \|_2 & \geqslant &
    \|(\tmmathbf{I}_n -\tmmathbf{\phi}_{\pi^{\ast}})\tmmathbf{\mu}_0 \|_2 -
    \|(\tmmathbf{\phi}_{\tilde{\pi}^{\ast}} -\tmmathbf{I}_n)\tmmathbf{\mu}_0
    \|_2 \nonumber\\
    & \overset{(i)}{\geqslant} & cn^{1 / 2} (k_0 - k)^{1 / 2} k_0^{- 1 / 2 -
    1 / d} - Cn^{1 / 2} k_0^{1 / 2} K^{- 1 / 2} \nonumber \\
    & \overset{\text{Assumption
    \ref{AS:hyperpara}}}{\geqslant} & cn^{1 / 2} (k_0 - k)^{1 / 2} k_0^{- 1 / 2
    - 1 / d},  \label{EQ:faimiu2lowerbound}
  \end{eqnarray}
  where $(i)$ uses Equation (\ref{EQ:pi1underfitted}) and Lemma
  \ref{LM:bestapp}. Besides, under $\mathcal{E}_n$, we have
  \begin{equation}
    \|(\tmmathbf{\phi}_{\tilde{\pi}^{\ast}}
    -\tmmathbf{\phi}_{\pi^{\ast}})\tmmathbf{\epsilon}\|_2 \leqslant C
    (k^{\dagger})^{1 / 2} \log^{1 / 2} (n) . \label{EQ:faiepsilonupper}
  \end{equation}
  Together with Assumption \ref{AS:hyperpara}, we conclude that
  \begin{eqnarray*}
    \tmmathbf{y}^T  (\tmmathbf{\phi}_{\pi^{\ast}}
    -\tmmathbf{\phi}_{\pi_0^{\ast}}) \tmmathbf{y} & \leqslant & - c n (k_0 -
    k) k_0^{- 1 - 2 / d} .
  \end{eqnarray*}
  Plugging the result into Equation (\ref{EQ:likelihoodratioexpre}), we have
  \begin{eqnarray*}
    \frac{\mathbb{P} (\tmmathbf{y} \mid \pi^{\ast}, \mathcal{X}, \mathcal{S})
    \lambda^{| \pi^{\ast} |}}{\mathbb{P} (\tmmathbf{y} \mid \pi^{\ast}_0,
    \mathcal{X}, \mathcal{S}) \lambda^{| \pi^{\ast}_0 |}} & \leqslant & \exp
    \left[ (k_0 - k) \left\{ \log (\lambda^{- 1}) + \frac{d \log (n + 1)}{2} -
    c k_0^{- 1 - 2 / d} n \right\} \right]\\
    & \overset{\text{Assumption \ref{AS:hyperpara}}}{\leqslant} & \exp (- c
    k_0^{- 1 - 2 / d} n) .
  \end{eqnarray*}
  \noindent\tmtextbf{Cases when} $\pi^{\ast} \in \Pi_2^{\ast}$:
  
  From Equation (\ref{EQ:Rsquaredif}), we can see that
  \[ \tmmathbf{y}^T  (\tmmathbf{\phi}_{\pi^{\ast}}
     -\tmmathbf{\phi}_{\pi_0^{\ast}}) \tmmathbf{y} \leqslant cn k_0 K^{- 1} +
     Ck^{\dagger} \log (n) . \]
  Thus, together with Assumption \ref{AS:hyperpara}, we have
  \begin{eqnarray*}
    \frac{\mathbb{P} (\tmmathbf{y} \mid \pi^{\ast}, \mathcal{X}, \mathcal{S})
    \lambda^{| \pi^{\ast} |}}{\mathbb{P} (\tmmathbf{y} \mid \pi^{\ast}_0,
    \mathcal{X}, \mathcal{S}) \lambda^{| \pi^{\ast}_0 |}} & \leqslant & \exp
    \left[ - \left\{ \log (\lambda^{- 1}) + \frac{d \log (n + 1)}{2} \right\}
    + cn k_0 K^{- 1} + Ck^{\dagger} \log (n) \right]\\
    & \overset{\text{Assumption \ref{AS:hyperpara}}}{\leqslant} & \exp [- c
    \log (\lambda^{- 1})] .
  \end{eqnarray*}
  
  \noindent\tmtextbf{Cases when} $\pi^{\ast} \in \Pi_3^{\ast}$:
  
  Using the similar arguments as cases of $\Pi_1^{\ast}$, it is easy to
  see that there exists a constant $c$, such that $\epsilon_1 (\pi^{\ast},
  \pi_0^{\ast}) K^2 > c (k_0 - k) k_0^{- 1} K^2$ holds for any $\pi^{\ast} \in
  \Pi_3^{\ast}$. Following the similar argument as cases of $\Pi_1^{\ast}$,
  we obtain
  \[ \begin{array}{ll}
       \frac{\mathbb{P} (\tmmathbf{y} \mid \pi^{\ast}, \mathcal{X},
       \mathcal{S}) \lambda^{| \pi^{\ast} |}}{\mathbb{P} (\tmmathbf{y} \mid
       \pi^{\ast}_0, \mathcal{X}, \mathcal{S}) \lambda^{| \pi^{\ast}_0 |}} &
       \leqslant
     \end{array} \exp (- c k_0^{- 1 - 2 / d} n) . \]
  \noindent\tmtextbf{Cases when} $\pi^{\ast} \in \Pi_4^{\ast}$:
  
  Write $\tilde{\pi}^{\ast} = \pi^{\ast} \cap \pi^{\ast}_0$, following the
  same arguments in (\ref{EQ:Rsquaredif}), we have
  \begin{eqnarray*}
    \tmmathbf{y}^T  (\tmmathbf{\phi}_{\pi^{\ast}}
    -\tmmathbf{\phi}_{\pi_0^{\ast}}) \tmmathbf{y} & \leqslant & cn k_0 K^{- 1}
    + Ck^{\dagger} \log (n) - \|(\tmmathbf{\phi}_{\tilde{\pi}^{\ast}}
    -\tmmathbf{\phi}_{\pi^{\ast}})\tmmathbf{y}\|^2_2\\
    & \leqslant & cn k_0 K^{- 1} + Ck^{\dagger} \log (n) -
    \{\|(\tmmathbf{\phi}_{\tilde{\pi}^{\ast}}
    -\tmmathbf{\phi}_{\pi^{\ast}})\tmmathbf{\mu}_0 \|_2
    -\|(\tmmathbf{\phi}_{\tilde{\pi}^{\ast}}
    -\tmmathbf{\phi}_{\pi^{\ast}})\tmmathbf{\epsilon}\|\}^2 .
  \end{eqnarray*}
  On the other hand, since $\pi^{\ast} \in \Pi_4^{\ast}$ and from Lemmas
  \ref{LM:bestapp} and \ref{LM:underfitted}, we can see
  \begin{eqnarray*}
    \|(\tmmathbf{\phi}_{\tilde{\pi}^{\ast}}
    -\tmmathbf{\phi}_{\pi^{\ast}})\tmmathbf{\mu}_0 \|_2 & \geqslant &
    \|(\tmmathbf{I}_n -\tmmathbf{\phi}_{\pi^{\ast}})\tmmathbf{\mu}_0 \|_2 -
    \|(\tmmathbf{\phi}_{\tilde{\pi}^{\ast}} -\tmmathbf{I}_n)\tmmathbf{\mu}_0
    \|_2\\
    & \geqslant & c n^{1 / 2} k_0^{1 / 2} K^{- 1 / 2} \log^{\alpha_0 / 2} (n)
    - Cn^{1 / 2} k_0^{1 / 2} K^{- 1 / 2},
  \end{eqnarray*}
  and
  \[ \|(\tmmathbf{\phi}_{\tilde{\pi}^{\ast}}
     -\tmmathbf{\phi}_{\pi^{\ast}})\tmmathbf{\epsilon}\|_2 \leqslant
     \|\tmmathbf{\phi}_{\tilde{\pi}^{\ast}} \tmmathbf{\epsilon}\|_2 +
     \|\tmmathbf{\phi}_{\pi^{\ast}} \tmmathbf{\epsilon}\|_2 \leqslant C
     (k^{\dagger})^{1 / 2} \log^{1 / 2} (n) \]
  under $\mathcal{E}_n$. According to Assumption \ref{AS:hyperpara} and using the
  fact that $0 < \alpha_{\dagger} \leqslant \alpha_b$, we have
  \[ n^{1 / 2} k_0^{1 / 2} K^{- 1 / 2} \log^{\alpha_0 / 2} (n) \gg
     (k^{\dagger})^{1 / 2} \log^{1 / 2} (n), \]
  leading to
  \[ \{\|(\tmmathbf{\phi}_{\tilde{\pi}^{\ast}}
     -\tmmathbf{\phi}_{\pi^{\ast}})\tmmathbf{\mu}_0 \|_2
     -\|(\tmmathbf{\phi}_{\tilde{\pi}^{\ast}}
     -\tmmathbf{\phi}_{\pi^{\ast}})\tmmathbf{\epsilon}\|_2 \}^2 \geqslant c
     n^{1 / 2} k_0^{1 / 2} K^{- 1 / 2} \log^{\alpha_0 / 2} (n) . \]
  Combining the results, we obtain
  \begin{eqnarray*}
    \tmmathbf{y}^T  (\tmmathbf{\phi}_{\pi^{\ast}}
    -\tmmathbf{\phi}_{\pi_0^{\ast}}) \tmmathbf{y} & \leqslant & cn k_0 K^{- 1}
    + Ck^{\dagger} \log (n) - c n k_0 K^{- 1} \log^{\alpha_0} (n)\\
    & \overset{}{\leqslant} & - c n k_0 K^{- 1} \log^{\alpha_0} (n) .
  \end{eqnarray*}
  Plugging this result into Equation (\ref{EQ:likelihoodratioexpre}), we have
  \begin{eqnarray*}
    \frac{\mathbb{P} (\tmmathbf{y} \mid \pi^{\ast}, \mathcal{X}, \mathcal{S})
    \lambda^{| \pi^{\ast} |}}{\mathbb{P} (\tmmathbf{y} \mid \pi^{\ast}_0,
    \mathcal{X}, \mathcal{S}) \lambda^{| \pi^{\ast}_0 |}} & = & \exp \left\{
    \frac{n\tmmathbf{y}^T (\tmmathbf{\phi}_{\pi^{\ast}}
    -\tmmathbf{\phi}_{\pi_0^{\ast}})\tmmathbf{y}}{2 \sigma^2 (n + 1)} \right\}
    \leqslant \exp \{ - c n k_0 K^{- 1} \log^{\alpha_0} (n) \} .
  \end{eqnarray*}
  Combining the results, the proposition is proved.
\end{proof}

\newpage
\bibliographystyle{apalike} 
\bibliography{ref}

\end{document}